\documentclass{article}
\usepackage{matharticle}
\usepackage{paralist}
\usepackage{fullpage}
\usepackage{placeins}

\newcommand{\fS}{\mathfrak{S}}

\newcommand{\Cov}{\mathrm{Cov}}
\newcommand{\Sgood}{S_g}
\newcommand{\Sbad}{S_b}
\newcommand{\wgood}{w_g}
\newcommand{\wbad}{w_b}
\newcommand{\wtilde}{{\widetilde{w}}}
\newcommand{\subsimplex}{{\Gamma}}
\newcommand{\Otilde}{\widetilde{O}}
\newcommand{\cN}{\mathcal N}
\newcommand{\e}{\epsilon}
\newcommand{\mper}{\; .}

\newcommand{\tmax}{\tau_{\max}}
\newcommand{\tO}{\widetilde{O}}
\usepackage{tikz}
\newcommand{\cX}{\mathcal{X}}
\newcommand{\algname}{\textsc{QUEScoreFilter} }
\newcommand{\sgalgname}{\textsc{s.g.-QUEScoreFilter} }

\newcommand{\QUE}{\textsc{QUE}}

\DeclareMathOperator{\Tr}{Tr}
\DeclareMathOperator{\Id}{Id}

\usepackage{float}
\usepackage{caption}
\usepackage{subcaption}

\title{Quantum Entropy Scoring for Fast Robust Mean Estimation and Improved Outlier Detection}

\author{Yihe Dong\thanks{\protect\url{yihe.dong@microsoft.com} Microsoft Research} \and Samuel B. Hopkins\thanks{\protect\url{hopkins@berkeley.edu} UC Berkeley, supported by a Miller Postdoctoral Fellowship.} \and Jerry Li\thanks{\protect\url{jerrl@microsoft.com} Microsoft Research. Part of this work was accomplished when SBH and JL were visiting the Simons Institute for Theory of Computing.} }

\begin{document}
\maketitle

\begin{abstract}
  We study two problems in high-dimensional robust statistics: \emph{robust mean estimation} and \emph{outlier detection}.
In robust mean estimation the goal is to estimate the mean $\mu$ of a distribution on $\R^d$ given $n$ independent samples, an $\e$-fraction of which have been corrupted by a malicious adversary.
In outlier detection the goal is to assign an \emph{outlier score} to each element of a data set such that elements more likely to be outliers are assigned higher scores.
Our algorithms for both problems are based on a new outlier scoring method we call \QUE-scoring based on \emph{quantum entropy regularization}.
For robust mean estimation, this yields the first algorithm with optimal error rates and nearly-linear running time $\tO(nd)$ in all parameters, improving on the previous fastest running time $\tO(\min(nd/\e^6, nd^2))$.
For outlier detection, we evaluate the performance of \QUE-scoring via extensive experiments on synthetic and real data, and demonstrate that it often performs better than previously proposed algorithms.
Code for these experiments is available at \url{https://github.com/twistedcubic/que-outlier-detection}.


\end{abstract}

\newpage
\setcounter{tocdepth}{1}
\tableofcontents

\newpage


\section{Introduction}
\label{sec:intro}

We study outlier-robust statistics in high dimensions, focusing on the question: \emph{can theoretically sound outlier robust algorithms have practical running times for large, high-dimensional data sets?}
We address two related problems: \emph{robust mean estimation}, which is primarily theoretical, and an applied counterpart, \emph{outlier detection}.

\textbf{Robust mean estimation} Our main theoretical contribution is the first nearly-linear time algorithm for robust mean estimation with nearly-optimal error.
Here the goal is to estimate the mean $\mu \in \R^d$ of a $d$-dimensional distribution $D$ given $\e$-\emph{corrupted} samples $X_1,\ldots,X_n$ -- that is, i.i.d. samples, an unknown $\e$-fraction of which have been maliciously corrupted.
Under (for instance) the assumption that the covariance of $D$ is bounded by $\Id$, it has been long known to be possible in exponential time to estimate $\mu$ by $\hat{\mu}$ having $\|\mu - \hat{\mu}\|_2 \leq O(\sqrt{\e})$.
In particular, this rate of error is independent of $d$.

Polynomial-time algorithms provably achieving such $d$-independent error became known only recently, starting with the works \cite{diakonikolas2019robust,lai2016agnostic}.
Until our work, the running time of algorithms with provably $d$-independent error remained suboptimal by polynomial factors in $d$ or $\e$: the fastest running time achieved before this work was $\tO(\min(n d^2, n d /\e^6))$ \cite{cheng2019high,diakonikolas2019robust,lai2016agnostic,diakonikolas2017being}.
(Here $\tO(\cdot)$ notation hides logarithmic factors in $n$ and $d$).
While these running times represent a dramatic improvement over previous exponential-time algorithms, there are still many interesting regimes where the additional runtime overheads these algorithms incur render them impractically slow.
\emph{We give the first algorithm for robust mean estimation with running time $\tO(nd)$ which achieves error $\|\mu - \hat{\mu} \|_2 \leq O(\sqrt \e)$.} Note that this running time is nearly-linear in the input size $nd$.
Similar to prior works, our algorithm has information-theoretically optimal sample complexity and nearly-optimal error rates in both the bounded-covariance and sub-Gaussian regimes.

\textbf{Outlier detection}
Our main applied contribution is a new algorithm for high-dimensional outlier detection, which we assess via experiments on both synthetic and real data \footnote{Code is available at \url{https://github.com/twistedcubic/que-outlier-detection}.}.
Our goal is to take a dataset $X_1,\ldots,X_n \in \R^d$ and assign to each $X_i$ an \emph{outlier score} $\tau_i \geq 0$, so that higher scores $\tau_i$ are assigned to points $X_i$ more likely to be outliers.
Of course, what constitutes an outlier varies across applications, so no single algorithm for outlier detection is likely to be the best in all domains.
We show that our method performs well in settings where individual outliers are difficult to pick out on their own (by, say, their $\ell_2$ norms or their distances to nearby points), but still collectively bias empirical statistics such as the mean and covariance.

We compare our method to baselines based on PCA and Euclidean distances, as well as more sophisticated algorithms from existing literature based on nearest-neighbor distances.
Our algorithm has nearly-linear running time in theory, and simple implementations in practice incur minimal overhead beyond standard spectral methods, allowing us to run on $1024$-dimensional data with no special optimizations and $8192$-dimensional data with a fast approximate implementation.
It can therefore be used in practice to complement existing approaches to outlier detection in exploratory data analysis.

\subsection{What is an outlier and why are they hard to find?}
\label{sec:props-of-outliers}
For us, an outlier is an element of a data set which was generated according to a different process than the majority of the data.
For instance, we may imagine that our samples $X_1,\ldots,X_n$ were sampled i.i.d. from a distribution $(1-\e) D + \e N$ over $\R^d$, where $D$ is the distribution of inliers, $N$ is the distribution of outliers, and $\e > 0$ is a small number -- that is, we imagine that a \emph{constant fraction} of our data may be outliers.

For this discussion, we also informally imagine that $N$ is sufficiently distinct from $D$ that the set of outliers could be approximately identified by brute-force search over subsets of $(1-\e)n$ samples, if given unlimited computational resources.
Otherwise, outlier detection is not a meaningful problem, and robust mean estimation is easy (because the empirical mean will be a good estimator).
Under these circumstances, what makes identifying outliers and estimating the mean in their presence difficult?
Chiefly:

\emph{Outliers may not be identifiable in isolation.}
  On its own, a typical outlier $X_i \sim N$ may look much like a typical inlier $X_j \sim D$.
  For instance, it could be $\|X_i\|_2 \approx \|X_j\|_2$, and $X_i,X_j$ may have similar distance to the nearest few neighboring samples, especially in high dimensions where samples are far apart.

\emph{Outliers still introduce bias, collectively}
  Even if individual outliers look innocuous, the collective effect a modified $\e$-fraction of samples $X_i$ can still substantially change the empirical distribution of $X_1,\ldots,X_n$.
  As a result, even simple statistical tasks like estimating the mean or covariance of $D$ require sophisticated estimators: naively pruning individual outliers and then employing standard empirical estimators typically leads to far-suboptimal error rates.
  For example, an $\e$-fraction of $X_1,\ldots,X_n$ which are all slightly biased in a single direction may shift the empirical mean of $X_1,\ldots,X_n$, but this bias will be difficult to detect by looking at small numbers of samples at once.
  This also demonstrates that successful outlier detection can require global geometric information about a high-dimensional dataset, such as whether or not a direction exists in which many (say, $\e n$) samples are unusually biased.

\emph{Outliers may be inhomogeneous.}
  Outliers need not exhibit unusual bias in only one direction, or all have the same norm, or lie in a single cluster.
  Rather, if a dataset exhibits several forms of corruption, there may be as many different-looking kinds of outliers.
  In the theoretical robust mean estimation setting, the adversary producing $\e$-corrupted samples may corrupt $\e n / 10$ samples by biasing them in some direction, another $\e n /10$ samples by unusually enlarging their norms, and so forth.

Since robust mean estimation involves a malicious adversary, all of the above phenomena must be addressed by our robust mean estimation algorithm.
In the empirical section of this paper, we focus on designing an outlier detection method suited to situations where at least one of them occurs -- in other cases, existing methods (such as those based on Euclidean norms or local neighborhoods of individual samples \cite{campos2016evaluation}) may be more appropriate.

\subsection{\QUE: Quantum Entropy Scoring}
Recent innovations in robust mean estimation \cite{lai2016agnostic,diakonikolas2019robust} rely on the following crucial observation about $\e$-corrupted samples $X_1,\ldots,X_n$ from a distribution $D$ with covariance $\Sigma \preceq \Id$.
Namely:
\emph{any subset $S \subseteq \{X_1,\ldots,X_n\}$ of samples which shift the empirical mean by distance more than $\sqrt{\e}$ in some direction $v$ also introduce an eigenvalue of magnitude greater than $1$ to the empirical covariance.}

In robust mean estimation, this leads to (amongst others) the \emph{filter} algorithm of \cite{diakonikolas2019robust,diakonikolas2017being}, one of the first to achieve dimension-independent error rates.
Roughly speaking, the algorithm iterates the following until the empirical covariance $\overline{\Sigma}$ has small spectral norm: (1) compute the top eigenvector $v$ of the empirical covariance of $\overline{\Sigma}$, then (2) throw out samples $X_i$ whose projections $|\iprod{X_i - \overline{\mu},v}| \gg 1$ is unusually large, where $\overline{\mu}$ is the empirical mean of the corrupted dataset.
For outlier detection this suggests a natural scoring rule -- let the outlier score $\tau_i$ of sample $X_i$ be proportional to $|\iprod{X_i - \overline{\mu},v}|$.

The main drawback of these algorithms is that they do not adequately account for inhomogeneity of outliers.
For the filter, this leads to a worst-case running time of $\tO(nd^2)$, because the filter operation (which can be implemented in $\tO(nd)$ time) may have to be repeated as many as $d$ times if the adversary introduces outliers lying in $d$ orthogonal directions.
The rule $\tau_i = |\iprod{X_i - \bar{\mu},v}|$ may miss outliers causing a large eigenvalue of $\overline{\Sigma}$, but in a direction orthogonal to the top eigenvector $v$.

In the opposite extreme, if outliers are maximally inhomogeneous -- no group of them is unusually biased in some shared direction $v$ -- then the only way they can bias the empirical mean is for the individual $\ell_2$ norms $\|X_i - \overline{\mu}\|_2$ to be larger than typical.
This suggests a different scoring rule: $\tau_i = \|X_i - \overline{\mu}\|_2$.
This approach, however, breaks down in the situation we started with, that groups of outliers are biased in a shared direction but they do not have larger norms than good samples.

\emph{Our main conceptual contribution is an approach to utilize information about outliers beyond what is available in the top eigenvector of the empirical covariance $\overline{\Sigma}$ and in individual $\ell_2$ norms.}
Appropriately adapted to their respective settings, this leads to our algorithms for both robust mean estimation and outlier detection.

Our first observation is that \emph{any eigenvalue/eigenvector $\lambda,v$ -- not just the top ones -- of the empirical covariance with $\lambda \gg 1$ must be due to outliers}.
We therefore consider the intermediate goal of finding a \emph{distribution} over directions $v \in \R^d$ containing information about as many outlier directions as possible.
We formalize this as the following entropy-regularized convex program over $d \times d$ positive semidefinite matrices:
\begin{align}
\label{eq:entropy}
\max_{U \in \R^{d \times d}} \alpha \cdot \iprod{U,\overline{\Sigma}} + S(U) \text{ such that } U \succeq 0, \tr(U) = 1 \; ,
\end{align}
where $\alpha \geq 0$ is some constant and $\iprod{A,B} = \tr( A B^\top)$ denotes the trace inner product of matrices.
Here, $S(U) = -\iprod{U, \log U}$ is the \emph{quantum entropy} (also known as the \emph{von Neumann entropy}) of the matrix $U$.
If $U = \sum_{i=1}^d \mu_i v_i v_i^\top$ is the eigendecomposition of $U$, since it has $\tr(U)= 1$ we may interpret it as a distribution over orthonormal vectors $v_1,\ldots,v_d$ with weights $\mu_1,\ldots,\mu_d$ and hence with entropy $S(U)$.
Under this interpetation, $\iprod{U,\overline{\Sigma}} = \E_{v_i \sim \mu} \iprod{v_i,\overline{\Sigma} v_i}$.
As $\alpha$ varies, \eqref{eq:entropy} trades off optimizing for a distribution supported on many distinct directions for a distribution supported on eigenvectors of $\overline{\Sigma}$ with large eigenvalues.
The optimizer of \eqref{eq:entropy} takes the form $U = \exp(\alpha \cdot \overline{\Sigma}) / \tr \exp (\alpha \cdot \overline{\Sigma})$ where $\exp(\cdot)$ is the matrix exponential function.
\begin{definition}
  Let $U = \exp(\alpha \cdot \overline{\Sigma})/ \tr \exp(\alpha \cdot \overline{\Sigma})$ be the optimizer of \eqref{eq:entropy}, for some data set $\cX = X_1,\ldots,X_n \in \R^d$ where $\overline{\Sigma}$ is the covariance of $\cX$.
  The quantum entropy (\QUE) scores with parameter $\alpha$ are given by  $\tau_i = (X_i - \overline{\mu})^\top U (X_i - \overline{\mu})$, where $\overline{\mu}$ is the mean of $\cX$.
\end{definition}
Intuitively, the \QUE{} scores will penalize any point which is causing a large eigenvalue in any direction, which should allow us to find more outliers than the naive spectral scores presented above.
\QUE{} scores also interpolate between two more naive scoring rules: when $\alpha = 0$ we have $U = \Id / d$ and so $\tau_i = \tfrac{1}{d} \|X_i - \overline{\mu}\|_2^2$ is the $\ell_2$ norm (up to a scaling), while when $\alpha \rightarrow \infty$ we have $U \rightarrow vv^\top$ where $v$ is the top eigenvector of $\overline{\Sigma}$, recovering naive spectral scoring.
\emph{In both experiments and theory we find that choosing $\alpha$ strictly between $0$ and $\infty$ outperforms either of the extreme choices.}

\QUE{} scores are also appealing from a computational perspective: we show that a list of approximate \QUE{} scores $\tau_i' = (1 \pm 0.01)\tau_i$ can be computed from $X_1,\ldots,X_n$ in nearly-linear time, by appropriate use of Johnson-Lindenstrauss sketching and efficient computation of the matrix exponential by series expansion.
This is crucial to both the nearly-linear running time of our algorithm for robust mean estimation and to the scalability of our outlier detection method.

In Section~\ref{sec:robust-mean} we describe refinements of \QUE{} scoring which fit it into the matrix multiplicative weights framework \cite{arora2012multiplicative}, leading to our nearly-linear time algorithms for robust mean estimation.
We give two very similar algorithms, one for when the distribution of inliers is only assumed to have bounded covariance, and one when the inliers are assumed to be subgaussian.
The resulting algorithms are conceptually similar to the following modification of the filter mentioned above: until $\|\overline{\Sigma}\|_2 \leq O(1)$, compute \QUE{} scores $\tau_i$, throw out data points $X_i$ with $\tau_i \gg 1$, and repeat.
(To obtain provable guarantees, our final algorithms are somewhat more complex: in some iterations we use \QUE{} scores based on certain reweightings of the data learned in previous iterations.)

In Section~\ref{sec:experiments} we describe experiments validating the \QUE{} scoring rule on both synthetic and real data sets.
We show that it performs especially well by comparison to local-neighborhood methods and to scoring based on only the top eigenvector in data sets where the inliers are close to isotropic (or can be made so by applying data whitening procedures) and in which there are heterogeneous outliers.

\subsection{Related work}

\textbf{Robust mean estimation:} The study of robust statistics and in particular robust mean estimation began with major works by Anscombe, Huber, Tukey and others in the 1960s \cite{anscombe1960rejection,tukey1960survey,huber1992robust,tukey1975mathematics}.
The literature on polynomial-time algorithms for robust statistics has exploded in recent years, following works by Diakonikolas et al and Lai, Rao and Vempala giving the first polynomial-time algorithms for robust mean estimation with dimension-independent (or nearly dimension-indepedent) error \cite{diakonikolas2019robust,lai2016agnostic}.
A full survey is beyond our scope here -- see e.g. the recent theses \cite{li2018principled,steinhardt2018robust} for a thorough account.
Particularly relevant to our work is the recent work of Cheng, Diakonikolas, and Ge who design an algorithm for robust mean estimatin with running time $\tO(nd/\e^6)$ -- the first to achieve nearly linear time for constant $\e$ -- by appeal to nearly linear time solvers for packing and covering semidefinite programs \cite{cheng2019high}.
Our algorithms carry two advantages over this prior work: first, our algorithm runs in nearly linear time for any choice of $\e = \e(n,d)$, and second, because we avoid the $1/\e^6$ scaling and appeal to semidefinite programming, our theoretical ideas lead to a practical method for outlier detection.
The techniques of Diakonikolas et al. were later extended to robust covariance estimation \cite{cheng2019faster}; it remains an interesting direction to extend our techniques to covariance estimation.

\emph{Concurrent work:}
After this manuscript was initially submitted, we became aware of the concurrent work \cite{lecue2019robust}, which also obtains a nearly-linear time algorithm for robust mean estimation of distributions with bounded covariance.
The algorithm of \cite{lecue2019robust} also obtains \emph{subgaussian confidence intervals} (see e.g. \cite{lugosi2019sub}), which the algorithm in this work does not.
By contrast, the algorithms in our work also obtain improved rates of error with respect to $\e$ when the underlying distribution is sub-Gaussian, and our method is sufficiently practical that we are able to implement parts of it to run our experiments on outlier detection.
(The method of \cite{lecue2019robust} relies on nearly-linear time solvers for packing/covering semidefinite programs, which are not yet practical.)
Finally, implicit in the work \cite{lecue2019robust} is a reduction from arbitrary $\e$ to the case $\e = 1/100$; we describe this reduction and some consequences in Appendix~\ref{app:reduction}.

\textbf{Outlier detection}
Detection of outliers goes back nearly to the beginning of statistics itself \cite{hawkins1980identification}.
Even restricting to the high dimensional case it has a literature too broad to survey here.
Much recent work has focused on so-called \emph{local outlier factor}-based methods, which assign outlier scores based on the local density of other samples near each $X_i$ -- see e.g. \cite{knorr1997unified,knox1998algorithms} and further references in \cite{campos2016evaluation}.
We find that QUE scoring compares favorably to such local methods in high-dimensional datasets like we describe in Section~\ref{sec:props-of-outliers} -- see Sections~\ref{sec:experiments} and~\ref{sec:local-methods} for details.


\subsection{Robust mean estimation: results and algorithm overview}
\label{sec:robust-mean}

We turn to our algorithm for robust mean estimation, deferring details to Sections~\ref{sec:prelims}---\ref{sec:subgaussian}.

\begin{definition}[$\e$-corrupted samples]
  Let $D$ be a distribution on $\R^d$.
  We say that $X_1,\ldots,X_n$ are an $\e$-corrupted set of samples from $D$ if they are first drawn i.i.d. from $D$, then modified by an adversary who may adaptively inspect all the samples, remove $\e n$ of them, and replace them with arbitrary vectors in $\R^d$.
\end{definition}
Note that $\e$-corruption is a \emph{stronger} outlier model than the $(1-\e)D + \e N$ mixture model we described in Section~\ref{sec:intro}; our algorithms also work in this milder mixture model.
Our main theoretical result is:
\begin{theorem}
\label{thm:main-intro}
  For every $n,d \in \N$ and $\e > 0$ there are algorithms \algname,\sgalgname with running time $\Otilde (n d)$, such that for every distribution $D$ on $\R^d$ with mean $\mu$ and covariance $\Sigma$, given $n$ $\e$-corrupted samples from $D$, \algname produces $\hat{\mu}$ such that $\|\hat{\mu} - \mu\|_2 \leq O(\sqrt{\e}) + \Otilde(\sqrt{d/n})$ if $\Sigma \preceq \Id$, and \sgalgname produces $\hat{\mu}$ such that $\|\hat{\mu} - \mu\|_2 \leq O(\e \sqrt{\log(1/\e)} + \sqrt{d/n})$ if $D$ is sub-Gaussian with $\Sigma = \Id$, all with probability at least $0.99$.
\end{theorem}
For the bounded covariance case, the $O(\sqrt \e)$ term information-theoretically optimal up to constant factors.
The other term, $\tO(\sqrt{d/n})$, is information-theoretically optimal up to the logarithmic factors in the $\tO(\cdot)$ even without corruptions.
For the sub-Gaussian case, the $O(\eps \sqrt{\log 1/\e})$ term is believed to be necessary for computationally efficient algorithms (see e.g the statistical-query lower bound \cite{diakonikolas2017statistical}), although that term can be made $O(\e)$ by using computationally-intractable estimators such as Tukey median, and the latter is information-theoretically optimal \cite{tukey1975mathematics}.
The $\sqrt{d/n}$ term is information-theoretically optimal even without corruptions.

In this section we discuss our algorithm for the bounded-covariance case $\Sigma \preceq \Id$ in the setting that the adversary may not remove samples, leaving technical details and the modifications necessary to handle removed samples and sub-Gaussian $D$ to Section~\ref{sec:subgaussian}.
\begin{definition}[Simplified robust mean estimation]
\label{def:robust-mean-simplified}
  Let $S = \{X_1,\ldots,X_n\} \subseteq \R^d$ be a dataset with the property that $S$ partitions into $S = S_g \cup S_b$ with $|S_b| \leq \e n$ and $\E_{i \sim S_g} (X_i - \mu_g)(X_i - \mu_g)^\top \preceq \Id$, where $\mu_g = \E_{i \sim S_g} X_i$.
  Given $S$, the goal is to find a vector $\hat{\mu}$ with $\|\mu_g - \hat{\mu}\|_2 \leq O(\sqrt \e)$.
\end{definition}

Like prior algorithms for robust mean estimation, ours maintains a \emph{weight vector} $w_1,\ldots,w_n \geq 0$ with $\sum w_i \leq 1$, initialized to $w_i = 1/n$.
The algorithm iteratively decreases the weight of points suspected to be outliers that are causing $\|\mu(w) - \mu_g\|_2$ to be large.\footnote{Some prior algorithms, e.g. the \emph{filter} of \cite{diakonikolas2019robust} instead iteratively throw out points suspected to be outliers. However, since those algorithms are (necessarily) randomized, they can also be viewed as weighting points, where the weight of $X_i$ is the probability it has not been thrown out. The algorithm we present here can also be implemented by throwing out points in a randomized fashion -- we discuss further in the appendix.}
A key insight of recent work on robust mean estimation is that it suffices to find weights $w$ which place almost as much mass on $S_g$ as does the uniform weighting and whose empirical covariance is small.
This is formalized in the following lemma.
For a weight vector $w$, let $|w| = \sum w_i$, $\mu(w) = \tfrac 1 {|w|} \sum w_i X_i$, and $M(w) = \tfrac 1 {|w|} \sum w_i (X_i - \mu(w))(X_i - \mu(w))^\top$.
Let $\| M \|_2$ be the spectral norm of a matrix $M$.
\begin{lemma}[Implicit in prior work]
\label{lem:geom-main}
  Let $S = \{X_1,\ldots,X_n\}$ be as in Definition~\ref{def:robust-mean-simplified}.
  Suppose that $w$ is a weight vector such that $\|M(w)\|_2 \leq O(1)$ and $w$ is \emph{mostly good}, by which we mean $|\tfrac 1 n \textbf{1}_{S_g} - w_g| \leq |\tfrac 1 n \textbf{1}_{S_b} - w_b|$, where $\textbf{1}_{S_g},\textbf{1}_{S_b}$ are the indicators of $S_g,S_b$ and $w_g,w_b$ are $w$ restricted to $S_g,S_b$ respectively.
  (Intuitively, $w$ is mostly good if it results by removing from the uniform weighting $\textbf{1}_S/n$ more weight from $S_b$ than from $S_g$.)
  Then $\|\mu(w) - \mu_g\|_2 \leq O(\sqrt \e)$.
\end{lemma}
Lemma~\ref{lem:geom-main} captures the following geometric intuition: if the bad points $S_b$ receive enough weight in $w$ to cause $\|\mu(w) - \mu_g\|_2 \gg \sqrt{\e}$, then an $O(\e)$-fraction of the mass of $w$ is on $X_i$ which are unusually correlated with the vector $\mu(w) - \mu_g$, which leads to a large maximum eigenvalue in $M(w)$.
Prior works employ a variety of methods to find a mostly good weight vector $w$ with $\|M(w)\|_2 \leq O(1)$.
Perhaps the simplest is the \emph{filter} of \cite{diakonikolas2019robust}, which iterates:
While $\|M(w)\|_2 \gg 1$, compute its top eigenvector $v$ and \emph{naive spectral scores} $\tau_i = \iprod{X_i - \mu(w),v}^2$.
Throw out $X_i$ with large $\tau_i$ and repeat.

The filter ensures that the weight vector it maintains is mostly good because (in an averaged sense) $\tau_i$ can be large only for $X_i$ which are corrupted.
This is because the (weighted) sum of all scores $\sum w_i \tau_i = \iprod{M(w), vv^\top} \gg 1$, while the contribution to this sum from $S_g$ has $\sum_{i \in S_g} w_i \tau_i \approx \iprod{\tfrac 1 n  \sum_{i \in S_g} (X_i - \mu_g)(X_i - \mu_g)^\top, vv^\top} \leq 1$.
(Here we ignore some details about centering $X_i$ at $\mu_g$ rather than $\mu(w)$.)
Thus, the $\tau_i$ from $S_b$ must make up almost all of $\sum w_i \tau_i$.
Simple approaches to removing or downweighting $X_i$ with large $\tau_i$ then remove strictly more weight from $S_b$ than from $S_g$.

However, filtering based on naive spectral scores alone faces a barrier to achieving nearly-linear running-time.
If the corruptions $S_b$ are split among many orthogonal directions, the naive spectral filter will have to find those directions one at a time.
Thus, it may require $\Omega(d)$ iterations (leading to $\Omega(nd^2)$ running time) to arrive at $w$ with $\|M(w)\|_2 \leq O(1)$.

Our main idea is that by replacing naive spectral scores with slightly modified \QUE{} scores, each iteration of the filter can take into account projections of each sample onto many large eigenvectors of $M(w)$.
We show that our modified \QUE{} scores $\tau_i$ maintain the property that $\sum_{i \in S_b} w_i \tau_i \gg \sum_{i \in S_g} w_i \tau_i$, and so downweighting according to $\tau_i$ removes more mass from $S_b$ than $S_g$.
However, filtering with \QUE{} scores makes faster progress than with naive spectral scores: roughly speaking, we show that only $O(\log d)^2$ rounds of filtering according to \QUE{} scores are required to find a mostly-good weight vector $w$ with $\|M(w)\|_2 \leq O(1)$.

The core of our algorithm is a subroutine, \textsc{DecreaseSpectralNorm}, to take a mostly good weight vector $w$ with $\|M(w)\|_2 \gg 1$ and in $O(\log d)$ rounds of \QUE{} filtering produce another mostly good $w'$ with $\|M(w')\|_2 \leq \tfrac 3 4 \|M(w)\|_2$.
Repeating this subroutine $O(\log d)$ times and then outputting the resulting $\mu(w)$ yields our main algorithm.
An outline of this subroutine is presented as Algorithm~\ref{alg:mmw-filter-main}.
We first establish a rigorous sense in which downweighting according to outlier scores $\tau_i$ makes progress: \emph{it decreases the weighted average of the scores while removing more weight from bad points than good}.
\begin{lemma}[Progress in one round of downweighting, informal]
\label{lem:1dfilter-main}
There is a downweighting algorithm which takes a density matrix $U$ and a mostly good weight vector $w$ and produces a mostly good weight vector $w'$ by downweighting points with large score $\tau_i = \iprod{X_i -\mu(w), U (X_i - \mu(w)}$ such that $\sum w_i' \tau_i \leq \tfrac 1 3 \sum w_i \tau_i$ so long as $\sum w_i \tau_i \gg 1$.
Furthermore, $M(w') \preceq M(w)$.
\end{lemma}
Let us give a geometric interpretation to Lemma~\ref{lem:1dfilter-main}: it establishes that if $\sum w_i \tau_i = \iprod{U,M(w)} \gg 1$ then the quadratic form of $M(w')$ decreases \emph{in the directions defined by $U$}, since
\begin{align}
\label{eq:progress}
\iprod{M(w'),U} \approx \sum w_i' \tau_i \leq \frac 1 3 \sum w_i \tau_i = \frac 1 3 \iprod{M(w),U}\mper
\end{align}
This guarantee becomes more meaningful as the entropy $S(U)$ increases, because it suggests the quadratic form of $M(w)$ has decreased in more directions.
To make this formal, we appeal to the matrix multiplicative weights framework.
\textsc{DecreaseSpectralNorm} applies downweighting iteratively using a sequence of entropy-maximizing density matrices $U_1,\ldots,U_{T}$ chosen according to the matrix multiplicactive weights update rule, leading to a series of mostly good weight vectors $w_1,\ldots,w_T$ such that $\|M(w_T)\|_2 \leq \tfrac 3 4 \|M(w_0)\|_2$.
We choose
\begin{align}
\label{eq:mmw-main}
U_t = \exp \Paren{\frac 1 {\|M(w)\|_2} \sum_{k=0}^{t-1} M(w_k)} \bigg/ \tr \exp \Paren{\frac 1 {\|M(w)\|_2} \sum_{k=0}^{t-1} M(w_k)} \; ,
\end{align}
where $w_0 = w$ is the input weight vector, $U_0 = \Id$, and $w_t$ results from applying the downweighting of Lemma~\ref{lem:1dfilter-main} to $w_{t-1}$ using $U_{t}$ (if $\iprod{M(w_{t-1}),U_{t}} \gg 1$).
The following lemma is a special case of the standard (local norm) \emph{regret bound} for matrix multiplicative weights.

\begin{lemma}[Special case of Theorem 3.1, \cite{allen2015spectral}]
\label{lem:mmw-main}
  For any $w_0,\ldots,w_T$, if $\alpha \leq 1/\|M(w_t)\|_2$ for all $t \leq T$, then
  \begin{align}
  \label{eq:mmw-main}
  \Norm{\sum_{t=0}^{T-1} M(w_t)}_2 \leq \sum_{t=0}^{T-1} \iprod{U_t, M(w_{t})} + \alpha \sum_{t=0}^{T-1} \iprod{U_t, M(w_{t})} \cdot \|M(w_{t})\|_2 + \frac{\log d}{\alpha}\mper
  \end{align}
\end{lemma}

Now we sketch the analysis of \textsc{DecreaseSpectralNorm}.
\begin{claim}[Informal]
If $w = w_0$ is mostly good, with $\|M(w_0)\|_2 \geq 100$, then \textsc{DecreaseSpectralNorm} produces mostly good $w_T$ with $\|M(w_T)\|_2 \leq \tfrac 3 4 \|M(w)\|_2$.
\end{claim}
\begin{proof}[Proof sketch]
Since $M(w_t) \preceq M(w_{t+1})$ by Lemma~\ref{lem:1dfilter-main}, we have $\|M(w_t)\|_2 \leq \|M(w_0)\|_2$ for all $t$, and hence $\alpha = 1/\|M(w_0)\|_2 \leq 1/\|M(w_t)\|_2$ for all $t$, so $w_0,\ldots,w_T$ and $U_0,\ldots,U_{T-1}$ satisfy the hypotheses of Lemma~\ref{lem:mmw-main}.
By our choice of $\alpha$ and $M(w_T) \preceq M(w_t)$ for all $t$, \eqref{eq:mmw-main} implies
\[
  T \cdot \|M(w_T)\|_2 \leq \Norm{\sum_{t=0}^{T-1} M(w_t)}_2 \leq 2 \sum_{t=0}^{T-1} \iprod{U_t,M(w_{t})} + \|M(w_0)\|_2 \cdot \log d\mper
\]
If $\iprod{U_t,M(w_{t-1})} \geq \|M(w_0)\|_2 /3 \gg 1$, then \textsc{DecreaseSpectralNorm} performs downweighting, and by Lemma~\ref{lem:1dfilter-main} and \eqref{eq:progress} (which we establish rigorously in supplemental material), $\iprod{M(w_{t}),U_t} \leq \tfrac 1 3 \iprod{M(w_{t-1}),U_t} \leq \tfrac 1 3 \|M(w_0)\|$.
Otherwise, by hypothesis $\iprod{M(w_{t}),U_t} = \iprod{M(w_{t-1}),U_t} \leq \|M(w_0)\|_2 /3$.
Using this bound and dividing by $T$, we obtain $\|M(w_T)\|_2 \leq (\tfrac 2 3 + \tfrac{\log d}{T}) \|M(w_0)\|_2$.
Choosing $T \geq 20 \log d$ completes the proof sketch.
\end{proof}

\emph{Running time:}
Our overall algorithm only requires $\log(nd)^{O(1)}$ iterations of \textsc{DecreaseSpectralNorm}, and the latter only requires $O(\log (d))$ iterations of downweighting, so we just have to implement downweighting in nearly-linear time.
We show in supplemental material that this can be done by avoiding representing any of the matrices $U_t$ explicitly in memory: instead, we maintain only low-rank sketches of them.
This leads to some approximation error in computing the \QUE{} scores, but we show that approximations to the \QUE{} scores suffice for all arguments above.

For remaining technical details and full proofs, see Sections 5-9 of supplemental materials.



\begin{algorithm}[ht]
\caption{\textsc{DecreaseSpectralNorm}}
\label{alg:mmw-filter-main}
  \begin{algorithmic}[1]
  \STATE \textbf{Input:} $X_1,\ldots,X_n$ as in Definition~\ref{def:robust-mean-simplified}, mostly good weight vector $w_0$.
  \STATE For iteration $t = 0, \ldots, O(\log d)$, if $\|M(w_t)\|_2 \leq \tfrac 3 4 \|M(w_0)\|_2$, output $w_t$ and halt. Otherwise, let $U_t$ as in \eqref{eq:mmw-main}. If $\iprod{U_t,M(w_{t-1})} \leq \tfrac 1 3 \|M(w_0)\|_2$, let $w_{t+1} = w_t$. Else let $w_{t+1}$ be the output of downweighting from Lemma~\ref{lem:1dfilter-main} with $U_t$.
  \STATE Output $w_T$.
  \end{algorithmic}
\end{algorithm}


\subsection{Outlier detection: algorithm and experimental results}
\label{sec:experiments}

In this section, we empirically evaluate outlier detection using \QUE{} scoring.
We must work with data containing well-defined and known inliers and outliers so that we can compare our results to ground-truth.
We generate such data sets in three distinct ways, leading to three main experiments.
(In supplemental material we also study some outlier-detection data sets appearing in prior work \cite{campos2016evaluation}.)

\emph{Synthetic:} We create synthetic data sets in $128$ dimensions and $10^3-10^4$ samples with an $\e$-fraction of inhomogeneous outliers in $k$ directions by sampling from a mixture of $k+1$ Gaussians $(1-\e) \cN(0,\Id) + \sum_{i=1}^k \e_i [\tfrac 1 2 \cN(C \sqrt{k/\e} \cdot e_i, \sigma^2 \Id)+ \tfrac 12 \cN(-C \sqrt{k/\e} \cdot e_i, \sigma^2 \Id)]$, where $e_1,\ldots,e_k$ are standard basis vectors, with $C \approx 1$ and $\sigma \ll 1$.
  The outliers are the samples from $\cN(\pm C \sqrt{k/\e} e_i, \sigma^2 \Id)$.
  By varying $\e,k$ and the distribution $\e_1,\ldots,\e_k$ of outlier weights, we demostrate in this simplified model how max-entropy outlier scoring improves on baseline algorithms in the presence of inhomogeneous outliers.
  We choose the scaling $\sqrt{k/\e} \cdot e_i$ because then standard calculations predict that if $\e_i \approx \e/k$ the outliers from $\cN(\pm C \sqrt{k/\e} e_i,\sigma^2 \Id)$ will contribute an eigenvalue greater than $1$ to the overall empirical covariance.

\emph{Mixed -- word embeddings:} We create a data set consisting of word embeddings drawn from several sources.
  Inliers are the $100$-dimensional GloVe embeddings (\cite{pennington2014glove}) of the words in a random $\approx 10^3$ word long section of a novel (we use \emph{Sherlock Holmes}) and outliers are embeddings of the first paragraphs of $k$ featured Wikipedia articles from May 2019 \cite{wikipedia_2013}.

\emph{Perturbed -- images:} We create a data set consisting of CIFAR10 images some of which have artificially-introduced \emph{dead pixels}.
  Inliers are $\approx 4500$ random CIFAR images $X \in \{1,\ldots,256\}^{1024}$ (restricted to the red color channel).
  Outliers are $\approx 500$ random CIFAR images, partitioned into groups $S_1,\ldots,S_k$, such that for each group $i$ a random coordinate $p_i \in \{1,\ldots,1024\}$ and a random value $c_i \in \{1,\ldots,256\}$ is chosen and for each $X \in S_i$ we set $X_{p_i} = c_i$.

\textbf{Metric:} All the methods we evaluate produce a vector of \emph{scores} $\tau_1,\ldots,\tau_n \in \R$.
We use the standard \emph{ROCAUC} metric to compare these scores to a ground-truth partition $S = S_g \cup S_b$ into inlier and outlier sets.
$\text{ROCAUC}(\tau_1,\ldots,\tau_n,S_b,S_g) = \Pr_{i \sim S_b, j \sim S_g}(\tau_i \geq \tau_j)$ is simply the probability that a randomly chosen outlier is scored higher than a random inlier.

\textbf{Baselines:} We compare \QUE{} scoring to the following other scoring rules. $\ell_2$: $\tau_i = \|X_i - \overline{\mu}\|$ is the distance of $X_i$ to the empirical mean; \emph{top eigenvector naive spectral:} $\tau_i = \iprod{X_i - \overline{\mu}, v}^2$ where $v$ is the top eigenvector of the empirical covariance; \emph{$k$-nearest neighbors (k-NN)} \cite{knn2000,campos2016evaluation} and \emph{local outlier factor (LOF)} \cite{lof2000,campos2016evaluation} methods: $\tau_i$ is a function of the distances to its $k$ nearest neighbors; \emph{isolation forest and elliptic envelope:} standard outlier detection methods as implemented in scikit-learn \cite{rousseeuw1999fast, liu2008isolation, pedregosa2011scikit}.

\textbf{Whitening:}
Scoring methods based on the projection of data points $X_i$ onto large eigenvectors of the empirical covariance work best when those eigenvectors correspond to directions in which many outliers lie.
In particular, if $\Sigma_g$, the covariance of $S_g$, itself has large eigenvalues then such spectral methods perform poorly.
We assume access to a \emph{whitening transformation} $W \in \R^{d \times d}$, which captures a small amount of prior knowledge about the distribution of inliers $S_g$.
For best performance $W$ should approximate $W^* = (\Sigma_g)^{-1/2}$ since $W^* X_i$ form an isotropic set of vectors.
Of course, to compute $W^*$ exactly would require knowing which points are inliers, but we find that relatively naive approximations suffice.
In particular, if a clean dataset $Y_1,\ldots,Y_m$ whose distribution is similar to the distribution of inliers is available, its empirical covariance can be used to find a good whitening transformation $W$.
In our synthetic data we use $W = \Id$.
In our word embeddings experiment, we obtain $W$ using the empirical covariance of the embedding of another random section of Sherlock Holmes.
In our CIFAR-10 experiment, we obtain $W$ from the empirical covariance of a fresh sample of $\approx 5000$ randomly chosen images from CIFAR-10.

\begin{algorithm}[]
\caption{\QUE-Scoring for Outlier Detection}
  \label{alg:max-entropy}
  \begin{algorithmic}[1]
  \STATE \textbf{Input:} dataset $X_1,\ldots,X_n \in \R^d$, optional whitening transformation $W \in \R^{d \times d}$, scalar $\alpha > 0$.
  \STATE Let $X_i' = W X_i$ be whitened data, $\overline{\mu} = \tfrac 1 n \sum_{i=1}^n X_i'$ and $\overline{\Sigma} = \tfrac 1 n \sum_{i=1}^d (X_i' - \overline{\mu})(X_i' - \overline{\mu})^\top$.
  \STATE For $i \leq n$, let $\tau_i = ({X_i'}^\top \exp(\alpha \overline{\Sigma} / \|\overline{\Sigma}\|_2) X_i') / \Tr \exp(\alpha \overline{\Sigma} / \|\overline{\Sigma}\|_2)$. Return $\tau_1,\ldots,\tau_n$.
  \end{algorithmic}
  \emph{Note on $\alpha$:} in both synthetic and real data we find that $\alpha = 4$ is a good rule-of-thumb choice, consistently resulting in improved scores over baseline methods.
\end{algorithm}

\textbf{High-dimensional scaling:} Implementing Algorithm~\ref{alg:max-entropy} by explicitly forming the matrix $\overline{\Sigma}$ and performing a singular value decomposition (SVD) to compute $\exp(\alpha \overline{\Sigma})$ is feasible on relatively low-dimensional data ($d \approx 100$). See Section~\ref{sec:local-methods} for discussion and results of a nearly-linear time implementation.

\begin{figure}[ht]
\label{fig:vs-tau0}
	\centering
	\begin{subfigure}{.3\textwidth}
		\includegraphics[width=\textwidth]{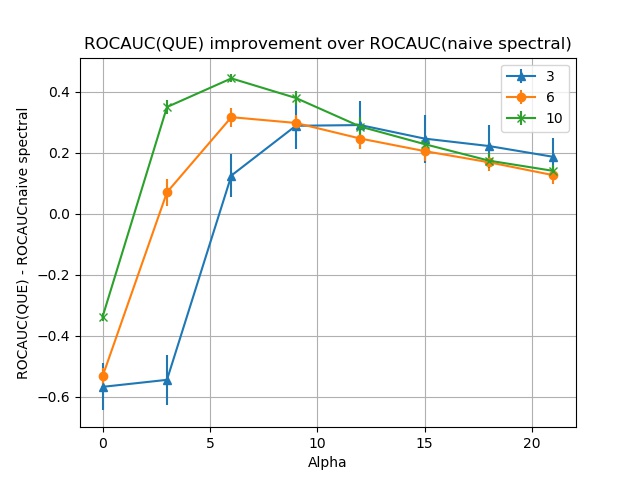}
		\caption{synthetic}
	\end{subfigure}
	\begin{subfigure}{.3\textwidth}
		\includegraphics[width=\textwidth]{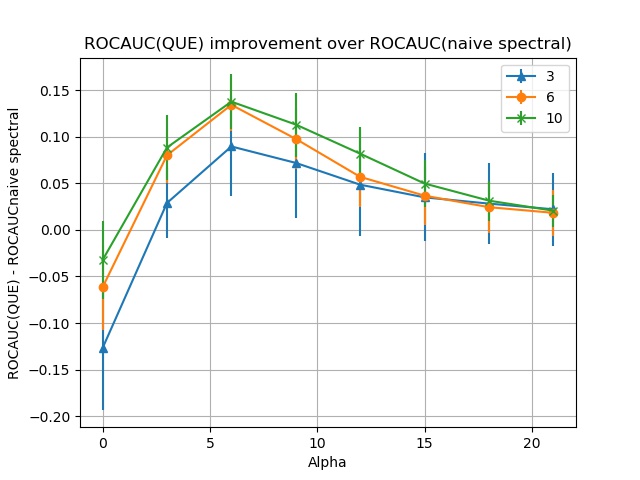}
		\caption{whitened CIFAR-10}
	\end{subfigure}
	\begin{subfigure}{.3\textwidth}
		\includegraphics[width=\textwidth]{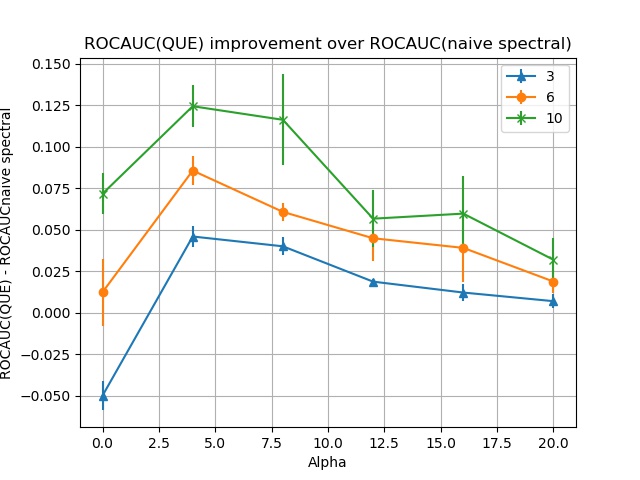}
		\caption{whitened word embeddings}
	\end{subfigure}
	\begin{subfigure}{.3\textwidth}
		\includegraphics[width=\textwidth]{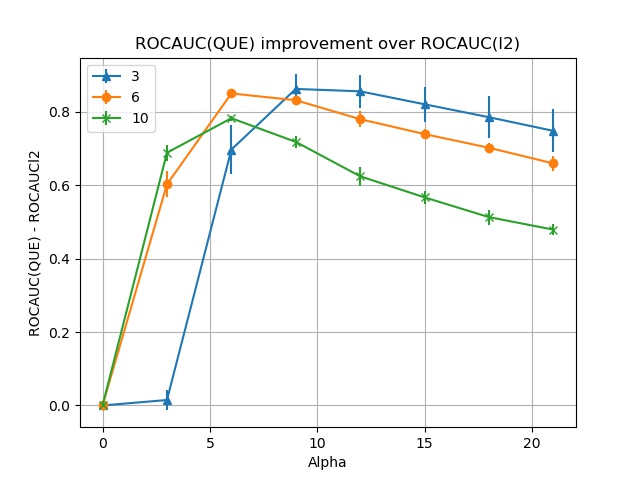}
		\caption{synthetic}
	\end{subfigure}
	\begin{subfigure}{.3\textwidth}
		\includegraphics[width=\textwidth]{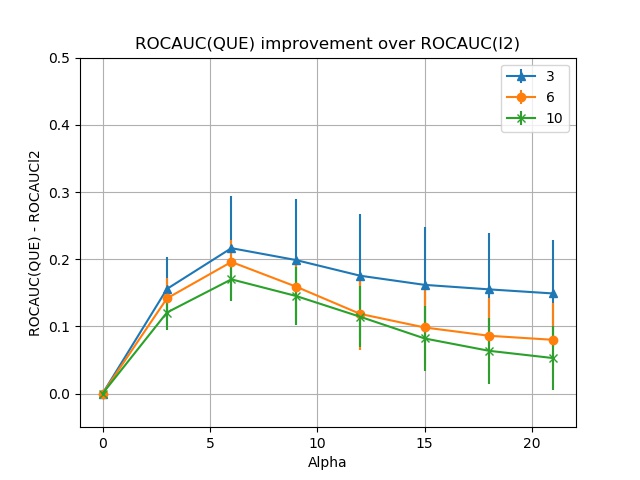}
		\caption{whitened CIFAR-10}
	\end{subfigure}
	\begin{subfigure}{.3\textwidth}
		\includegraphics[width=\textwidth]{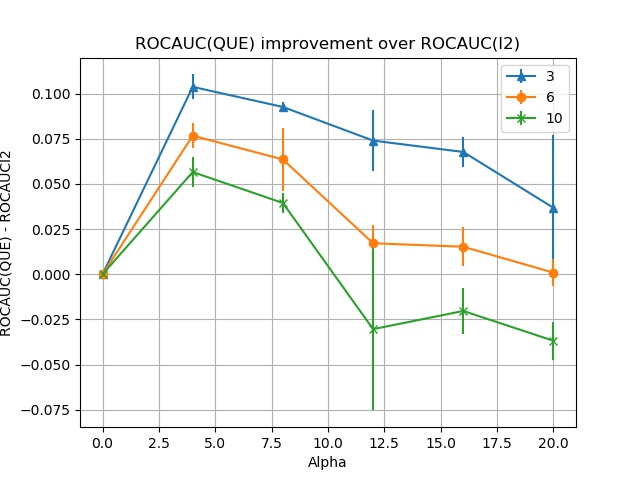}
		\caption{whitened word embeddings}
	\end{subfigure}
	\begin{subfigure}{.3\textwidth}
		\includegraphics[width=\textwidth]{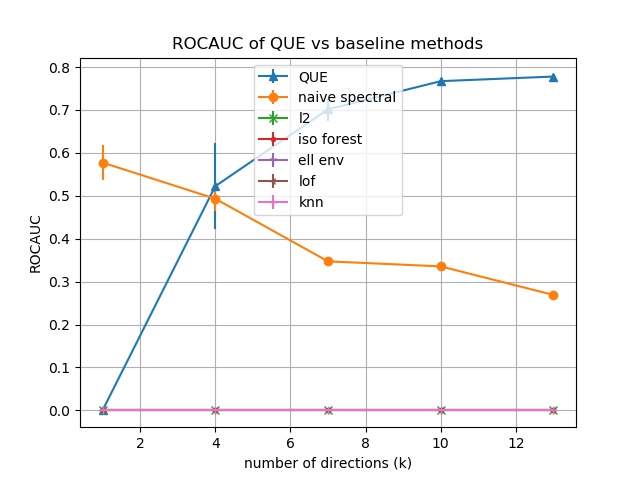}
		\caption{synthetic}
	\end{subfigure}
	\begin{subfigure}{.3\textwidth}
		\includegraphics[width=\textwidth]{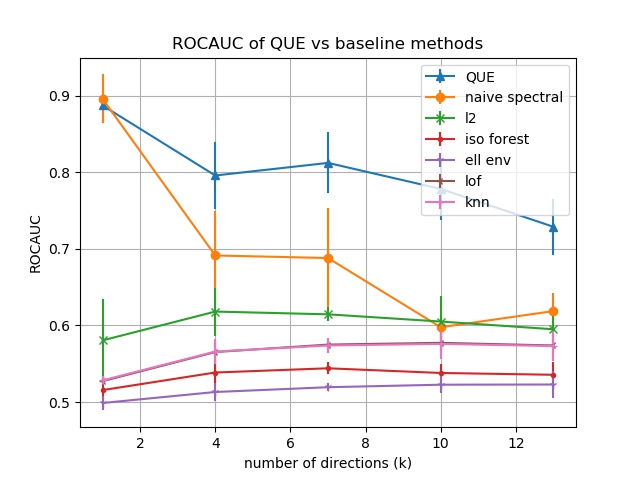}
		\caption{whitened CIFAR-10}
	\end{subfigure}
	\begin{subfigure}{.3\textwidth}
		\includegraphics[width=\textwidth]{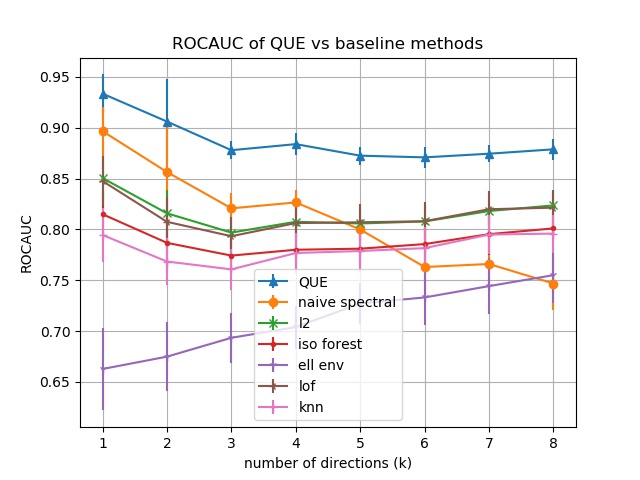}
		\caption{whitened word embeddings}
	\end{subfigure}
        \caption{\textbf{(a-f)}: We plot the difference between ROCAUC performance of \QUE{} and naive spectral (a-c), $\ell_2$ scoring (d-f) on all three data sets, as $\alpha$ varies. Error bars represent one empirical standard deviation in 20 trials. Note that in all three cases the mean improvement in ROCAUC score given by \QUE{} is at least one standard deviation above $0$ for a wide range of $\alpha$. Observe also that in synthetic data (which most closely parallels theory) the optimal $\alpha$ decreases with increasing number of outlier directions, in accord with the need to find a higher-entropy solution to \eqref{eq:entropy}. \textbf{(g-i)} We plot ROCAUC scores of \QUE{} (with $\alpha = 4$) and a variety of other methods as the number of outlier directions increases. Error bars represent one standard deviation over $3-4$ trials. Number of trials is small due to large running time requirements of Scikit-learn methods IsolationForest and EllipticEnvelope. The methods "lof" and "knn" are based on nearest-neighbor distances \cite{campos2016evaluation}. All except spectral methods perform poorly on synthetic data; as $k$ increases the performance gap between \QUE{} and naive spectral scoring grows. In all plots $\e = 0.2$. Experiments were generated on a quad-core $2.6$Ghz machine with 16GB RAM and an NVIDIA P100 GPU.}
\end{figure}

\FloatBarrier

\section{Robust mean estimation: results and preliminaries}
\label{sec:prelims}
For two functions $f, g$, we say $f = \Otilde(g)$ if $f = O(g \log^c g)$ for some universal constant $c > 0$.
We similarly define $\widetilde{\Omega}$ and $\widetilde{\Theta}$.
For vectors $v \in \R^d$, we let $\norm{\cdot}_2$ denote the usual $\ell_2$ norm, and $\Iprod{\cdot, \cdot}$ denote the usual inner product between vectors.
Let $\one_m \in \R^m$ denote the $m$-dimensional all-ones vector.

For matrices $A, M \in \R^{d \times d}$ we let $\Norm{M}_2$ denote its spectral norm, we let $\Norm{M}_F$ denote its Frobenius norm, and we let $\Iprod{M, A} = \tr (M^\top A)$ denote the trace inner product between matrices.
For any symmetric matrix $A \in \R^{d \times d}$, let $\exp (A)$ denote the usual matrix exponential of $A$.
Finally, for scalars $x, y \in \R$, and any $\alpha > 0$, we say that $x \approx_\alpha y$ if $\frac{1}{1 + \alpha} x \leq y \leq (1 + \alpha) x$.

We say a distribution $D$ over $\R^d$ is isotropic if $\Cov_{X \sim D} [X] = \Id$.
We say a univariate distribution $D$ with mean $\mu$ is sub-gaussian with variance proxy $s^2$ if 
\[
\E_{X \sim D} \Brac{ \Paren{X - \mu}^k } \leq \E_{X \sim \normal (0, s^2)} \Brac{X^k}
\]
for all $k$ even.
We say a distribution $D$ over $\R^d$ and mean $\mu$ is sub-gaussian with variance proxy $\Sigma \preceq I$, if for all unit vectors $v$, the distribution of $\Iprod{v, X - \mu}$ is sub-Gaussian with variance proxy $v^\top \Sigma v$.
Intuitively, a sub-Gaussian distribution is simply any distribution which concentrates as well as a Gaussian.

\subsection{Our results}
With this terminology in place, we are now ready to state our main results on robust mean estimation.
Our first result is for robust mean estimation under the assumption of bounded covariance:
\begin{theorem}
\label{thm:main-bounded-cov}
Let $D$ be a distribution on $\R^d$ with mean $\mu$ and covariance $\Sigma \preceq I$.
Let $\eps > 0$ be sufficiently small, and let $\delta > 0$.
Let $S$ be an $\eps$-corrupted set of samples from $D$ of size $n$.
There is an algorithm $\textsc{QUEScoreFilter}(S, \delta, \e)$ which takes $S$ and $\delta$, and outputs $\widehat{\mu}$ so that with probability $1 - \delta - \exp (-\eps n)$, we have 
\[
\Norm{\mu - \widehat{\mu}}_2 \leq O \Paren{\sqrt{\eps} + \sqrt{\frac{d}{n \delta}} + \sqrt{\frac{d (\log d + \log 1 / \delta)}{n}}} \; .
\]
Moreover, the algorithm runs in time $\Otilde(n d \log 1 / \delta)$.
\end{theorem}
\noindent
We make two observations about this problem.
First, it is well-understood (see e.g.~\cite{steinhardt2018robust,li2018principled}) that $\Omega (\sqrt{\eps})$ error is unavoidable for this problem, no matter how many samples are given.
Second, observe that the rate $O(\sqrt{d / n})$ is necessary for this problem even without corruptions.
Thus, up to log factors, and the dependence on $\delta$, this guarantee is information-theoretically optimal.
We note also that the algorithm does not need to know $\e$ exactly; any upper bound will suffice (with commensurate weakening in the error rate).

We also prove a strong statement for the case of robust mean estimation for sub-gaussian distributions:
\begin{theorem}
\label{thm:main-gaussian}
Let $D$ be an isotropic sub-gaussian distribution with variance proxy $I$ and mean $\mu$.
Let $\eps > 0$ be sufficiently small, and let $\delta > 0$.
Let $S$ be an $\eps$-corrupted set of samples from $D$ of size $n$.
There is an algorithm $\textsc{s.g.-QUEScoreFilter}(S, \delta, \eps)$ which takes $S, \delta$, and $\eps$, and outputs $\widehat{\mu}$ so that with probability $1 - \delta$, we have 
\[
\Norm{\mu - \widehat{\mu}}_2 \leq O \Paren{\eps \sqrt{\log 1/ \eps} + \sqrt{\frac{d + \log 1/\delta}{n}}} \; .
\]
Moreover, the algorithm runs in time $\Otilde(n d \log 1 / \delta)$.
\end{theorem}
\noindent
It is suspected, based on statistical-query lower bounds, that $\Omega(\eps \sqrt{\log 1 / \eps})$ error is incurred by any computationally-efficient algorithm in this setting, although $\Theta(\epsilon)$ is the minimax optimal dependence of the error rate on $\e$ \cite{diakonikolas2017statistical,tukey1975mathematics}.
Moreover, $\Omega \Paren{\sqrt{(d + \log 1/\delta) / n}}$ is the minimax rate for mean estimation for Gaussians without noise.
Thus, the error guarantees of this algorithm are minimax optimal up to constants and the factor $\sqrt{\log(1/\epsilon)}$.

This algorithm assumes that the distribution is isotropic.
There is evidence that such an assumption is necessary to get error beyond $\sqrt{\eps}$ using computationally efficient (i.e. poly-time) algorithms~\cite{hopkins2019hard}.
In our theorem statement above we also assume that the variance proxy is at most $I$.
This is done for simplicity: it is easily verifiable that our algorithm works (with an appropriate scaling in front of the error guarantee) if the variance proxy is PSD upper bounded by $\sigma^2 I$ for any $\sigma^2$.

Before we describe our techniques, we require a few additional algorithmic tools, which we describe below.
\subsection{Soft selection of subsets of points}
In our presentation of our filtering algorithm for robust mean estimation, it will be convenient for us to work with a ``soft'' version of the filter.
Instead of wholly removing points that we deem suspicious, we will maintain a set of weights for each point, and downweight those that we find suspicious.
In this section, we establish notation for dealing with such operations.
However, we briefly remark that, as we will explain later in Appendix~\ref{app:random-filter}, the same results (up to log factors) can be established using ``hard'' filtering more akin to the algorithms presented in prior work, e.g.~\cite{diakonikolas2017being}.

Throughout this paper, we will let $\Delta_n$ denote the simplex in $n$ dimensions, and we let 
\[
\subsimplex_n = \{w \in \R^n: w_i \geq 0, \sum w_i \leq 1 \} \; .
\]
For $w \in \subsimplex_n$, let $|w| = \sum w_i$ be its $\ell_1$ norm.
For $w, w' \in \subsimplex_n$, we say $w' \leq w$ if $w'_i \leq w_i$ for all $i = 1, \ldots, n$.
Let $S = \{X_1, \ldots, X_n\}$ be a (multi)-set of $n$ points.
For any $w \in \subsimplex_n$, we let $\mu(w) = \mu(S, w) = \frac{1}{|w|} \sum_{i = 1}^n w_i X_i$, and we let
\[
M(w) = M(S, w) = \sum_{i = 1}^n w_i (X_i - \mu(w)) (X_i - \mu(w))^\top \; .
\]
Typically when the set $S$ is understood, we will omit the dependence on $S$ in the notation.
For any set $T \subseteq S$, we let $\mu(T) = \mu(w)$ where $w \in \subsimplex_n$ is the vector $w_i = 1/n$ for $i \in T$ and $w_i = 0$ otherwise, and similarly we let $M(T) = M(w)$.

\subsection{Naive pruning}
One primitive we will require will be the ability to removes points which are ``obviously'' outliers.
It is well-known that there exist randomized nearly-linear time algorithms for achieving this.
For completeness we prove this lemma in Appendix~\ref{sec:app-prelims}.
\begin{lemma}[folklore]
\label{lem:naive-prune}
There is an algorithm \textsc{NaivePrune} with the following guarantees.
Let $\eps > 1/2$, and let $\delta > 0$.
Let $S \subset \R^d$ be a set of $n$ points so that there exists a ball $B$ of radius $r$ and a subset $S' \subseteq S$ so that $|S'| \geq (1 - \eps)n$, and $S' \subset B$.
Then $\textsc{NaivePrune}(S, r, \delta)$ runs in time $O(n d \log 1 / \delta)$ and with probability $1 - \delta$ outputs a set of points $T \subseteq S$ so that $S' \subseteq T$, and $T$ is contained in a ball of radius $4r$.
\end{lemma}
\noindent
In the case where the output of $\textsc{NaivePrune}$ satisfies the conditions of the lemma, we say that $\textsc{NaivePrune}$ \emph{succeeds}.

\subsection{The one-dimensional filter}
\label{sec:1dfilter}
An important algorithmic primitive for us will be an univariate soft outlier removal step.
The sub-problem considered here is as follows: we are given a set of nonegative scores $\tau_1, \ldots, \tau_m$, with the guarantee that there is a small subset $S \subseteq [m]$ so that $\sum_{i \in S} \tau_i > \frac{1}{2} \sum_{i = 1}^m \tau_i$, that is, they contribute a majority of the mass of the points.
The goal is to then either downweight (or remove) the overall set of scores in such a way so that more mass from $S$ is removed than from outside of $S$, or alternatively, more points are removed from $S$ than from outside $S$.
An algorithm for achieving this via downweighting has already been described in~\cite{steinhardt2018robust}, and a randomized algorithm that achieves the same sorts of guarantees with high probability by removing points is implicit in the filtering algorithm of~\cite{diakonikolas2019robust,diakonikolas2017being} (e.g. in Algorithm 3 in Appendix A of~\cite{diakonikolas2017being}).
In this paper, we will require a slight strengthening of these algorithms.
We require that not only do we remove more weight from the bad points than the good points, but we also decrease the overall sum by a constant factor.
We observe that while we will present a method for acheving this via downweighting, one can achieve the same guarantee (with high probability) by removing points.
In the main text, we choose to present the soft downweighting method for robust mean estimation for simplicity.
See Appendix~\ref{app:random-filter} for details.

Formally, we describe an algorithm \textsc{1DFilter} and prove the following guarantee for the algorithm.
The algorithm and its analysis are fairly straightforward so we defer the formal descriptions and proofs to Appendix~\ref{app:1dfilter}.
\begin{theorem}
\label{thm:1D}
Let $\eta \in (0, 1/2)$, let $b \geq 2 \eta$, and let $w_1, \ldots, w_m$ and $\tau_1, \ldots, \tau_m$ be non-negative numbers so that $\sum_{i = 1}^m w_i \leq 1$.
Let $\tmax = \max_{i \in [m]} \tau_i$.
Suppose there exist two disjoint sets $S_g, S_b$ so that $S_g \cup S_b = [m]$, and moreover,
\[
\sum_{i \in \Sgood} w_i \tau_i \leq \eta \sigma\; \mbox{, where } \; \sigma = \sum_{i = 1}^n w_i \tau_i \; .
\]
Then $\textsc{1DFilter} (w, \tau, b)$ runs in time $O\Paren{\Paren{1 + \log \frac{\tmax}{b \sigma}} m}$ and outputs $0 \leq w' \leq w$ so that:
\begin{itemize}
\item more weight is removed from $S_b$ than $S_g$, i.e. $\sum_{i \in S_g} w_i - w_i' \leq \sum_{i \in S_b} w_i - w_i'$, and
\item the weighted sum of the $\tau$ has decreased, i.e. $w'$ satisfies
\begin{equation}
\label{eq:1D-final}
\sum_{i = 1}^m w'_i \tau_i \leq b \sigma  \; .
\end{equation}
\end{itemize}
\end{theorem}
\noindent
In particular, note that if $b = \Omega (1)$ and $\tmax / \sigma \leq m^{O(1)}$, then this algorithm runs in nearly linear time.

\paragraph{Randomized outlier removal}
As mentioned previously, there is also a randomized strategy that avoids downweighting and achieves the same guarantee with high probability (up to logarithmic factors in runtime).
Our overall robust mean estimation algorithm (for both settings presented in the paper) can be instantiated using this algorithm rather than~\textsc{1DFilter}.
While as far as we know this yields no theoretical improvements for robust mean estimation (indeed, our analysis of it proves bound which are worse by logarithmic factors than our analysis of soft downweighting), it is much closer to the practical outlier detection method used in the experiments in Section~\ref{sec:experiments} and also to prior algorithms presented in~\cite{diakonikolas2017being}, and may be of instructive value.
For this reason we describe this algorithm in Appendix~\ref{app:random-filter}.


\subsection{Matrix Multiplicative weights}
\label{sec:mmw}
We will use the following form of the MMW update, which is essentially the same as presented in~\cite{allen2015spectral}. 
In each iteration $t = 0, \ldots, T$, the player chooses an action $U_k \in \Delta_{d \times d}$, receives a gain matrix $F_t \in \R^{d \times d}$, and receives reward $\iprod{F_k, U_k}$.
Then the player sees $F_k$.
In~\cite{allen2015spectral}, they demonstrate that if the player plays according to the entropy regularizer (or equivalently, matrix multiplicative weights), namely, 
\begin{equation}
\label{eq:mmw_update}
U_k = \exp \Paren{cI + \alpha \sum_{t = 0}^{k - 1} F_t} \; ,
\end{equation}
where $c$ is a constant ensuring that $\tr (X_k) = 1$, and $\alpha$ satisfies $\alpha F_t \preceq I$ for all $0 = 1, \ldots, T - 1$, then we have that for any $U \in \Delta_{n \times n}$,
\begin{equation}
\label{eq:mmw_regret}
\sum_{t = 0}^{T - 1} \iprod{F_t, U - U_t} \leq \alpha \sum_{t = 0}^{T - 1} \iprod{U_t, |F_t|} \cdot \| F_t \|_2 + \frac{\log n}{\alpha} \; .
\end{equation}
Here, for any symmetric matrix $A = \sum_{i = 1}^d \lambda_i v_i v_i^\top$, we let $|A|$ denote $|A| = \sum_{i = 1}^d |\lambda_i| v_i v_i^\top$.
Equivalently, by rearranging terms, and taking a supremum over $U$ of~\eqref{eq:mmw_regret}, we obtain that the update satisfies
\begin{equation}
\label{eq:mmw_regret}
\Norm{\sum_{t = 0}^{T - 1} F_t}_2 \leq \sum_{t = 0}^{T - 1} \iprod{U_t, F_t} + \alpha \sum_{t = 0}^{T - 1} \iprod{U_t, |F_t|} \cdot \| F_t \|_2 + \frac{\log n}{\alpha} \; .
\end{equation}


\section{An MMW algorithm for robust mean estimation with bounded covariance}
\label{sec:ideal-bounded-cov}
In this section we describe the algorithm which achieves Theorem~\ref{thm:main-bounded-cov}.
We first identify a deterministic condition on the set of inliers under which our algorithm is guaranteed to be correct.
It is a very mild condition: at a high level, it simply states that the empirical mean of the samples is converging to the true mean, and the empirical covariance is bounded.

\begin{definition}
We say a set of points $S$ is $(\gamma_1, \gamma_2)$-good with respect to a distribution $D$ with mean $\mu$ and covariance $\Sigma \preceq \Id$ if the following two properties hold:
\begin{itemize}
\item $\| \mu(S) - \mu \|_2 \leq \gamma_1$, and
\item $\| \Cov (S) \|_2 \leq \gamma_2$.
\end{itemize} 
\end{definition}

\noindent
The following is a generalization of Lemma A.18 in~\cite{diakonikolas2017being}, which states that, with high probability, any set of i.i.d. points from a distribution with bounded covariance will contain a large set which is good with respect to that distribution.
For completeness we prove this lemma in Appendix~\ref{app:ideal-bounded-cov}.

\begin{lemma}
\label{lem:second-moment-conc}
Let $\eps \in [0, 1/2)$, and let $n$ be a positive integer.
Let $D$ be a distribution with mean $\mu$ and covariance $\Sigma \preceq \Id$, and let $X_1, \ldots, X_n$ be independent draws from $D$.
Then, with probability $1 - \delta - \exp(-\eps n)$, there exists a set $S \subseteq \{X_1, \ldots, X_n\}$ so that the following two conditions are simultaneously satisfied:
\begin{itemize} 
\item $|S| \geq (1 - \eps) n$, and
\item $S$ is $(\gamma_1, \gamma_2)$-good with respect to $D$, where 
\begin{equation}
\label{eq:conc-second-moments}
\gamma_1 = \frac{1}{1 - \eps}\cdot \Paren{ \sqrt{\frac{2d}{n \delta}} + \sqrt{c \eps} } \; ,~\mbox{and}~ \gamma_2 = \frac{1}{1 - \eps} \cdot \frac{d (\log d + \log 2 / \delta)}{c' \eps n} \; ,
\end{equation}
for some universal constants $c, c' > 0$.
\end{itemize}
\end{lemma}
\noindent
In particular, observe that for constant $\delta$ and for $n = \Omega (d \log d / \eps)$, we have that $\gamma_1 = O(\eps)$ and $\gamma_2 = O(1)$.

\subsection{The good set}
Throughout we will let $S = \Sgood \cup \Sbad \setminus S_r$, where $\Sgood$  is $(\gamma_1, \gamma_2)$-good with respect to $D$, and $|\Sbad|, |S_r| \leq \eps |S|$.
For any $w \in \subsimplex_n$, we let $\wgood \in \subsimplex_{|\Sgood|}$ denote the restriction of $w$ to the indices in $\Sgood$, and similarly define $\wbad \in \subsimplex_{|\Sbad|}$.

Our set of weights of interest will be slightly different than those considered in prior papers, but morally captures the same concept, up to issues of reweighting.
We will always guarantee that the weights we consider lie within the following set:
\[
\fS_{n, \eps} = \Cbrac{w \in \Gamma_n : w \leq \frac{1}{n} \one_n ~\mbox{ and }~ \Abs{\frac{1}{n} \one_{|\Sgood|} - w_g } \leq \Abs{\frac{1}{n} \one_{|\Sbad|} - w_b}} \; .
\]
Intuitively, weights in the set $\fS_{n, \eps}$ are what happens when we start with the uniform weighting $\tfrac 1 n \cdot \one$ and remove weight from points in the data set, always removing at least as much mass from the bad set as we do from the good set.

\subsection{Geometric lemmata}
We first prove the following sequence of structural lemmata.
The first, which is implicit in the earlier work, and which is in some sense the fundamental geometric fact which guides our algorithmic design, gives an upper bound on the deviation between the weighted empirical mean of the data set and the true mean of the distribution in terms of the spectral norm of the weighted covariance of the dataset.
\begin{lemma}
\label{lem:mean-to-variance}
Let $S, \gamma_1, \gamma_2$ be as above, and let $w \in \fS_{n, \eps}$.
Then
\[
  \| \mu(w) - \mu \|_2 \leq \frac{1}{1 - 2\eps} \cdot \Paren{\sqrt{\eps \gamma_2} + (1 + \eps) \gamma_1 + 2 \sqrt{\eps \Norm{M(w)}_2}}  \; .
\]
\end{lemma}
\begin{proof}
Let $\rho = \mu(w) - \mu$.
We have the following sequence of identities:
\begin{align*}
\| \mu(w) -  \mu \|_2^2 &= \Iprod{\mu(w) - \mu, \mu(w) - \mu} \\
&= \frac{1}{|w|} \sum_{i = 1}^n w_i \Iprod{X_i - \mu, \rho} \\
&= \frac{1}{|w|} \Paren{\sum_{i \in \Sgood} w_i \Iprod{X_i - \mu, \rho} + \sum_{i \in \Sbad} w_i \Iprod{X_i - \mu, \rho}} \\
&= \frac{1}{|w|} \Paren{ \underbrace{\sum_{i \in \Sgood} \frac{1}{n} \Iprod{X_i - \mu, \rho}}_{W_1} - \underbrace{\sum_{i \in \Sgood} \Paren{\frac{1}{n} - w_i} \Iprod{X_i - \mu, \rho}}_{W_2} + \underbrace{\sum_{i \in \Sbad} w_i \Iprod{X_i - \mu, \rho}}_{W_3} } \; .
\end{align*}
\noindent
We now upper bound each term separately.
By Cauchy-Schwarz, we have
\begin{align*}
\left| W_1 \right| &\leq (1 - \eps) \left\| \frac{1}{(1 - \eps) n} \sum_{i \in \Sgood} X_i - \mu \right\|_2 \| \rho \|_2 \\
&\leq \gamma_1 \| \rho \|_2 \; ,
\end{align*}
by the goodness of $\Sgood$.

We now turn our attention to $W_2$ and $W_3$.
Both bounds will follow from the following claim:
\begin{claim}
\label{claim:sos-bound}
Let $w', \alpha \in \subsimplex_n$ be so that $|w'| \leq \eps$, $w'_i \leq \frac{1}{n}$ for all $i \in [n]$, and $w' \leq \alpha$.
Then, for any $v \in \R^d$, we have
\[
\left| \sum_{i = 1}^n w'_i \Iprod{X_i - \mu, v} \right| \leq \sqrt{\eps \| M(\alpha) \|_2} \| v \|_2  + \eps \Norm{\mu(\alpha) - \mu}_2 \Norm{v}_2 \; .
\]
\end{claim}
\begin{proof}
We have
\begin{align*}
\left| \sum_{i = 1}^n w'_i \Iprod{X_i - \mu, v} \right| &= \left| \sum_{i = 1}^n w'_i \Iprod{X_i - \mu(\alpha), v} + |w'| \iprod{\mu(\alpha) - \mu, v} \right| \\
&\leq  \left| \sum_{i = 1}^n w'_i \Iprod{X_i - \mu(\alpha), v} \right| + \eps \Abs{\iprod{\mu(\alpha) - \mu, v}} \\
&\leq \left| \sum_{i = 1}^n w'_i \Iprod{X_i - \mu(\alpha), v} \right| + \eps \Norm{\mu(\alpha) - \mu}_2 \Norm{v}_2 \; .
\end{align*}
By H\"{o}lder's inequality, we have
\begin{align*}
\Paren{\sum_{i = 1}^n w'_i \Iprod{X_i - \mu(\alpha), v}}^2 &\leq \Paren{\sum_{i = 1}^n \frac{(w'_i)^2}{\alpha_i}} \cdot \Paren{\sum_{i = 1}^n \alpha_i \Iprod{X_i - \mu(\alpha), v}^2} \\
&\stackrel{(a)}{\leq} |w'| \cdot \sum_{i = 1}^n \alpha_i \Iprod{X_i - \mu(\alpha), v}^2 \\
&\stackrel{(b)}{\leq} \eps \| M(\alpha) \|_2 \| v \|_2^2 \; .
\end{align*}
Here (a) follows since $w_i' \leq \alpha_i$, and (b) follows from the definition of spectral norm, and the assumption on $|w'|$.
Thus, by taking square roots and combining terms, we have
\[
\left| \sum_{i = 1}^n w'_i \Iprod{X_i - \mu, v} \right| \leq \sqrt{\eps \| M(\alpha) \|_2} \| v \|_2  + \eps \Norm{\mu(\alpha) - \mu}_2 \Norm{v}_2 \; ,
\]
as claimed.
\end{proof}
\noindent
With this claim, we can now bound $W_2$ and $W_3$.
To bound $W_2$, let $w'_i = \frac{1}{n} - w_i$ for $i \in \Sgood$ and $w_i = 0$ otherwise, and let $\alpha_i = \frac{1}{n}$ if $i \in S_g$, and $0$ otherwise.
Then, applying the claim with $v = \rho$ yields that
\[
|W_2| \leq \sqrt{\eps \| M(\alpha) \|_2} \| \rho \|_2  + \eps \Norm{\mu(S_g) - \mu}_2 \Norm{\rho}_2 \leq \sqrt{\eps \gamma_2} \cdot \Norm{\rho}_2 + \eps \gamma_1 \Norm{\rho}_2 \; .
\]
Similarly, to bound $W_3$, let $w'_i = w_i$ if $i \in \Sbad$ and $w'_i = 0$ otherwise, and let $\alpha_i = w_i$ for all $i \in S$.
Again, letting $v = \rho$, we get that
\[
|W_3| \leq \sqrt{\eps \| M(w) \|_2} \| \rho \|_2  + \eps \| \rho \|_2^2 \; ,
\]
as well.
Combining these three bounds, and using the fact that $|w| \leq 1$, yields that
\[
\| \rho \|_2^2 \leq \sqrt{\eps \gamma_2} \| \rho \|_2 + (1 + \eps) \gamma_1 \norm{\rho}_2 + 2 \sqrt{\eps \| M(w) \|_2} \| \rho \|_2  + 2 \eps \| \rho \|_2^2 \; .
\]
Simplifying this expression then yields the desired bound on $\| \rho \|_2$.
\end{proof}

\noindent
We also require the following linear algebraic fact:
\begin{lemma}
\label{lem:psd-ordering}
Let $w', w \in \subsimplex_n$ so that $w' \leq w$.
Then $|w'| M(w') \preceq |w| M(w) $.
\end{lemma}
\begin{proof}
We first observe that if $w' \leq w$, then 
\[
\sum_{i = 1}^n w_i' (X_i - \mu(w)) (X_i - \mu(w))^\top \preceq \sum_{i = 1}^n w_i (X_i - \mu(w)) (X_i - \mu(w))^\top = M(w) \; .
\]
To complete the argument, observe that 
\begin{align*}
\sum_{i = 1}^n w_i' (X_i - \mu(w)) (X_i - \mu(w))^\top &= \sum_{i = 1}^n w_i' (X_i - \mu(w')) (X_i - \mu(w'))^\top + |w'| (\mu(w') - \mu(w)) (\mu(w') - \mu(w))^\top \\
&\succeq \sum_{i = 1}^n w_i' (X_i - \mu(w')) (X_i - \mu(w'))^\top = M(w') \; ,
\end{align*}
and so $M(w') \preceq M(w)$, as claimed.
\end{proof}

\subsection{General algorithm description, bounded second moment}
In this section we describe the algorithm we would like to run via matrix multiplicative weights, and demonstrate that it will terminate in a small number of iterations, when given good approximations to the entropic scores.

The algorithm, which we call \algname, proceeds in epochs, and takes as input a corrupted dataset $S$, and a \emph{score oracle} $\cO$.
Initially, in epoch $s = 0$, we let $w^{\Paren{0}} = \frac{1}{n} \one_n$.
Then, in epoch $s$, the algorithm proceeds iteratively as follows.
First, approximately compute $\lambda^{\Paren{s}} \approx_{0.1} \| M (w^{\Paren{s}}) \|_2$, and if $\lambda^{\Paren{s}} \leq 100 \gamma_2$, then we terminate and output $\mu(w^{\Paren{s}})$.

Otherwise, we let $w^{\Paren{s}}_0 = w^{\Paren{s}}$.
Then, in iteration $t = 0, \ldots, T_s$, we first approximately compute $\lambda^{\Paren{s}}_t \approx_{0.1} \| M(w^{\Paren{s}}_t) \|_2$.
If $\lambda^{\Paren{s}}_t \leq \frac{2}{3} \lambda^{\Paren{s}}_0$, we terminate the epoch and let $w^{\Paren{s + 1}} = w^{\Paren{s}}_t$.
Otherwise, we let $U^{\Paren{s}}_t$ be prescribed by the MMW update with parameter $\alpha^{\Paren{s}} = \frac{1}{1.1 \cdot \lambda^{\Paren{s}}_0}$, where the loss is given as follows.
At time $t$, and for all $i \in [n]$, we let $\tilde{\tau}_{t,i}^{\Paren{s}} = \cO(S, w_0^{\Paren{s}}, \ldots, w_n^{\Paren{s}})$ be the set of scores that the oracle produces.
We then compute $\sum_{i} w_{t,i}^{(s)} \tilde{\tau}_{t, i}^{\Paren{s}}$.
If $\sum_{i} \tilde{\tau}_{t, i}^{\Paren{s}} \leq \frac{1}{5} \lambda^{\Paren{s}}_0$, then let $w^{\Paren{s}}_{t + 1} = w^{\Paren{s}}_{t}$.
Otherwise, let $w^{\Paren{s}}_{t + 1} = \textsc{1DFilter} (w^{\Paren{s}}_t, \tilde{\tau}^{\Paren{s}}_t, 1/4)$.
In either case, the algorithm receives the gain matrix $F_t = M(w^{\Paren{s}}_{t + 1})$.
The formal pseudocode for this algorithm is given in Algorithm~\ref{alg:mmw-filter}.

\begin{algorithm}[ht]
\caption{MMW-based filtering method for robust mean estimation with bounded second moments}
  \label{alg:mmw-filter}
  \begin{algorithmic}[1]
  \STATE \textbf{Input:} dataset $S \subset \R^d$ of size $n$, parameters $\gamma_1, \gamma_2$, score oracle $\cO$
  \STATE Let $w^{\Paren{0}} = \frac{1}{n} \one_n$.
  \FOR{epoch $s = 0, \ldots, O(\log \kappa)$}
  	\STATE Let $\lambda^{\Paren{s}} \approx_{0.1} \Norm{M(w^{\Paren{s}})}_2$
  	\IF{$\lambda^{\Paren{s}} \leq 100 \gamma_2$}
  		\RETURN $\mu (w^{\Paren{s}})$
  	\ENDIF
  	\STATE Let $w^{\Paren{s}}_0 = w^{\Paren{s}}$
  	\STATE Let $\alpha^{\Paren{s}} = \frac{1}{1.1 \cdot \lambda^{\Paren{s}}_0}$
  	\FOR{iteration $t = 0, \ldots, O(\log d)$}
  		\STATE Compute $\lambda^{\Paren{s}}_t \approx_{0.1} \| M(w^{\Paren{s}}_t) \|_2$.
  		\IF{$\lambda^{\Paren{s}}_t \leq \frac{2}{3} \lambda^{\Paren{s}}_0$} 
  			\STATE terminate epoch
  		\ENDIF
  		\STATE Let $U^{\Paren{s}}_t$ be given by MMW update with parameter $\alpha^{\Paren{s}}$
  		\STATE For $i = 1, \ldots, n$, let $\tilde{\tau}_{t,i}^{\Paren{s}} = \cO(S, w_0^{\Paren{s}}, \ldots, w_n^{\Paren{s}})$
  		\IF{$\sum_{i} \tilde{\tau}_{t, i}^{\Paren{s}} \leq \frac{1}{5} \lambda^{\Paren{s}}_0$} \label{line:bad-certificate}
  			\STATE Let $w_{t + 1}^{\Paren{s}} = w_t^{\Paren{s}}$.
  		\ELSE
  			\STATE Let $w^{\Paren{s}}_{t + 1} = \textsc{1DFilter} (w^{\Paren{s}}_t, \tilde{\tau}^{\Paren{s}}_t, 1/4)$.
  		\ENDIF
  		\STATE Let feedback matrix to MMW update be $F_t = M(w^{\Paren{s}}_{t + 1})$
  	\ENDFOR
  	\STATE Let $w^{\Paren{s + 1}} = w^{\Paren{s}}_t$
  \ENDFOR
  \end{algorithmic}
\end{algorithm}
\noindent
\paragraph{The score oracles}
There are two important score oracles for our purposes.
The first is the \emph{exact} score oracle $\cO_{\mathrm{exact}}$, whose output is
\begin{equation}
\label{eq:tau-definition}
\tau^{\Paren{s}}_{t, i} = \Paren{X_i - \mu \Paren{w_{t,i}^{\Paren{s}}}}^\top U^{\Paren{s}}_t \Paren{X_i - \mu \Paren{w_{t,i}^{\Paren{s}}}} \; ,
\end{equation}
where $U_t^{\Paren{s}}$ is as in Algorithm~\ref{alg:mmw-filter}, namely,
\begin{equation}
\label{eq:ut-def}
U_{t}^{\Paren{s}} = \exp \Paren{c \Id - \alpha \sum_{j = 0}^{t - 1} M(w_{j}^{\Paren{s}})} \; ,
\end{equation}
where $c$ is chosen so that $\tr (U_t^{(s)}) = 1$.

Filtering using these scores corresponds to the intuition that we are using $U^{\Paren{s}}_t$ as a high entropy certificate.
Ignoring runtime, this set of scores would be the most natural set of scores for the algorithm.
However, computing this score is quite inefficient (polynomial but super-linear).
Thus, we also require an \emph{approximate} score oracle, which is an oracle $\cO_{\mathrm{approx}}$ whose output $\tilde{\tau}_{t, i}^{\Paren{s}}$ satisfies that 
\begin{equation}
\label{eq:approx-tau-definition}
\tilde{\tau}_{t, i}^{\Paren{s}} \approx_{0.1} \tau_{t, i}^{\Paren{s}}
\end{equation}
for all $t, i, s$, where $\tau_{t, i}^{\Paren{s}}$ is defined in~\eqref{eq:tau-definition} (the choice of $0.1$ in the approximation here as well as for $\lambda^{(s)}_{t}$ is arbitrary; any constant sufficiently small will suffice).
In Section~\ref{sec:approx-scores} we will demonstrate how to implement such an approximate score oracle in (randomized) nearly-linear time by sketching.

\subsection{Correctness of \algname}
The remainder of this section is dedicated to a proof of correctness of \algname when instantiated with $\cO_{\mathrm{approx}}$.
Formally, we show:
\begin{theorem}
\label{thm:naive-cov}
Let $D$ be a distribution with covariance $\Sigma \preceq I$ and mean $\mu$.
Let $\eps < c$, where $c$ is a universal constant, and let $\gamma_1, \gamma_2 > 0$.
Let $S$ be a dataset so that $S = \Sgood \cup \Sbad \setminus S_r$ so that $\Sgood$ is $(\gamma_1, \gamma_2)$-good with respect to $D$, and $|\Sbad|, |S_r| \leq \eps |S|$.
Suppose moreover that $\| X_i \|_2 \leq \kappa$ for all $i \in S$.
Then $\algname(S, \cO_{\mathrm{approx}},\delta)$ terminates after at most $S = O(\log \kappa)$ epochs, and outputs a $w \in \fS_{n, \eps} $ so that 
\[
\| \mu(w) - \mu \|_2 \leq O(\sqrt{\eps \gamma_2} + \gamma_1) \; ,
\]
with probability at least $1-\delta$.
Moreover, each epoch runs for at most $O(\log d)$ iterations, requires $O(\log d)$ calls to $\cO_{\mathrm{approx}}$, and requires $\Otilde(nd + n \log \kappa) \cdot \log(1/\delta)$ additional computation.
\end{theorem}
\noindent
Randomness in \algname is limited to the choice of random starting point when using the power method to compute $\lambda^{(s)}$ and the use of the randomized $\cO_{\mathrm{approx}}$.
Standard arguments (taking a union bound over calls to $\cO_{\mathrm{approx}}$, which is called with $\delta/(\poly \log(d,\kappa))$ show that the randomized steps of the algorithm succeed jointly with probability at least $1-\delta$; in the following analysis we implicitly condition on that event and leave details to the reader.

We first prove correctness.
The main lemma is the following per-epoch guarantee:
\begin{lemma}
\label{lem:epoch-inv}
The following invariants always hold.
For all epochs $s$, we have:
\begin{itemize}
\item $w^{s} \in \fS_{n, \eps}$, and
\item If $\| M\Paren{w^{\Paren{s}}} \|_2 > \frac{100}{1.1} \gamma_2$ and $\|M \Paren{w^{\Paren{s}}}\|_2 > \frac {100}{1.1} \gamma_1^2$, then epoch $s$ finishes after $O(\log d)$ iterations, and outputs $w^{\Paren{s + 1}}$ so that $\| M\Paren{w^{\Paren{s + 1}}} \|_2 \leq \frac{2}{3} \| M\Paren{w^{\Paren{s}}} \|_2$.
\end{itemize}
\end{lemma}
\noindent We first show how the lemma implies the theorem.
\begin{proof}[Proof of Theorem~\ref{thm:naive-cov} given Lemma~\ref{lem:epoch-inv}]
We first prove correctness.
The bound on the $\ell_2$ norm of $X_i$ for all $i \in S$ immediately also implies that $\Norm{M(w)}_2 \leq O(\kappa^2)$ for all $w \in \subsimplex_n$, and in particular for $w = \frac{1}{n} \one_n$.
By Lemma~\ref{lem:epoch-inv}, after $S = O(\log \kappa)$ epochs the algorithm must terminate, and moreover every epoch can run for at most $O(\log d)$ iterations.
Lemma~\ref{lem:epoch-inv} additionally implies that if $w$ is the output of $\algname (S)$, then $w \in \fS_{n, \eps} $, and moreover, $\| M(w) \|_2 \leq 110 \max(\gamma_2,\gamma_1^2)$.
Therefore, by Lemma~\ref{lem:mean-to-variance}, we have $\| \mu(w) - \mu \|_2 \leq O(\sqrt{\eps \gamma_2} + \gamma_1)$, from which the desired conclusion immediately follows.

We now prove the runtime bound.
In every iteration, besides the call the the oracle, the only costly operations are the approximate top eigenvalue computations and running \textsc{1DFilter}.
However, the approximate top eigenvalue computations can be done in time $\Otilde (n d)$ via power method since we only ask for a constant multiplicative approximation, and the bound on $\Norm{X_i}_2$ implies that $\textsc{1DFilter}$ runs in $O(n \log \kappa)$ time.
\end{proof}
\noindent
We now prove Lemma~\ref{lem:epoch-inv}.
\begin{proof}
Clearly these invariants hold at the beginning of the algorithm.
Since in the remainder of the proof we will only deal with a single epoch $s$, for conciseness we will omit the superscript.
We will require the following claim:
\begin{claim}
\label{claim:score-bound}
Suppose $w \in \fS_{n, \eps}$ so that $\| M(w) \|_2 \geq 100 \max(\gamma_2,\gamma_1^2)$, and let $U \in \Delta_{d \times d} $.
Let $\tau_i = (X_i - \mu(w))^\top U (X_i - \mu(w))$, and $\tilde{\tau}_i$ be so that $\tilde{\tau}_i \approx_{0.1} \tau_i$ for all $i$.
Suppose that $\lambda \approx_{0.1} \Norm{M(w)}_2$, and $\sum_{i = 1}^n w_i \tilde{\tau}_i \geq \frac{1}{5} \lambda$.
Then, if $w' = \textsc{1DFilter} (w, \tilde{\tau}, 1/4)$, then $w \in \fS_{n, \eps}$, and $\Iprod{M(w'), U} \leq 0.31 \Iprod{M(w), U}$.
\end{claim}
\begin{proof}
We first show that $\sum_{i \in \Sgood} w_i \tau_i \leq c \sum_{i = 1}^n w_i \tau_i$ for some universal constant $c \leq 0.11$.
Let $\wtilde_i = \frac{1}{n}$ if $i \in \Sgood$ and $\wtilde_i = 0$ otherwise.
Then, we have
\begin{align*}
\sum_{i \in \Sgood} w_i \tau_i &= \Iprod{\sum_{i \in \Sgood} w_i (X_i - \mu(w)) (X_i - \mu(w))^\top, U} \\
&\stackrel{(a)}{\leq} \Iprod{\sum_{i = 1}^n \wtilde_i (X_i - \mu(w)) (X_i - \mu(w))^\top, U } \\
&= \Iprod{\sum_{i = 1}^n \wtilde_i (X_i - \mu(\wtilde)) (X_i - \mu(\wtilde))^\top, U} + |\wtilde| \cdot (\mu(\wtilde) - \mu(w))^\top U (\mu(\wtilde) - \mu(w)) \\
&\stackrel{(b)}{\leq} (1 - \eps) \Iprod{M(\wtilde), U} +\| \mu(\wtilde) - \mu(w) \|_2^2 \\
&\stackrel{(c)}{\leq} 2 \gamma_2 + \| \mu(\wtilde) - \mu(w) \|_2^2 \\
&\leq 2 \gamma_2 + 2 \| \mu(\wtilde) - \mu \|_2^2 + 2 \| \mu(w) - \mu \|_2^2 \\
&\stackrel{(d)}{\leq} 2 \gamma_2 + 3 \gamma_1^2 + 9 (\gamma_2 + \eps \Norm{M(w)}_2)\\
& \stackrel{(e)}{\leq} \frac{1}{60} \| M(w) \|_2 \\
&\leq \frac{1.1}{12} \lambda \leq \frac{1.1^2}{12} \sum_{i = 1}^n w_i \tilde{\tau}_i \; , \numberthis \label{eq:score-bound}
\end{align*}
for $\eps$ sufficiently small and $\| M (w) \|_2 > \frac{100}{1.1} \gamma_2$.
Here (a) follows since $w_i \leq \tilde{w_i}$ for $i \in S_g$, (b) follows since $\| U \|_2 \leq 1$, (c) follows from $\eps$-goodness of $\Sgood$, and (d) follows from $\eps$-goodness of $\Sgood$ and Lemma~\ref{lem:mean-to-variance}.
Step (e) follows by hypothesis.
Therefore overall we have $\sum_{i \in S_g} w_i \tilde{\tau}_i \leq \frac{1.1^3}{12} \sum_{i = 1}^n w_i \tilde{\tau}_i$.
Thus Theorem~\ref{thm:1D} applies, and $w'$ satisfies $\sum_{i = 1}^n w'_i \tilde{\tau}_i \leq \frac{1}{4} \sum_{i = 1}^n w_i \tilde{\tau}_i$ as well as $w \in \fS_{n,\eps}$.
Applying the guarantee that $\tilde{\tau}_i = (1 \pm 0.1) \tau_i$ again yields that
\begin{align*}
\Iprod{M(w'), U} & \leq \Iprod{U, \sum_{i =1}^n w_i' (X_i - \mu(w))(X_i - \mu(w))}\\
& = \sum_{i = 1}^n w'_i \tau_i \\
& \stackrel{(a)}{\leq} 1.1 \sum_{i = 1}^n w'_i \tilde{\tau}_i\\
& \leq \frac{1.1}{4} \sum_{i = 1}^n w_i \tilde{\tau}_i\\
& \leq \frac{1.1^2}{4} \sum_{i = 1}^n w_i \tau_i\\
& = \frac{1.1^2}{4} \Iprod{M(w), U}\\
&\leq 0.31 \Iprod{M(w), U} \; .
\end{align*}
Here (a) follows from $M(w') \preceq \sum_{i=1}^n w_i'(X_i - \mu(w))(X_i - \mu(w))^\top$.
This completes the proof of the claim.
\end{proof}
\noindent
Notice that Claim~\ref{claim:score-bound} immediately implies that the first invariant $w \in \fS_{n,\eps}$ always holds.
We now turn to proving the second invariant.
Let $T$ be the number of iterations that the epoch runs for.
Observe that for all $t = 0, \ldots, T$, we have that $M(w_t) \preceq M(w_0) \preceq \frac{1}{\alpha} I$, and so we are indeed in the setting of the guarantee in Section~\ref{sec:mmw}.
Thus, by~\eqref{eq:mmw_regret}, since $F_t = M(w_{t + 1})$, we obtain the following regret bound:
\begin{align*}
\left\| \sum_{t = 0}^{T - 1} M(w_{t+1}) \right\|_2 &\leq\sum_{t = 0}^{T - 1} \Iprod{M(w_{t + 1}), U_t} + \alpha \sum_{t = 0}^{T - 1} \Iprod{U_t, M (w_{t + 1})} \Norm{M(w_{t + 1})}_2 + \frac{\log d}{\alpha} \\
&\leq 2 \sum_{t = 0}^{T - 1} \Iprod{M(w_{t + 1}), U_t} + \| M(w_0) \|_2 \cdot \log d\; , \numberthis \label{eq:MMW-regret}
\end{align*}
where the second inequality follows by our choice of $\alpha$.
We claim that for all $t > 0$, we must have $\Iprod{M(w_t), U_t} \leq 0.31 \| M(w_0) \|_2$.
There are two cases. 
If we enter the if statement in Line~\ref{line:bad-certificate}, then 
\[
\Iprod{M_{t+1}, U_t} = \sum_{i = 1}^n w_i \tau_i \leq 1.1 \sum_{i = 1}^n w_i \tilde{\tau}_i \leq \frac{1.1}{5} \lambda \leq 0.31 \| M(w_0) \|_2 \; ,
\]
and $M(w_{t + 1}) = M(w_t)$, so this is clearly satisfied.
Otherwise, the desired bound follows by Claim~\ref{claim:score-bound}.
Thus overall, by~\eqref{eq:MMW-regret} we have that
\begin{equation}
\label{eq:MMW-regret-2}
\left\| \sum_{t = 0}^{T - 1} M(w_t) \right\|_2 \leq T \cdot 0.62 \| M(w_0) \|_2 + \| M(w_0) \|_2 \cdot \log d \; .
\end{equation}
By Lemma~\ref{lem:psd-ordering}, we further have that $M(w_{t + 1}) \preceq M(w_t)$ for all $t = 0, \ldots, T - 1$, and so this implies that 
\[
T \| M (w_T) \|_2 \leq 0.62 \| M(w_0) \|_2 + \| M(w_0) \|_2 \cdot \log d \; .
\]
Simplifying both sides yields that if $T = C \log n$ for some sufficiently large constant $C$, then $\| M(w_T) \|_2 \leq \frac{2}{3} \| M(w_0) \|_2$.
Thus after $O(\log d)$ iterations, we must terminate.
\end{proof}


\section{An MMW algorithm for robust mean estimation for sub-gaussian distributions}
\label{sec:subgaussian}
In this section we give an analog of the result in Section~\ref{sec:ideal-bounded-cov} but but in the setting where the distribution $D$ is subgaussian, and has identity covariance.
We again first identify a deterministic condition for the inlers under which our algorithms will succeed.
In this case, we need a stricter analog of $\eps$-goodness. 
Specifically, we will require:
\begin{definition}[Subgaussian goodness]
\label{def:sg-goodness}
Let $D$ be a distribution with covariance $\Id$ and mean $\mu$.
We say a set of points $S$ is \emph{$(\eps, \gamma_1, \gamma_2, \beta_1, \beta_2)$-subgaussian good} (or $(\eps, \gamma_1, \gamma_2, \beta_1, \beta_2)$-s.g. good for short) with respect to $D$ if there exist universal constants $C_1, C_2$ so that the following inequalities are satisfied:
\begin{itemize}
\item $\Norm{\mu(S) - \mu} \leq \gamma_1$ and $\Norm{\frac{1}{|S|} \sum_{i \in S} \Paren{X_i - \mu(S)} \Paren{X_i - \mu(S)}^\top - \Id}_2 \leq \gamma_2$, and
\item For any subset $T \subset S$ so that $|T| = 2 \eps |S|$, we have
\begin{align*}
\Norm{\frac{1}{|T|} \sum_{i \in T} X_i - \mu} \leq \beta_1 \; ,~\mbox{and}~ \Norm{\frac{1}{|T|} \sum_{i \in T} \Paren{X_i - \mu(S)} \Paren{X_i - \mu(S)}^\top - \Id}_2 \leq \beta_2 \; .
\end{align*}
\end{itemize}
\end{definition}
\noindent
For conciseness, when the parameters $\eps, \gamma_1, \gamma_2, \beta_1, \beta_2$ are understood, we will omit them and refer to the data set as s.g.-good.

We have the following concentration inequality.
The proof is very similar to that of Lemmata 2.1.8 and 2.1.9 in~\cite{li2018principled}.
For completeness we include a proof of this lemma in Appendix~\ref{sec:app-subgaussian}.
\begin{lemma}
\label{lem:conc-gaussian}
Let $X_1, \ldots, X_n \sim D$, where $D$ is subgaussian with variance proxy $1$.
Then, for any $\eps > 0$ sufficiently small, we have that $S = \{X_1, \ldots, X_n\}$ is $\Paren{\eps, \gamma_1, \gamma_2 }$-s.g. good with probability $1 - \delta$, where
\begin{align}
\gamma_1 = O \Paren{\sqrt{\frac{d + \log 1 / \delta}{n}}} \; &,~\gamma_2 = O \Paren{ \max \Paren{ \sqrt{\frac{d + \log 1 / \delta}{n}}, \frac{d + \log 1 / \delta}{n}}} \; , \mbox{and} \label{eq:conc-gaussian} \\
\beta_1 = O \Paren{\sqrt{\log 1 / \eps} + \sqrt{\frac{d + \log 1 / \delta}{\eps n}}} \; ,~&\mbox{and}~ \beta_2 = O \Paren{\log 1 / \eps + \frac{d + \log 1 / \delta}{\eps n}} \; . \label{eq:conc-gaussian-2}
\end{align}
\end{lemma}
\noindent
In particular, note that when $n = \Omega \Paren{\frac{d + \log 1 / \delta}{\eps^2 \log 1 / \eps}}$, then Lemma~\ref{lem:conc-gaussian} implies that $n$ i.i.d. samples from an isotropic sub-gaussian distribution is $(\eps, O(\eps \sqrt{\log 1 / \eps}), O(\eps \log 1 / \eps), O(\sqrt{\log 1 / \eps}), O(\log 1/\eps))$-s.g. good with probability $1 - \delta$.
We will also require the following simple consequences of subgaussian goodness.
\begin{fact}
\label{fact:sg-convexity}
Let $D$ be an isotropic distribution.
Let $S$ be $(\eps, \gamma_1, \gamma_2, \beta_1, \beta_2)$-s.g. good w.r.t. $D$.
Then:
\begin{itemize}
\item
for all $w' \in \subsimplex_n$ with $w' \leq \frac{1}{n} \one_n$ and $|w'| \leq 2 \eps$, and for all unit vectors $v \in \R^d$, we have
\[
\sum_{i = 1}^n w'_i \iprod{X_i - \mu, v}^2 \leq 2 \eps \beta_2 + O \Paren{\eps \log 1 / \eps }\; .
\]
\item if $w \in \subsimplex_n$ satisfies $w \leq \frac{1}{n} \one_n$ and $\Abs{\frac{1}{n} \one_n - w} \leq 2 \eps$, then 
\begin{align}
\Norm{\sum_{i = 1}^n w_i \Paren{X_i - \mu} (X_i - \mu)^\top - \Id}_2 &\leq \gamma_2 + \gamma_1^2 + 2 \eps \beta_2 + O(\eps \log 1 / \eps) \; ,~\mbox{and} \label{eq:sg-convex-1} \\
\Norm{\sum_{i = 1}^n w_i \Paren{X_i - \mu(w)} (X_i - \mu(w))^\top - \Id}_2 &\leq \gamma_2 + 2 \gamma_1^2 + 4 \eps^2 \beta_1^2 + 2 \eps \beta_2 + O(\eps \log 1 / \eps) \; . \label{eq:sg-convex-2}
\end{align}
\end{itemize}
\end{fact}
\begin{proof}
We first prove the first claim.
Let $w'' \in \Gamma_n$ be anything so that $w' \leq w'' \leq \frac{1}{n} \one_n $ and $|w''| = \eps$.
Then since all quantities on the LHS of the expression are nonnegative, we have that 
\begin{equation}
\label{eq:w'-to-w''}
\sum_{i = 1}^n w'_i \iprod{X_i - \mu, v}^2 \leq \sum_{i = 1}^n w''_i \iprod{X_i - \mu, v}^2 \; .
\end{equation}
Now let $A = \{w'' \in \Gamma_n: |w''| = \eps\}$.
This set is clearly convex, and moreover, by inspection, the vertices of $A$ are exactly given by $\frac{1}{n} \one_{T}$ where $|T| = 2 \eps n$.
Thus, by convexity, the maximum of the RHS of~\eqref{eq:w'-to-w''} over $w'' \in A$ is obtained by $w'' = \frac{1}{n} \one_T$ for some $T$ with $|T| = 2 \eps n$.
But then we have
\begin{align*}
\frac{1}{n} \sum_{i \in T} \iprod{X_i - \mu, v}^2 &= 2 \eps \cdot \frac{1}{|T|} \sum_{i \in T} \Paren{\iprod{X_i - \mu, v}^2 - 1} + \eps \\
&\leq 2 \eps \Norm{\frac{1}{|T|} \sum_{i \in T} \Paren{X_i - \mu} \Paren{X_i - \mu}^\top - \Id}_2 + \eps \\
&\leq 2 \eps \beta_2 + O(\eps \log 1 / \eps) \; ,
\end{align*}
by the s.g.-goodness of $S$.
This completes the proof of the first bullet point.

We now turn our attention to the second claim.
We have that 
\begin{align*}
\Norm{\sum_{i = 1}^n w_i \Paren{X_i - \mu} (X_i - \mu)^\top - \Id}_2 &= \Norm{\sum_{i = 1}^n \frac{1}{n} \Paren{X_i - \mu} (X_i - \mu)^\top - \Id + \sum_{i = 1}^n \Paren{\frac{1}{n} - w_i} \Paren{X_i - \mu} (X_i - \mu)^\top }_2 \\
&\leq \Norm{\sum_{i = 1}^n \frac{1}{n} \Paren{X_i - \mu} (X_i - \mu)^\top - \Id }_2 + 2 \eps \beta_2 + O(\eps \log 1 / \eps) \; ,
\end{align*}
by the first claim.
Further expanding, we have 
\begin{align*}
\Norm{\sum_{i = 1}^n \frac{1}{n} \Paren{X_i - \mu} (X_i - \mu)^\top - \Id }_2 &= \Norm{\sum_{i = 1}^n \frac{1}{n} \Paren{X_i - \mu(S)} (X_i - \mu(S))^\top - \Id + |w| \Paren{\mu(S) - \mu}\Paren{\mu(S) - \mu}^\top }_2 \\
&\leq \Norm{\sum_{i = 1}^n \frac{1}{n} \Paren{X_i - \mu(S)} (X_i - \mu(S))^\top - \Id}_2 + \Norm{\mu(S) - \mu}_2^2 \\&
\leq \gamma_2 + \gamma_1^2 \; .
\end{align*}
Putting it all together yields~\eqref{eq:sg-convex-1}.
To prove~\eqref{eq:sg-convex-2}, simply observe that 
\begin{align*}
\Norm{\sum_{i = 1}^n w_i \Paren{X_i - \mu} (X_i - \mu)^\top - \sum_{i = 1}^n  w_i \Paren{X_i - \mu(w)} (X_i - \mu(w))^\top }_2 &= |w| \Norm{\Paren{\mu(w) - \mu}\Paren{\mu(w) - \mu}^\top }_2 \\
&\leq (\gamma_1 + 2 \eps \beta_1)^2 \\
&\leq 2 \gamma_1^2 + 4 \eps^2 \beta_1^2 \; .
\end{align*}
In the first inequality we have used the definition of subgaussian goodness and convexity.
\end{proof}
\noindent
As a result, we also have the following tail bound on mean shifts caused by small subsets of points:
\begin{corollary}
\label{cor:sg-convexity-mean}
Let $D$ be an isotropic distribution.
Let $S$ be $(\eps, \gamma_1, \gamma_2, \beta_1, \beta_2)$-s.g. good w.r.t. $D$.
Then:
\begin{itemize}
\item for all $w' \in \subsimplex_n$ with $w' \leq \frac{1}{n} \one_n$ and $|w'| \leq 2 \eps$, we have
\[
\Norm{\sum_{i = 1}^n w'_i \Paren{X_i - \mu}}_2 \leq 2 \eps \sqrt{ \beta_2} + O \Paren{ \eps \sqrt{\log 1 / \eps} }\; ,~\mbox{and}
\]
\item if $w \in \subsimplex_n$ satisfies $w \leq \frac{1}{n} \one_n$ and $\Abs{\frac{1}{n} \one_n - w} \leq 2 \eps$, then 
\[
\Norm{\mu(w) - \mu}_2 \leq \frac{1}{1 - \eps} \Paren{\gamma_1 + 2 \eps \sqrt{\beta_2} + O(\eps \sqrt{\log 1/\eps}) }\; .
\]
\end{itemize}
\end{corollary}
\begin{proof}
We first prove the first claim.
Fix any unit vector $v \in \R^d$. 
Then we have
\begin{align*}
\Paren{\sum_{i = 1}^n w'_i \iprod{X_i - \mu, v}}^2 &\stackrel{(a)}{\leq} |w'| \sum_{i = 1}^n w_i' \iprod{X_i - \mu, v}^2 \\
&\stackrel{(b)}{\leq} 2 \eps (2 \eps \beta_2 + O \Paren{\eps \log 1 / \eps }) = 4 \eps^2 \beta_2 + O(\eps^2 \log 1 / \eps) \; ,
\end{align*}
where (a) follows from Cauchy-Schwarz, and (b) follows from Fact~\ref{fact:sg-convexity}.
By taking square roots and a supremum over all unit vectors $v$, we obtain the desired conclusion.

We now prove the second claim.
We expand
\begin{align*}
|w| \Norm{\mu(w) - \mu}_2 &= \Norm{\sum_{i = 1}^n w_i \Paren{X_i - \mu}}_2 \\
&= \Norm{\sum_{i = 1}^n \frac{1}{n} \Paren{X_i - \mu} + \sum_{i = 1}^n \Paren{\frac{1}{n} - w_i} \Paren{X_i - \mu}}_2 \\
&\leq \gamma_1 + 2 \eps \sqrt{\beta_2} + O(\eps \sqrt{\log 1/\eps}) \; ,
\end{align*}
where the last line follows from subgaussian goodness, and applying the first claim with $w'_i = \frac{1}{n} - w_i$.
\end{proof}
\subsection{Mean deviations to moment bounds}
As before, we will require a lemma which relates the mean shift caused by a small fraction of points to spectral deviations.
However, because in this case we will assume that our data is subgaussian, we will be able to prove stronger statements, which will in turn allow us to achieve much better error.
We first record the following simple fact, which states that if we have a set of weights that puts almost all of its mass on a good set, then the restriction of that set of weights to the good set satisfies the conditions of the lemmata proved in the above section.
\begin{fact}
Let $\eps < 1/2$, and suppose $S = S_g \cup S_b \setminus S_r$, where $S_g$ is $(\eps, \gamma_1, \gamma_2, \beta_1, \beta_2)$-s.g. good, and $|S_b|, |S_r| \leq \eps |S|$.
Let $w \in \fS_{n, \eps}$.
Then $w_g \leq \frac{1}{|S_g|} \one_{S_g}$ and $\Abs{\frac{1}{|S_g|} \one_{S_g} - w_g} \leq 2 \eps$.
\end{fact}
\noindent
We first show that, no matter what, the smallest eigenvalue of the empirical covariance we choose cannot be too small.
Formally, for the remainder of the section, let
\begin{equation}
\label{eq:xi-def}
\xi = \xi (\eps, \gamma_1, \gamma_2, \beta_1, \beta_2) = \gamma_2 + 2 \gamma_1^2 + 4 \eps^2 \beta_1^2 + 2 \eps \beta_2 + O(\eps \log 1 / \eps) \; .
\end{equation}
Then, we have:
\begin{lemma}
\label{lem:sg-PSD-ordering}
Let $\gamma_1, \gamma_2, \beta_1, \beta_2 > 0$.
Suppose $S = S_g \cup S_b \setminus S_r$, where $S_g$ is $(\eps, \gamma_1, \gamma_2, \beta_1, \beta_2)$-s.g. good, and $|S_b|, |S_r| \leq \eps |S|$.
Let $w \in \fS_{n, \eps}$.
Then
\[
\sum_{i \in S_g \cap S} w_i \Paren{X_i - \mu(w)} \Paren{X_i - \mu(w)}^\top \succeq \Paren{1 - \xi} \Id \; .
\]
\end{lemma}
\begin{proof}
Let $w' \in \fS_{n, \eps}$ be defined by $w'_i = w_i$ if $i \in S_g \cap S$ and $w'_i = 0$ otherwise.
By Lemma~\ref{lem:psd-ordering}, we know that 
\begin{align*}
\sum_{i \in S_g \cap S} w_i \Paren{X_i - \mu(w)} \Paren{X_i - \mu(w)}^\top &\succeq \sum_{i \in S_g \cap S} w_i \Paren{X_i - \mu(w')} \Paren{X_i - \mu(w')}^\top \\
&\succeq \Paren{1 - \xi} \Id \; ,
\end{align*}
by~\eqref{eq:sg-convex-2} of Fact~\ref{fact:sg-convexity}.
\end{proof}
\noindent
We now show the following:
\begin{lemma}
\label{lemma:s.g.-geometry}
Let $\gamma_1, \gamma_2 > 0$.
Suppose $S = S_g \cup S_b \setminus S_r$, where $S_g$ is $(\eps, \gamma_1, \gamma_2, \beta_1, \beta_2)$-s.g. good, and $|S_b|, |S_r| \leq \eps |S|$.
Let $w \in \fS_{n, \eps}$, and let $\lambda = \Norm{M(w_t) - \Id}_2$.
Then
\[
\Norm{\mu(w) - \mu}_2 \leq \frac{1}{1 - \eps} \cdot \Paren{2 \gamma_1 + \sqrt{\eps (\lambda +\xi)} + O(\eps \sqrt{\log 1 / \eps})} \; .
\]
\end{lemma}
\noindent
Before we prove this lemma, observe that if $\gamma_1 = O(\eps \sqrt{\log 1 / \eps}), \gamma_2 = O(\eps \log 1 / \eps), \beta_1 = O(\sqrt{\log 1/\epsilon}), \beta_2 = O(\log(1/\epsilon))$, and $\eps \leq 1/2$ then the RHS of the lemma simplifies to $O(\eps \sqrt{\log 1 / \eps})$.
\begin{proof}[Proof of Lemma~\ref{lemma:s.g.-geometry}]
Let $\rho = \mu(w) - \mu$.
As before, we have the following sequence of identities:
\begin{align*}
|w| \cdot \Norm{\mu(w) - \mu}_2^2 &= |w| \cdot \Iprod{\mu(w) - \mu, \rho} \\
&= \sum_{i = 1}^n w_i \Iprod{X_i - \mu, \rho} \\
&= \underbrace{\sum_{i \in S_g \cap S} w_i \Iprod{X_i - \mu, \rho}}_{W_0} + \underbrace{\sum_{i \in S_b} w_i \Iprod{X_i - \mu, \rho}}_{W_1} \; .
\end{align*}
We treat the two terms on the RHS separately.
We first consider $W_0$.
We continue expanding, and observe:
\begin{align*}
\Abs{\sum_{i \in S_g \cap S} w_i \Iprod{X_i - \mu, \rho}}
&= \Abs{\sum_{i \in S_g \cap S} \frac{1}{n} \Iprod{X_i - \mu, \rho} + \sum_{i \in S_g \cap S} \Paren{w_i - \frac{1}{n}} \Iprod{X_i - \mu, \rho}} \\ 
&\leq \Abs{\sum_{i \in S_g} \frac{1}{n} \Iprod{X_i - \mu, \rho}} + \Abs{\sum_{i \in S_r} \frac{1}{n} \Iprod{X_i - \mu, \rho}} + \Abs{\sum_{i \in S_g \cap S} \Paren{\frac{1}{n} - w_i} \Iprod{X_i - \mu, \rho}} \\
&\stackrel{(a)}{\leq} \Abs{\sum_{i \in S_g} \frac{1}{n} \Iprod{X_i - \mu, \rho}} + 4 \eps \sqrt{\beta_2}\|\rho\|_2 +O \Paren{ \eps \sqrt{\log 1 / \eps}} \Norm{\rho}_2 \\
&\stackrel{(b)}{\leq} \Paren{\gamma_1 + 4 \eps \sqrt{\beta_2} + O\Paren{ \eps \sqrt{\log 1 / \eps}}} \Norm{\rho}_2 \; , \numberthis \label{eq:w0-bound}
\end{align*}
where (a) follows from two applications of Corollary~\ref{cor:sg-convexity-mean}, and (b) follows from Cauchy-Schwarz and subgaussian goodness.

We now turn our attention to bounding $W_1$.
We have
\begin{align*}
\Abs{W_1} &\leq \Abs{\sum_{i \in S_b} w_i \iprod{X_i - \mu(w), \rho} } + \sum_{i \in S_b} w_i \| \rho \|_2^2 \\
&\leq \Abs{\sum_{i \in S_b} w_i \iprod{X_i - \mu(w), \rho} } + \eps \Norm{\rho}_2^2 \; .
\end{align*}
Focusing in on the first term in the RHS, we have
\begin{align*}
\Paren{\sum_{i \in S_b} w_i \iprod{X_i - \mu(w), \rho} }^2 &\stackrel{(a)}{\leq} \Paren{\sum_{i \in S_b} w_i } \sum_{i \in S_b} w_i \iprod{X_i - \mu(w), \rho}^2
&\stackrel{(b)}{\leq} \eps \sum_{i \in S_b} w_i \iprod{X_i - \mu(w), \rho}^2 \; , \numberthis \label{eq:W1-holders}
\end{align*}
where (a) follows from Cauchy-Schwarz.
and (b) follows since $w$ places at most $\eps$ mass on $S_b$.
Now observe that
\begin{align*}
\Abs{\sum_{i \in S_b} w_i \Paren{\iprod{X_i - \mu(w), \rho}^2}} &= \Abs{\sum_{i = 1}^n w_i \Paren{\iprod{X_i - \mu(w), \rho}^2} - \sum_{i \in S_g \cap S} w_i \Paren{\iprod{X_i - \mu(w), \rho}^2}} \\
&\leq \Norm{M(w_t) - \Id}_2 \Norm{\rho}_2^2 + \Abs{\sum_{i \in S_g \cap S} w_i \Paren{\iprod{X_i - \mu(w), \rho}^2 - \Norm{\rho}_2^2}} + O(\e \|\rho\|_2^2) \\
&\leq \Norm{M(w_t) - \Id}_2 \Norm{\rho}_2^2 + \Norm{\sum_{i \in S_g \setminus S_r} w_i \Paren{X_i - \mu(w)} \Paren{X_i - \mu(w)}^\top - \Id}_2 \Norm{\rho}_2^2 + O(\e \|\rho\|_2^2) \\
&\leq \Paren{\Norm{M(w_t) - \Id}_2 + \xi + O(\e) } \Norm{\rho}_2^2 \; ,
\end{align*}
by Fact~\ref{fact:sg-convexity}, where we take the convention that $w_i = 0$ for $i \in S_r$.
Hence, combining terms, recalling the definition of $\lambda = \Norm{M(w_t) - \Id}_2$, and taking square roots, we have
\[
|W_1| \leq O \Paren{\sqrt{\eps (\lambda+ \xi)}} \cdot \| \rho \|_2 + \eps \Norm{\rho}_2 \; .
\]
Combining this with~\eqref{eq:w0-bound} yields 
\begin{align*}
\Norm{\rho}_2^2 &\leq \Abs{W_0} + \Abs{W_1} \\
&\leq \Paren{2 \gamma_1 + \sqrt{\eps (\lambda+ \xi)}  + O(\eps \sqrt{\log 1 / \eps})} \Norm{\rho}_2 + \eps \Norm{\rho}_2^2 \; .
\end{align*}
Solving for $\Norm{\rho}_2$ yields the desired claim.
\end{proof}

\subsection{Algorithm description}
The algorithm is quite similar to the algorithm presented in Section~\ref{sec:ideal-bounded-cov} for the bounded covariance case.
The formal pseudocode is presented in Algorithm~\ref{alg:mmw-filter2}.
For any $w \in \subsimplex_n$, let $\mu(w)$ and $M(w)$ be as in Section~\ref{sec:ideal-bounded-cov}.
However, we will require a slightly stronger notion of score oracle than before.

Recall that before, given a dataset $S$, and a sequence of weights $w_0, \ldots, w_{t - 1}$, the score oracle is asked to produce multiplicative approximations to $\tau_{t, i}$ where $\tau_{t, i}$ is defined as~\eqref{eq:tau-definition}.
One consequence of this is that this allows us to produce multiplicative approximations to $\Iprod{M(w_t^{\Paren{s}}), U_t^{\Paren{s}}} = \sum_{i = 1}^n w_{t, i} \tau_{t, i}$.
However, we will require multiplicative approximations to $\Iprod{M(w_t^{\Paren{s}}) - \Id, U_t^{\Paren{s}}}$, which cannot be obtained black-box via multiplicative approximations to the original scores.

To rectify this, we say that an algorithm $\cO^*$ is an \emph{augumented score oracle} if it takes a dataset $S$ and a sequence of weights $w_0, \ldots, w_{t - 1}$, and outputs $\tilde{\tau}_{t, i}$ for all $i = 1, \ldots, n$, but also an overall score $q_t$ which is intended to approximate $\Iprod{M(w_t^{\Paren{s}}) - \Id, U_t^{\Paren{s}}}$.

Given $S$ and such an oracle $\cO^*$, the algorithm again proceeds in epochs.
Initially, we let $w^{\Paren{0}} = \frac{1}{n} \one_n$.
In epoch $s = 0, \ldots, L - 1$, we proceed as follows.
First, compute $\lambda^{(s)} \approx_{0.1} \Norm{M(w^{\Paren{s}}) - \Id}_2$. If $\lambda^{(s)} \leq O \Paren{\xi}$, we terminate and output $\mu(w^{\Paren{s}})$.

Otherwise, we let $w^{\Paren{s}}_0 = w^{s}$.
Then, in iteration $t = 0, \ldots, T_s - 1$, we first (approximately) compute $\lambda^{\Paren{s}}_t \approx_{0.1} \| M(w^{\Paren{s}}_t) - \Id \|_2$. 
If $\lambda_t^{\Paren{s}} \leq \frac{1}{2} \lambda_0^{\Paren{s}}$, we terminate and let $w^{\Paren{s + 1}} = w_t^{\Paren{s}}$.
Otherwise, we let $U_t^{\Paren{s}}$ be prescribed by the MMW update with parameter $\alpha = 1 / (1.1 \cdot \lambda^{\Paren{s}})$.
Then, produce the gain matrix is given as follows.

At time $t$, run $\cO^*$ given $S$ and the sequence of weights $w_{0}^{\Paren{s}}, \ldots, w_{t - 1}^{\Paren{s}}$ to obtain scores $\tilde{\tau}_{t, i}^{\Paren{s}}$ as well as an overall score $\tilde{q}_t^{\Paren{s}}$. 
Then, check if $\tilde{q}_t^{\Paren{s}} \leq \frac{1}{5} \lambda_0^{\Paren{s}}$.
If so, then we let $w^{\Paren{s}}_{t + 1} = w^{\Paren{s}}_t$.

Otherwise, sort the $\tilde{\tau}^{\Paren{s}}_{t, i}$ in descending order.
WLOG assume that $\tilde{\tau}^{\Paren{s}}_{t, 1} \geq \tilde{\tau}^{\Paren{s}}_{t, 2} \geq \ldots \geq \tilde{\tau}^{\Paren{s}}_{t, n}$.
Let $m$ be the smallest integer so that $\sum_{i \leq m} w_{t, i}^{\Paren{s}} \geq 2 \eps$.
Then, run $\textsc{1DFilter}$ on $\tilde{\tau}^{\Paren{s}}_{t, 1}, \ldots, \tilde{\tau}^{\Paren{s}}_{t, m}$ with corresponding weights $w_{t, 1}^{\Paren{s}}, \ldots, w_{t, m}^{\Paren{s}}$, to obtain a new set of weights $w'_1, \ldots, w'_m$, and let $w_{t + 1, i}^{\Paren{s}}$ be defined by
\begin{equation}
\label{eq:subgaussian-update}
w_{t + 1, i}^{\Paren{s}} = \left\{ \begin{array}{ll}
         w_{t, i}^{\Paren{s}} & \mbox{if $i > m$};\\
        w'_i & \mbox{if $i \leq m$}.\end{array} \right.
\end{equation}
That is, we find the largest $2 \eps$-percentile of the scores weighted by the current weights, and run the univariate filter on these set of weights, leaving the other weights unchanged.
Finally, we output the gain matrix $F^{\Paren{s}}_t = M(w^{\Paren{s}}_{t + 1}) - \Id$.
Notice that this matrix may not be PSD.

\begin{algorithm}[ht]
\caption{MMW-based filtering method for robust mean estimation for subgaussian distributions}
  \label{alg:mmw-filter2}
  \begin{algorithmic}[1]
  \STATE \textbf{Input:} dataset $S \subset \R^d$ of size $n$, parameters $\gamma_1, \gamma_2$, augmented score oracle $\cO$
  \STATE Let $C > 0$ be a sufficiently large universal constant.
  \STATE Let $w^{\Paren{0}} = \frac{1}{n} \one_n$.
  \FOR{epoch $s = 0, \ldots, O(\log \kappa)$}
  	\STATE Let $\lambda^{\Paren{s}} \approx_{0.1} \Norm{M(w^{\Paren{s}}) - \Id}_2$
  	\IF{$\lambda^{\Paren{s}} \leq C \cdot \xi$}
  		\RETURN $\mu (w^{\Paren{s}})$
  	\ENDIF
  	\STATE Let $w^{\Paren{s}}_0 = w^{\Paren{s}}$
  	\STATE Let $\alpha^{\Paren{s}} = \frac{1}{1.1 \cdot \lambda^{\Paren{s}}_0}$
  	\FOR{iteration $t = 0, \ldots, O(\log d)$}
  		\STATE Compute $\lambda^{\Paren{s}}_t \approx_{0.1} \| M(w^{\Paren{s}}_t) - \Id \|_2$.
  		\IF{$\lambda^{\Paren{s}}_t \leq \frac{1}{2} \lambda^{\Paren{s}}_0$} 
  			\STATE terminate epoch
  		\ENDIF
  		\STATE Let $U^{\Paren{s}}_t$ be given by MMW update with parameter $\alpha^{\Paren{s}}$
  		\STATE For $i = 1, \ldots, n$, let $\tilde{\tau}_{t,i}^{\Paren{s}}, \tilde{q}_t^{(s)} = \cO(S, w_0^{\Paren{s}}, \ldots, w_n^{\Paren{s}})$
  		\IF{$\tilde{q}_t^{(s)} \leq \frac{1}{1.1 \cdot 5} \lambda^{\Paren{s}}_0$} \label{line:bad-certificate}
  			\STATE Let $w_{t + 1}^{\Paren{s}} = w_t^{\Paren{s}}$.
  		\ELSE
  			\STATE Sort the $\tilde{\tau}_{t, i}^{(s)}$ in descending order. 
  			\STATE WLOG assume that $\tilde{\tau}_{t, 1}^{(s)} \geq \tilde{\tau}_{t, 2}^{(s)} \geq \ldots \geq \tilde{\tau}_{t, n}^{(s)}$
  			\STATE Let $m$ be the smallest integer so that $\sum_{i \leq m} w_{t, i}^{\Paren{s}} \geq 2 \eps$.
  			\STATE Let $w' = \textsc{1DFilter} ((w^{\Paren{s}}_{t, 1}, \ldots,w^{\Paren{s}}_{t, m}) , (\tilde{\tau}^{\Paren{s}}_{t, 1}, \ldots, \tilde{\tau}^{\Paren{s}}_{t, m}), 1/4)$.
  			\STATE Let $w^{(s)}_{t + 1}$ be as defined in~\eqref{eq:subgaussian-update}.
  		\ENDIF
  		\STATE Let feedback matrix to MMW update be $F_t = M(w^{\Paren{s}}_{t + 1})$
  	\ENDFOR
  	\STATE Let $w^{\Paren{s + 1}} = w^{\Paren{s}}_t$
  \ENDFOR
  \end{algorithmic}
\end{algorithm}

\paragraph{The score oracles}
The exact score oracle $\cO^*_{\mathrm{exact}}$ would, given $S$, and given $w_0^{\Paren{s}}, \ldots, w_{t - 1}^{\Paren{s}}$, would output scores $\tau_{t, i}^{\Paren{s}}$ given by~\eqref{eq:tau-definition} with $U_{t}^{\Paren{s}}$ as given by Algorithm~\ref{alg:mmw-filter2}.
The exact score oracle would also output
\begin{equation}
\label{eq:q-def}
q_t^{(s)} = \Iprod{M(w_t^{(s)}) - \Id, U_{t}^{(s)}} \; .
\end{equation}
Observe that the fact that the $F^{\Paren{s}}_t$ includes a negative identity term does not affect these scores at all, and indeed we can take $U_t^{\Paren{s}}$ to be as in~\eqref{eq:ut-def},
since the parameter $c$ is chosen in any case to normalize $U_{t}^{\Paren{s}}$ to have trace $1$.

As before, we cannot access these exact scores in nearly-linear time, so instead we ask for approximations.
Specifically, we will assume an approximate augmented score oracle $\cO^*_{\mathrm{approx}}$, which given $S$ and $w_0^{\Paren{s}}, \ldots, w_{t - 1}^{\Paren{s}}$, output scores $\tilde{\tau}_{t, i}^{\Paren{s}}$ so that $\tilde{\tau}_{t, i}^{\Paren{s}} \approx_{0.1} \tau_{t, i}^{\Paren{s}}$ for all $i = 1, \ldots, n$, as well as $\tilde{q}_t^{(s)}$ satisfying
\[
\Abs{\tilde{q}_t^{(s)} - q_t} \leq 0.1 \cdot q_t + 0.05 \cdot \Norm{M(w_t) - \Id}_2 \; .
\]
As before, the choice of constants here is arbitrary, and any constants sufficiently small will work.
In Section~\ref{sec:approx-scores} we construct such an approximate augmented score oracle in nearly-linear time.

\subsection{Correctness of \sgalgname}
The rest of this section is dedicated to the proof of the following theorem:
\begin{theorem}
\label{thm:subgaussian-filter}
Let $D$ be a subgaussian isotropic distribution on $\R^d$ with mean $\mu$.
Let $\eps < c$, where $c$ is a universal constant, let $\gamma_1, \gamma_2, \beta_1, \beta_2 > 0$, and let $\xi$ be as in~\eqref{eq:xi-def}.
Let $S$ be a dataset, $|S| = n$, so that $S = \Sgood \cup \Sbad \setminus S_r$ so that $\Sgood$ is $(\eps, \gamma_1, \gamma_2, \beta_1, \beta_2)$-s.g. good with respect to $D$, and $|\Sbad|, |S_r| \leq \eps |S|$.
Suppose that $\Norm{X_i}_2 \leq \kappa_1$ for all $i = 1, \ldots, n$, and moreover $\Norm{M(\frac{1}{n} \one_n) - \Id} \leq O(\eps \log 1 / \eps + \eps \kappa_2)$.
Then $\sgalgname(S, \eps, \cO_{\mathrm{approx}})$ terminates after at most $O(\log \kappa_2)$ epochs, and outputs a $w \in \fS_{n, \eps} $ so that
\[
\| \mu(w) - \mu \|_2 \leq O \Paren{\gamma_1 + \eps \sqrt{\log 1 / \eps} + \sqrt{\eps \xi }} \; .
\]
Moreover, each epoch runs for at most $O(\log d)$ iterations, requires $O(\log d)$ calls to $\cO_{\mathrm{approx}}$, and requires $\Otilde(nd + n \log \kappa_1)$ additional computation.

\end{theorem}
Our main lemma is the following:
\begin{lemma}
\label{lem:epoch-inv-subgaussian}
The following invariants always hold.
There exists some universal constant $C > 0$ so that for all epochs $s$, we have:
\begin{itemize}
\item $w^{\Paren{s}} \in \fS_{n, \eps}$, and
\item If 
\[
\Norm{M(w^{\Paren{s}}) - \Id}_2 > C \cdot \xi \; ,
\] 
then epoch $s$ terminates after $O(\log d)$ iterations, and outputs $w^{\Paren{s + 1}}$ so that $\Norm{M(w^{\Paren{s}}) - \Id}_2 \leq \frac{3}{4} \Norm{M(w^{\Paren{s}}) - \Id}_2$.
\end{itemize} 
\end{lemma}
\noindent
We first demonstrate how this lemma proves Theorem~\ref{thm:subgaussian-filter}.
\begin{proof}[Proof of Theorem~\ref{thm:subgaussian-filter} given Lemma~\ref{lem:epoch-inv-subgaussian}]
By our condition on $M(\tfrac{1}{n} \one_n)$, after at most $s = O(\log \kappa_2 )$ iterations, we must have that 
\[
\Norm{M(w^{\Paren{s}}) - \Id}_2 \leq C \cdot \xi \; .
\]
Since $w^{\paren{s}} \in \fS_{n, \eps}$, Lemma~\ref{lemma:s.g.-geometry} implies that for $\eps < c$ sufficiently small, we have
\[
\Norm{\mu(w) - \mu}_2 \leq O \Paren{\gamma_1 + \eps \sqrt{\log 1 / \eps} + \sqrt{\eps \xi}} \; ,
\]
as claimed.

We now turn to bounding the runtime.
As in Theorem~\ref{thm:naive-cov}, it is clear that we make at most $\log d$ calls to the oracle every epoch, and \textsc{1DFilter} runs in time $O(n \log \kappa_1)$.
Moreover, the approximate eigenvalue computations can still be done in $\Otilde(n d)$ time since we may run power method on $M(w_s^{(t)}) - \Id$, as we can evaluate matrix-vector multiplications against this matrix in $O(nd)$ time.
This completes the proof.
\end{proof}
The remainder of the section is dedicated to the proof of this lemma.
As in the previous section, for simplicity of notation, as we will only consider a fixed epoch $s$, we will drop the superscripts.

The proof of Lemma~\ref{lem:epoch-inv-subgaussian} breaks down into two parts.
First, we will show that assuming we have not yet made sufficient progress, we remain in the regime where the filter is guaranteed to make progress, i.e., the majority of the mass of the $\tau_i$ are from bad points.
This is captured in the following lemma:
\begin{lemma}
\label{lem:sg-stay-good}
At time $t$, suppose that $\Iprod{M_t - \Id, U_t} > \frac{1}{1.1 \cdot 5} \lambda_0$.
Then $w_{t + 1} \in \fS_{n, \eps}$ and $\Iprod{F_{t}, U_t} \leq 0.34 \cdot \Iprod{F_{t - 1}, U_t}$.
\end{lemma}
\noindent
Then, we will show that this implies that the regret bounds of MMW guarantee that we make constant progress in logarithmically many iterations:
\begin{lemma}
\label{lem:sg-inv-2}
Suppose for all $t = 0, \ldots, T - 1$, Lemma~\ref{lem:sg-stay-good} holds, where $T = O(\log d)$.
Then, $\Norm{M(w_T) - \Id}_2 \leq 0.63 \cdot \Norm{M(w_0) - \Id}_2$.
\end{lemma}
\begin{proof}[Proof of Lemma~\ref{lem:sg-stay-good}]
Recall that $m$ was chosen to be the $2 \eps$-percentile of the scores under the weighting given by $w$,
and as before, for simplicity assume that the $\tilde{\tau}_{t, i}$ are in decending order.
The main work in this proof will be to show that the scores $\tilde{\tau}_{t, 1}, \ldots, \tilde{\tau}_{t, m}$ and weights $w_{t, 1}, \ldots, w_{t, m}$ satisfy the conditions of Theorem~\ref{thm:1D}.

The condition that $\Iprod{M_t - \Id, U_t} > \frac{1}{1.1 \cdot 5} \lambda_0$ implies that in this case, we will run the univariate filter.
Let $S'_g = S_g \cap [m]$ and let $S'_b = S_b \cap [m]$, and let $w'_g$ and $w'_b$ the restriction of $w_t$ to $S'_g$ and $S'_b$, respectively.
Observe that $\sum_{i \in S'_b} w_i \leq \sum_{i \in S_b} w_i \leq \eps$, and therefore $\sum_{i \in S'_g} w_i \in [\eps, 2 \eps]$.
We then have
\begin{align*}
\sum_{i \in S'_g} w_{t, i} \tau_{t, i} &= \sum_{i \in S'_g} w_{t, i} \Iprod{\Paren{X_i - \mu(w_t)} \Paren{X_i - \mu(w_t)}^\top, U_t} \\
&\stackrel{(a)}{\leq} 2 \sum_{i \in S'_g} w_{t, i} \Iprod{\Paren{X_i - \mu} \Paren{X_i - \mu}^\top, U_t} + 2 |w'_g| \Iprod{(\mu - \mu(w_t))(\mu - \mu(w_t))^\top, U_t} \\
&\stackrel{(b)}{\leq} \xi + 2 |w'_g| \Iprod{(\mu - \mu(w_t))(\mu - \mu(w_t))^\top, U_t} \\
&\leq \xi + 4 \eps \Norm{\mu - \mu(w_t)}_2^2 \\
&\stackrel{(c)}{\leq} \xi + 4 \eps \Paren{\gamma_1 + O(\eps \sqrt{\log 1 / \eps}) + \sqrt{\eps (\lambda_t + \xi)} }^2 \\
&\leq \xi + 8 \eps \gamma_1 + 8 \eps^2 (\lambda_t + \xi) \\
&\leq \frac{1}{30} \Iprod{M(w_t) - \Id, U_t} \; . \numberthis \label{eq:sgp-bound}
\end{align*}
where (a) follows since for any vectors $x, y, z$, we have $(x - y)(x - y)^\top \preceq 2 (x - z)(x - z)^\top + 2 (y - z)(y - z)^\top$, (b) follows from Fact~\ref{fact:sg-convexity}, (c) follows from Lemma~\ref{lemma:s.g.-geometry}, and the last line follows from our assumption on $\Iprod{M(w_t) - \Id, U_t}$.
From this we conclude that $\sum_{i \in S'_g} w_{t, i} \tilde{\tau}_{t, i} \leq \frac{1}{29} \Iprod{M(w_t) - \Id, U_t}$.

Note that as a consequence of this, we have that $\tilde{\tau}_{t, m} \leq \xi + 8 \eps \gamma_1 + 8 \eps (\lambda_t + \gamma_2)$, since that is an upper bound on the average value of the $\tilde{\tau}_{t, i}$ for $i \in S'_g$, as $|w'_g| \geq \eps$.

We now turn to lower bound the contribution from $S_b'$.
First observe that
\begin{align*}
\sum_{i \in S_b} w_i \tau_{t, i} &= \sum_{i \in S_b} w_{t, i} \Iprod{\Paren{X_i - \mu(w_t)} \Paren{X_i - \mu(w_t)}^\top, U_t} \\
&= \sum_{i \in S_b} w_{t, i} \Iprod{\Paren{X_i - \mu(w_t)} \Paren{X_i - \mu(w_t)}^\top - \Id, U_t} - |w_b| \\
&= |w_t| \Iprod{M(w_t) - \Id, U_t} - \sum_{i \in S_g \cap S} w_{t, i} \Iprod{\Paren{X_i - \mu(w_t)} \Paren{X_i - \mu(w_t)}^\top - \Id, U_t} - |w_b| \\
&\stackrel{(a)}{\geq} \frac{999}{1000} \Iprod{M(w_t) - \Id, U_t} - \xi \\
&\geq \frac{99}{100} \Iprod{M(w_t) - \Id, U_t} \; ,
\end{align*}
where (a) follows from Fact~\ref{fact:sg-convexity} and since $|w_b| \leq \eps$, and the last inequality follows by our assumption on $\Iprod{M(w_t) - \Id, U_t}$.
Therefore
\begin{align*}
\sum_{i \in S_b'} w_i \tau_{t, i} &= \sum_{i \in S_b} w_i \tau_{t, i} - \sum_{i \in S_b \setminus S_b'} w_i \tau_{t, i} \\
&\geq \frac{99}{100} \Iprod{M(w_t) - \Id, U_t} - \sum_{i \in S_b \setminus S_b'} w_i \tau_{t, i} \\
&\stackrel{(a)}{\geq} \frac{99}{100} \Iprod{M(w_t) - \Id, U_t} - 1.21 \cdot |w_b| \Paren{\xi + 8 \eps \gamma_1 + 8 \eps (\lambda_t + \gamma_2)} \\
&\stackrel{(b)}{\geq} \frac{49}{50} \Iprod{M(w_t) - \Id, U_t} \; , \numberthis \label{eq:sbp-bound}
\end{align*}
where (a) follows since $\tau_{t, i} \leq 1.1 \tilde{\tau}_{t, i} \leq \cdot 1.1^2 \cdot \tau_{t, m}$ for all $i \in S_b \setminus S_b'$, and (b) follows from our assumption on $\Iprod{M(w_t) - \Id, U_t}$.

Equations~\eqref{eq:sgp-bound} and~\eqref{eq:sbp-bound} together imply that our set of scores in this instance will be in the setting of Theorem~\ref{thm:1D}.
Thus Theorem~\ref{thm:1D} guarantees that the univariate filter outputs a set of weights $w'_1, \ldots, w'_m$  so that $\sum_{i = 1}^m w'_i \tilde{\tau}_{t, i} \leq \frac{1}{5} \sum_{i = 1}^m w_{t, i} \tilde{\tau}_{t, i}$.
Notice that in particular this implies that $w_{t + 1} \in \fS_{n, \eps}$, which proves one of the claims in the Lemma.
To complete the proof of the lemma, observe that
\begin{align*}
\Iprod{M(w_t) - \Id, U_t} - \Iprod{F_t, U_t} &= \sum_{i = 1}^m (w_{t, i} - w'_t) \Paren{\tau_{t, i} - 1} \\
&\stackrel{(a)}{\geq} \sum_{i = 1}^m (w_{t, i} - w'_t) \tau_{t, i} - 2 \eps \\
&\stackrel{(b)}{\geq} \frac{4}{1.21 \cdot 5} \sum_{i = 1}^m w_{t, i} \tau_{t, i} - 2 \eps \\
&\stackrel{(c)}{\geq} \frac{4}{1.21 \cdot 5} \frac{49}{50} \Iprod{M(w_t) - \Id, U_t} - 2 \eps \\
&\stackrel{(d)}{\geq} 0.63 \Iprod{M(w_t) - \Id, U_t} \; ,
\end{align*}
where (a) follows since $m$ is the $2 \eps$-percentile, (b) follows from the guarantee of Theorem~\ref{thm:1D}, (c) follows from~\eqref{eq:sbp-bound}, and (d) follows from our assumption on $\lambda_t$.
Rearranging terms completes the proof.
\end{proof}
\noindent
We now show that this is enough to guarantee Lemma~\ref{lem:sg-inv-2}, which guarantees we make constant multiplicative progress in every epoch.
\begin{proof}[Proof of Lemma~\ref{lem:sg-inv-2}]
Observe that if we terminate prematurely we clearly satisfy the lemma.
Thus we may assume we do not terminate until timestep $T - 1$.
Lemma~\ref{lem:sg-stay-good} then implies that no matter which update we do at time $t$ for $t = 0, \ldots, T - 1$, we have the guarantee that 
\begin{equation}
\label{eq:sg-inv-guarantee}
\Iprod{F_t, U_t} \leq 0.34 \Iprod{F_{t - 1}, U_t} \leq 0.34 \Iprod{F_0, U_t} \leq 0.34 \Norm{M(w_0) - \Id}_2 \; .
\end{equation}
Moreover, by Lemma~\ref{lem:psd-ordering}, we have that $M(w_t) - \Id \preceq M(w_0) - \Id$, and hence $\frac{1}{\alpha} \Paren{M(w_t) - \Id} \preceq I$ by our choice of $\alpha$.
Therefore by our regret bound, we have
\begin{equation}
\label{eq:sg-regret}
\Norm{\sum_{i= 0}^{T - 1} \Paren{M(w_t) - \Id}}_2 \leq \sum_{t = 0}^{T - 1} \Iprod{U_t, M(w_t) - \Id} + \sum_{t = 0}^{T - 1} \Iprod{U_t, \Abs{M(w_t) - \Id}} \frac{\Norm{M(w_t) - \Id}_2}{\Norm{M(w_0) - \Id}_2} + \log (d) \cdot \Norm{M(w_0) - \Id}_2 \; .
\end{equation}
By Lemma~\ref{lem:sg-PSD-ordering}, we know that for all $t = 0, \ldots, T - 1$, we must have
\begin{equation}
\label{eq:sg-psd-ordering-regret}
M(w_t) - \Id \succeq - \xi \cdot \Id \; .
\end{equation}
This will allow us to simplify a number of expressions.
In particular, since $M(w_t) - \Id \preceq M(w_0) - \Id$, this implies that all positive eigenvalues of $M(w_t) - \Id$ are smaller than $\lambda_0$, and~\eqref{eq:sg-psd-ordering-regret} implies that all negative eigenvalues are bounded in absolute value by $\lambda_0$, by our assumption on $\lambda_0$.
Hence
\begin{equation}
\label{eq:sg-ratio-bound}
\frac{\Norm{M(w_t) - \Id}_2}{\Norm{M(w_0) - \Id}_2} \leq 1 \; .
\end{equation}
Another implication is that 
\[
\Abs{M(w_t) - \Id} \preceq M(w_t) - \Id + 2 \xi \cdot \Id \; ,
\]
and hence
\begin{equation}
\label{eq:sg-removing-abs}
\Iprod{U_t, \Abs{M(w_t) - \Id}} \leq \Iprod{U_t, M(w_t) - \Id} + 2\xi \; .
\end{equation}
Finally, we observe that~\eqref{eq:sg-psd-ordering-regret} and Lemma~\ref{lem:psd-ordering} together imply that either 
\begin{align}
\Norm{M(w_T) - \Id}_2 &\leq O(\xi) \;, ~\mbox{or} \label{eq:sg-case1-LHS} \\
\Norm{\sum_{i= 0}^{T - 1} \Paren{M(w_t) - \Id}}_2 &\geq T \cdot \Norm{M(w_{T - 1}) - \Id}_2 \label{eq:sg-case2-LHS} \; .
\end{align}
In the case of~\eqref{eq:sg-case1-LHS}, we are clearly done, so we may assume that we are in the case of~\eqref{eq:sg-case2-LHS}.
Thus, plugging in~\eqref{eq:sg-ratio-bound},~\eqref{eq:sg-removing-abs}, and~\eqref{eq:sg-case2-LHS} into~\eqref{eq:sg-regret}, and dividing by $T$, we obtain
\begin{align*}
\Norm{M(w_{T - 1}) - \Id}_2 &\leq \frac{2}{T} \sum_{i = 0}^{T - 1} \Paren{\Iprod{U_t, M(w_t) - \Id}} + 2 \xi + \frac{\log n}{T} \Norm{M(w_0) - \Id}_2 \\
&\stackrel{(a)}{\leq} \Paren{0.68 + \frac{\log n}{T}} \Norm{M(w_0) - \Id}+2 + 2\Paren{\gamma_2 + 2 \gamma_1^2 + O(\eps \log 1 / \eps)} \\
&\stackrel{(b)}{\leq} \frac{3}{4} \Norm{M(w_0) - \Id}_2 \; ,
\end{align*}
where (a) follows from~\eqref{eq:sg-inv-guarantee}, and (b) follows since $T = \Theta (\log d)$, and since $\Norm{M(w_0) - \Id}_2$ is a large constant factor larger than $\gamma_2 + 2 \gamma_1^2 + O(\eps \log 1 / \eps)$, by assumption.
This completes the proof.
\end{proof}


\section{Fast approximate score oracles}
\label{sec:approx-scores}
In this section we describe how to implement the approximate score oracles and approximate augmented score oracles in nearly-linear time.
Recall that an approximate score oracle takes as input a set of points $S \subset \R^d$ of size $n$, and a sequence of weights $w_0, \ldots, w_{t - 1}, w_t$, and computes $\tilde{\tau}_t \in \R^n$, where for all $i = 1, \ldots, n$ we have $\tilde{\tau}_{t, i} \approx_{0.1} \tau_{t, i}$, where
\begin{equation}
\label{eq:tau-def2}
\tau_{t, i} = \Paren{X_i - \mu(w_t)}^\top U_t \Paren{X_i - \mu(w_t)} \; ,
\end{equation}
where
\[
U_t = \exp \Paren{c \Id + \alpha \sum_{i = 0}^{t - 1} M(w_i)} = \frac{\exp \Paren{\alpha \sum_{i = 0}^{t - 1} M(w_i)}}{\tr \exp \Paren{\alpha \sum_{i = 0}^{t - 1} M(w_i)}} \; ,
\]
and $\alpha > 0$ is a parameter.
Additionally, recall that in Section~\ref{sec:subgaussian} we additionally require that the score oracle be able to produce $\tilde{q}_t$ so that
\begin{equation}
\label{eq:q-def2}
\Abs{\tilde{q}_t - q_t} \leq 0.1 \cdot q_t + 0.05 \cdot \Norm{M(w_t) - \Id}_2 \; , \mbox{where} \; q_t = \Iprod{M(w_t) - \Id, U_t} \; .
\end{equation}
Our main result in this section is the following, which says that it is possible to achieve such approximations with high probability in nearly linear time:
\begin{lemma}
\label{lem:approx-scores}
Let $S = \{X_1, \ldots, X_n\} \subseteq \R^d$, let $\delta > 0$, and let $w_0, \ldots, w_t$ be as above.
Then there is an algorithm $\textsc{ApproxScores} (S, w_0, \ldots, w_t, \delta)$ which outputs $\tilde{\tau}_{t, i}$ and $\tilde{q}_t$ so that with probability $1 - \delta$, we have $\tilde{\tau}_{t,i} \approx_{0.1} \tau_{t,i}$ for all $i = 1, \ldots, n$, where $\tau_t, \tilde{q}_t$ are defined in~\eqref{eq:tau-def2} and~\eqref{eq:q-def2}, respectively.
Moreover, $\textsc{ApproxScores}$ runs in time $\Otilde (t n d \log 1 / \delta)$.
\end{lemma}
\noindent
If the output of \textsc{ApproxScores} satisfies the conditions of Lemma~\ref{lem:approx-scores}, we say that \textsc{ApproxScores} \emph{succeeds}.

\subsection{Tools from randomized numerical linear algebra}
We need a few tools which are standard in the design of fast algorithms based on matrix multiplicative weights.
The first is the standard Johnson-Lindenstrauss dimension reduction lemma:
\begin{lemma}[Johnson-Lindenstrauss lemma~\cite{johnson1986extensions}]
\label{lem:jl}
  Let $\Phi \in \R^{r \times d}$ be a matrix whose entries are i.i.d. samples from $\cN(0,1/r)$.
  For every vector $u \in \R^d$ and every $\e \in (0,1)$,
  \[
  \Pr \Brac{ (1- \e) \|u\|_2 \leq \|\Phi u\|_2 \leq (1+\e) \|u\|_2} \geq 1 - \exp(-\Omega(\e^2 r))\mper
  \]
\end{lemma}
\noindent
We also require the following, slightly stronger version of the JL guarantee, which states that it preserves matrix inner products:
\begin{lemma}
\label{lem:jl-trace}
  Let $A, U \in \R^{d \times d}$.
  Suppose $U = BB^\top$ for some symmetric $B$.
  Let $S \in \R^{r \times d}$ have i.i.d. entries from $\cN(0,1/r)$.
  There is a universal constant $c$ such that for all $\e > 0$,
  \[
  \Pr \Brac{|\iprod{A, BS^\top SB} - \iprod{A,U}| > \e \|A\|_2 \cdot \tr(U)} \leq 2 \exp(-cr \cdot \min(\e,\e^2))\mper
  \]
\end{lemma}
\begin{proof}
  Notice that $\E \Brac{\iprod{A, BS^\top SB} = \iprod{A,U}}$.
  By the Hanson-Wright inequality~\cite{rudelson2013hanson} together with standard arguments about averages of i.i.d. sub-exponential random variables, for every $t > 0$,
  \[
  \Pr \Brac{ |\iprod{ A, BS^\top SB} - \iprod{A,U} | > t} \leq 2 \exp(-\ell \cdot \Omega(\min(t^2/\|BAB\|_F^2, t/\|BAB\|_2 )))\mper
  \]
  To finish the proof it will be enough to show that
  \[
  \|BAB\|_F^2 \leq \|A\|_2^2 \cdot \tr(U)^2 \quad \text{and} \quad \|BAB\|_2 \leq \|A\|_2 \cdot \tr(U) \mper
  \]

  For the first statement, note that
  \[
  \|BAB\|_F^2 = \tr(AUAU) = \iprod{AUA, U} \leq \|AUA\|_2 \cdot \tr(U) \; .
  \]
  Since spectral norm is sub-multiplicative, $\|AUA\|_2 \leq \|A\|_2^2 \|U\|_2 \leq \|A\|_2^2 \cdot \tr(U)$.

  It remains to prove $\|BAB\|_2 \leq \|A\|_2 \cdot \tr (U)$.
  Again we have
  \[
  \|BAB\|_2 \leq \|B\|_2^2 \|A\|_2 = \|U\|_2 \|A\|_2 \leq \tr(U) \cdot \|A\|_2 \; .
  \]
  which finishes the proof.
\end{proof}
\noindent
We will also make use of Taylor series approximations to the matrix exponential function.
The next lemma helps to control the errors incurred by such approximations.

\begin{lemma}[folklore, see e.g.~\cite{steurer2010fast}]
\label{lem:taylor}
  For $\ell \in \N$, let $P_\ell (Y) = \sum_{j=0}^\ell \tfrac 1 {j!} (Y)^j$ be the degree-$\ell$ Taylor series approximation to $\exp(Y)$.
  For every $n \times n$ symmetric real matrix $Y \succeq 0$,
  \[
  (1-e^{-\ell}) \exp(Y) \preceq P_\ell(Y) \preceq (1+e^{-\ell}) \exp(Y)\mper
  \]
\end{lemma}

(Note that when implementing our outlier detction algorithms we approximate the matrix exponential by Chebyshev polynomials rather than Taylor series.)


\subsection{Efficient approximate score oracles}
With these tools in hand, we are now ready to describe \textsc{ApproxScores}.
For some sufficiently large constant $C > 0$, let $\delta' = \delta / 3$, let $r = C \log n / \delta'$, let $\ell = C \log d$, and let $S \in \R^{r \times d}$ be a matrix with i.i.d. entries from $\normal (0, 1/r)$.
The algorithm will form the $r \times d$ matrix 
\begin{equation}
\label{eq:Arl}
A_{r, \ell} = S \cdot P_\ell \Paren{ \frac{\alpha}{2} \sum_{t = 0}^{t - 1} M(w_t) } \; .
\end{equation}
The estimate for the scores will then be given by
\begin{equation}
\label{eq:approx-tau}
\tilde{\tau}_{t, i} = \frac{1}{\tr (A_{r, \ell} A_{r, \ell}^\top)}\Norm{ A_{r, \ell} (X_i - \mu(w_t)) }_2^2 \; ,
\end{equation}
and the estimate for $q_{t, i}$ will be given by
\begin{equation}
\label{eq:approx-q}
\tilde{q}_{t, i} = \sum_{i = 1}^n \Paren{\tilde{\tau}_{t, i} - 1} \; .
\end{equation}
The formal pseudocode is given in Algorithm~\ref{alg:approx-scores}.
\begin{algorithm}[ht]
\caption{Randomized nearly-linear time approximate score computation}
  \label{alg:approx-scores}
  \begin{algorithmic}[1]
  \STATE \textbf{Input:} dataset $S \subset \R^d$ of size $n$, weight vectors $w_0, \ldots, w_t$, failure probability $\delta > 0$, learning rate $\alpha > 0$.
  \STATE Let $C > 0$ be a universal constant sufficiently large
  \STATE Let $\delta' = \delta / 3$, let $r = C \log n / \delta'$, and let $\ell = C \log d$.
  \STATE Let $S \in \R^{r \times d}$ have entries drawn i.i.d. from $\normal (0, 1/r)$.
  \STATE Compute $A_{r, \ell}$ as defined as in~\eqref{eq:Arl}.
  \STATE For all $i = 1, \ldots, n$, let $\tilde{\tau}_{t, i}$ be as in~\eqref{eq:approx-tau}, and let $\tilde{q}_t$ be as in~\eqref{eq:approx-q}.
  \RETURN $\tilde{\tau}_t, \tilde{q}_t$
  \end{algorithmic}
\end{algorithm}
\noindent
We first demonstrate that \textsc{ApproxScores} indeed runs in the claimed runtime:
\begin{lemma}
\label{lem:approx-scores-runtime}
$\textsc{ApproxScores}(S, w_0, \ldots, w_t, \delta)$ runs in time $\Otilde(t n d \log 1 / \delta)$.
\end{lemma}
\begin{proof}
The main algorithmic work being done in \textsc{ApproxScores} is to form the matrix $A_{r, \ell}$.
We claim that this matrix can be formed in time $\Otilde(t n d \log 1/\delta)$.
Afterwards, notice that $\tr (A_{r, \ell} A_{r, \ell}^\top)$ can be computed in time $O(dr^2)$, and given that, each $\tilde{\tau}_{t, i}$ can be computed in time $O(n r)$.
Given the $\tilde{\tau}_{t, i}$, we then observe that $\tilde{q}_t$ can be computed in time $O(n)$.
Since $r = O(\log n / \delta)$ and $\ell = O(\log d)$, we conclude that the overall runtime of $\textsc{ApproxScores}$ is dominated by the time to form $A_{r, \ell}$, and so it runs in time $\Otilde(t n d \log 1/\delta)$.

We now demonstrate how to form $A_{r, \ell}$ efficiently.
Observe that for any vector $v \in \R^d$, we can evaluate $v^\top \sum_{i = 0}^{t - 1} M(w_i)$ in time $O(t n d)$.
By iterating this process, we can compute $v^\top \Paren{\sum_{i = 0}^{t - 1} M(w_i)}^j$ in time $O(j t n d)$, and thus we can compute $v^\top P_\ell \Paren{-\frac{\alpha}{2} \sum_{i = 0}^{t - 1} M(w_i)}$ in time $O(\ell^2 t n d)$.
Since $S$ has $r = O(\log n / \delta)$ rows, and $\ell = O(\log d)$, we can therefore form $A_{r, \ell}$ in time $\Otilde \Paren{t n d \log 1 / \delta}$ by forming each row of $A_{r, \ell}$.
\end{proof}
\noindent
Note that \textsc{ApproxScores} is an approximate augmented score oracle, and so it is also clearly an approximate score oracle.
In the remainder of this section, we show:
\begin{lemma}
\label{lem:approx-scores-correctness}
With probability $\geq 1-\delta$, the output of \textsc{ApproxScores} satisfies $\tilde{\tau}_{t, i} \approx_{0.1} \tau_{t, i}$ for all $i = 1, \ldots, n$, and $\tilde{q}_t \approx_{0.1} q_t$.
\end{lemma}
\begin{proof}
Let $M = \alpha \sum_{i = 0}^{t-1} M(w_t)/2$, and let $Y_t = \exp \Paren{2 M}$, so that $U_t = \frac{Y_t}{\tr(Y_t)}$, and so that $A_{r, \ell} = S \cdot P_\ell(M)$.
Then, by Lemma~\ref{lem:taylor} and our choice of $\ell$, we have $\Norm{Y_t^{1/2} - P_\ell (M)}_2 \leq \frac{0.01}{d} \cdot \Norm{Y_t^{1/2}}_2$, which immediately implies that $\Norm{Y_t - P_\ell (M)^2 }_2 \leq \frac{0.03}{d} \cdot \Norm{Y_t}_2$.
Notice that this implies that $\tr (P_\ell(M)^2) \approx_{0.03} \tr (Y_t)$.

We now condition on the event that the following three events hold simultaneously:
\begin{align}
\Norm{S P_{\ell} (M) (X_i - \mu(w_t))}_2^2 \approx_{0.01} \Norm{P_\ell (M) (X - \mu(w_t))}_2^2 ~&\mbox{for all $i = 1, \ldots, n$} \; , \label{eq:tau-cond} \\
\tr (P_\ell (M) S^\top S P_\ell (M)) &\approx_{0.01} \tr (P_\ell (M)^2) \; , \label{eq:tr-cond} \\
\Abs{\Iprod{M(w_t) - \Id, P_\ell (M) S^\top S P_\ell (M)} - \Iprod{M(w_t) - \Id, P_{\ell}(M)^2}} &\leq 0.01\cdot \tr(P_\ell(M)^2) \Norm{M(w_t) - \Id}_2  \; . \label{eq:iprod-cond}
\end{align}
By our choice of $\delta'$, Lemma~\ref{lem:jl}, and a union bound, we know that~\eqref{eq:tau-cond} holds with probability at least $1 - \delta / 3$.
By instantiating Lemma~\ref{lem:jl-trace} with $A = I$ and $A = M(w_t) - \Id$ respectively, we also know that~\eqref{eq:tr-cond} and~\eqref{eq:iprod-cond} each hold with probability at least $1 - \delta / 3$.
Thus, by a union bound, all three conditions hold simultaneously with probability at least $1 - \delta$.
We claim that conditioned on these three events, the conditions of the lemma are satisfied.
Indeed, we have
\begin{align*}
\tilde{\tau}_{t, i} &= \frac{1}{\tr (P_\ell (M) S^\top S P_\ell (M))} \Norm{ S \cdot P_{\ell} (M) (X_i - \mu(w_t)) }_2^2 \\
&\approx_{0.0121} \frac{1}{\tr (P_\ell (M)^2)} \Norm{ P_{\ell} (M) (X_i - \mu(w_t)) }_2^2 \\
&\approx_{0.0363} \frac{1}{\tr (Y_t)} \Norm{ P_{\ell} (M) (X_i - \mu(w_t)) }_2^2 \\
&\approx_{0.01} \frac{1}{\tr (Y_t)} (X_i - \mu(w)_t)^\top Y_t (X_i - \mu(w)_t) \\
&= \tau_{t, i} \; .
\end{align*}
Here the second line follows from~\eqref{eq:tau-cond} and~\eqref{eq:tr-cond}, the third line follows from our condition on the trace, and the final approximation follows since $P_\ell (M)$ approximates $Y_t^{1/2}$ in PSD order and they commute.
This proves the claim about the $\tilde{\tau}_{t,i}$.

To conclude, we observe that
\begin{align*}
\tilde{q}_{t, i} &= \frac{1}{\tr (P_\ell (M) S^\top S P_\ell (M))} \sum_{i = 1}^n \Paren{ \Norm{ S \cdot P_{\ell} (M) (X_i - \mu(w_t)) }_2^2 - \tr (P_\ell (M) S^\top S P_\ell (M))} \\
&= \frac{1}{\tr (P_\ell (M) S^\top S P_\ell (M))} \Iprod{M(w_t) - \Id, P_\ell (M) S^\top S P_\ell (M)} \\
&= \frac{1}{\tr (P_\ell (M) S^\top S P_\ell (M))} \Paren{\Iprod{M(w_t) - \Id, P_\ell (M)^2} + \eta} \; ,
\end{align*}
where $|\eta| \leq 0.01 \cdot\tr (P_\ell (M)) \cdot \Norm{M(w_t) - \Id}_2$ by~\eqref{eq:iprod-cond}.
We further have
\begin{align*}
& \frac{1}{\tr (P_\ell (M) S^\top S P_\ell (M))} \Iprod{M(w_t) - \Id, P_\ell (M)^2}\\
& \approx_{0.01} \frac{1}{\tr (P_\ell (M)^2)} \Iprod{M(w_t) - \Id, P_\ell (M)^2}\\
& = (1 \pm 0.1)  \Iprod{M(w_t) - \Id, U_t}  \pm 0.02 \|M(w_t) - \Id\|_2\\
&= (1 \pm 0.1) q_t  \pm 0.02 \|M(w_t) - \Id\|_2 \; .
\end{align*}
Therefore
\begin{align*}
\Abs{\tilde{q}_{t} - q_{t}} \leq 0.1 q_t + \frac{|\eta|}{\tr (P_\ell (M) S^\top S P_\ell (M))} + 0.02 \|M(w_t) - \Id\|_2 & \leq 0.1 \cdot q_t + 0.05 \frac{|\eta|}{\tr (Y_t)} + 0.02\|M(w_t) - \Id\|_2 \\
& \leq 0.1 \cdot q_t + 0.05 \Norm{M(w_t) - \Id}_2 \; ,
\end{align*}
which was the desired bound for $\tilde{q}_t$.
\end{proof}
\noindent
Lemmata~\ref{lem:approx-scores-runtime} and~\ref{lem:approx-scores-correctness} together immediately imply Lemma~\ref{lem:approx-scores}.

\section{Robust mean estimation: putting it all together}
\label{sec:all-together-now}
In this section, we formally combine the guarantees derived in the previous sections to prove Theorems~\ref{thm:main-bounded-cov} and~\ref{thm:main-gaussian}.

\subsection{Proof of Theorem~\ref{thm:main-bounded-cov}}
Given the machinery we've developed, the algorithm is straightforward to describe.
Given a corrupted dataset $S$ and $\delta > 0$, run $\textsc{NaivePrune} (S, \sqrt{4 dn / \delta}, \delta / 4)$  to obtain a pruned dataset $S'$.
Center all points in $S'$ with the empirical mean of $S'$.
Then, run $\algname (S', \textsc{ApproxScores})$, with $\kappa = \sqrt{4 dn / \delta}$, and the $\delta$ parameter in $\textsc{ApproxScores}$ set to $O(\delta / (\log \kappa \log d))$.
The formal pseudocode is presented in Algorithm~\ref{alg:fast-cov-rme}.

\begin{algorithm}[ht]
\caption{Nearly-linear time robust mean estimation under bounded second moments}
  \label{alg:fast-cov-rme}
  \begin{algorithmic}[1]
  \STATE \textbf{Input:} dataset $S \subset \R^d$ of size $n$, failure probability $\delta > 0$
  \STATE Let $S' = \textsc{NaivePrune} (S, \sqrt{4 dn / \delta}, \delta / 4)$.
  \STATE Let $\kappa = \sqrt{4 dn / \delta}$.
  \STATE Center all points in $S'$ at the empirical mean of $S'$.
  \STATE Let $\cO$ be $\textsc{ApproxScores}$ with failure probability $O(\delta / (\log \kappa \log d))$.
  \STATE Let $\widehat{\mu} = \algname(S', \cO)$.
  \RETURN $\widehat{\mu}$.
  \end{algorithmic}
\end{algorithm}
\noindent
We now prove correctness.
\begin{proof}[Proof of Theorem~\ref{thm:main-bounded-cov}]
Recall that by definition, we may assume that $S = T \cup S_b \setminus S_r$, where $T$ is a set of $n$ i.i.d. samples from $D$, and $|S_b|, |S_r| \leq \eps n$.
Let $\gamma_1, \gamma_2$ be as in~\eqref{eq:conc-second-moments}.
We condition on four events:
\begin{itemize}
	\item $\Norm{X_i - \mu}_2 \leq \sqrt{\frac{4 dn}{\delta}}$ for all $i \in T$,
	\item $\textsc{NaivePrune}(S', \sqrt{4 dn / \delta}, \delta / 4)$ succeeds,
	\item $T = S_g \cup T_b$, where $S_g$ is $(\gamma_1, \gamma_2)$-good with respect to $D$, and $|T_b| \leq \eps n$, and
	\item every time it is called, $\textsc{ApproxScores}$ suceeds.
\end{itemize}
By Chebyshev's inequality, and an union bound over all $n$ points in $T$, the first bullet point holds with probability at least $\delta / 4$.
By a further union bound and by adjusting constants in our choices of $\delta$, all four of these conditions hold simultaneously with probability at least $1 - \delta - \exp (-\eps n)$.
We now claim that, conditional on these four events, we output a $\mu(w)$ so that $\Norm{\mu - \mu(w)}_2 = O(\sqrt{\eps}) + \Otilde(\sqrt{d / (n \delta)})$.
Indeed, the first two conditions imply that \textsc{NaivePrune} does not throw away any points in $T$, and moreover, all points $X$ in the set $S'$ satisfy $\Norm{X_i}_2 \leq \sqrt{4 d n / \delta}$ after centering.
Thus, since the scores output by $\textsc{ApproxScores}$ satisfy the necessary conditions for Theorem~\ref{thm:naive-cov}, it follows that the final output satisfies $\Norm{\mu - \mu(w)}_2 = O(\sqrt{\eps}) + \Otilde(\sqrt{d / (n \delta)})$, as claimed.

We now turn to runtime.
Since each epoch runs for at most $O(\log d)$ iterations, and so we run for at most $O(\log \kappa \log d)$ iterations, the total time spent running $\textsc{ApproxScores}$ is at most $\Otilde (n d \log 1 / \delta)$.
Thus overall the algorithm runs in time $\Otilde(n d \log 1 / \delta)$, as claimed.
\end{proof}

\subsection{Proof of Theorem~\ref{thm:main-gaussian}}
Again, the algorithm is straightforward.
Given a corrupted dataset $S$, parameters $\eps >0$ and $\delta > 0$, run $\textsc{NaivePrune} (S, \sqrt{4 d \log (n / \delta)}, \delta / 4)$  to obtain a pruned dataset $S'$.
Center all points in $S'$ with the empirical mean of $S'$.
Then, as above, run $\sgalgname (S', \eps, \textsc{ApproxScores})$, with $\kappa = \sqrt{4 d \log (n / \delta)}$, and the $\delta$ parameter in $\textsc{ApproxScores}$ set to $O(\delta / (\log \kappa / \eps \log d))$.
The formal pseudocode is presented in Algorithm~\ref{alg:fast-gauss-rme}.

\begin{algorithm}[ht]
\caption{Nearly-linear time robust mean estimation for isostropic subgaussian distributions}
  \label{alg:fast-gauss-rme}
  \begin{algorithmic}[1]
  \STATE \textbf{Input:} dataset $S \subset \R^d$ of size $n$, failure probability $\delta > 0$, fraction of error $\eps$
  \STATE Let $S' = \textsc{NaivePrune} (S, \sqrt{4 d \log (n / \delta)}, \delta / 4)$.
  \STATE Let $\kappa = \sqrt{4 d \log (n / \delta)}$.
  \STATE Center all points in $S'$ at the empirical mean of $S'$.
  \STATE Let $\cO$ be $\textsc{ApproxScores}$ with failure probability $O(\delta / (\log \kappa / \eps \log d))$.
  \STATE Let $\widehat{\mu} = \sgalgname(S', \cO, \eps)$.
  \RETURN $\widehat{\mu}$.
  \end{algorithmic}
\end{algorithm}
\noindent
We now prove correctness.
The proof is very similar to the proof presented above.
\begin{proof}[Proof of Theorem~\ref{thm:main-gaussian}]
Recall that by definition, we may assume that $S = S_g \cup S_b \setminus S_r$, where $T$ is a set of $n$ i.i.d. samples from $D$, and $|S_b|, |S_r| \leq \eps n$.
Let $\gamma_1, \gamma_2, \beta_1, \beta_2$ be as in~\eqref{eq:conc-gaussian} and~\eqref{eq:conc-gaussian-2}, and let $\xi$ be as in~\eqref{eq:xi-def}.
We condition on four events:
\begin{itemize}
	\item $\Norm{X_i - \mu}_2 \leq \sqrt{4 d \log (n / \delta)}$ for all $i \in S_g$,
	\item $\textsc{NaivePrune}(S', \sqrt{4 d \log (n / \delta)}, \delta / 4)$ succeeds,
	\item $S_g$ is $(\eps, \gamma_1, \gamma_2, \beta_1, \beta_2)$-s.g. good with respect to $D$,
	\item every time it is called, $\textsc{ApproxScores}$ suceeds.
\end{itemize}
By standard concentration inequalities for chi-squared random variables, and an union bound over all $n$ points in $T$, the first bullet point holds with probability at least $\delta / 4$.
By a further union bound and by adjusting constants in our choices of $\delta$, all four of these conditions hold simultaneously with probability at least $1 - \delta$.
We now claim that, conditional on these four events, we output a $\mu(w)$ so that 
\begin{align}
\Norm{\mu - \mu(w)}_2 = O \Paren{\gamma_1 + \eps \sqrt{\log 1 / \eps} + \sqrt{\eps \xi} } \label{eq:sg-final-condition} \; .
\end{align}
Notice that in this case, by our choice of $\xi$, and for $\eps \leq 1/2$ we have that 
\begin{align*}
\gamma_1 + \eps \sqrt{\log 1 / \eps} + \sqrt{\eps \xi} &= \gamma_1 + \eps \sqrt{\log 1 / \eps} + \sqrt{\eps (\gamma_2 + 2 \gamma_1^2 + 4 \eps^2 \beta_1^2 + 2 \eps \beta_2 + O(\eps \log 1/ \eps))} \\
&\leq 2 \gamma_1 + \sqrt{\eps \gamma_2} + 2 \eps^{3/2} \beta_1 + 2 \eps \sqrt{\beta_2} + O(\eps \sqrt{\log 1 / \eps}) \; .
\end{align*}
Letting $A = C \frac{d + \log 1 / \delta}{n}$ for some constant $C$ sufficiently large, by~\eqref{eq:conc-gaussian} and~\eqref{eq:conc-gaussian-2}, we now have the following inequalities for each term in the above sum:
\begin{align*}
\gamma_1 &\leq A^{1/2} \\
\sqrt{\eps \gamma_2} &\leq \sqrt{\eps (A^{1/2} + A)} \leq \sqrt{\eps A^{1/2}} + \sqrt{\eps A} \stackrel{(a)}{\leq} \frac{1}{2} \eps + \frac{1}{2} A^{1/2} + A^{1/2} \\
\eps^{3/2} \beta_1 &\leq C \cdot \eps^{3/2} \sqrt{\log 1/\eps} + \eps A^{1/2} \\
\eps \sqrt{\beta_2} &\leq C \cdot \eps \sqrt{\log 1 / \eps} + A^{1/2} \; ,
\end{align*}
where (a) follows from the arithmetic mean-geometric mean inequality.
Thus, overall we conclude that, assuming~\eqref{eq:sg-final-condition}, we have
\[
\Norm{\mu - \mu(w)}_2 \leq C_1 \cdot \eps \sqrt{\log 1/\eps} + C_2 \cdot \sqrt{\frac{d + \log 1 / \delta}{n}} \; ,
\]
for some universal constants $C_1, C_2$ sufficiently large, as desired.
It now remains to demonstrate that~\eqref{eq:sg-final-condition} is satisfied.

The first two conditions imply that \textsc{NaivePrune} does not throw away any points in $S_g$, and moreover, all points $X$ in the set $S'$ satisfy $\Norm{X_i}_2 \leq \sqrt{4 d \log (n / \delta)}$ after centering.
By standard arguments, this implies that $\Norm{M(\tfrac{1}{n} \one_n) - \Id}_2 \leq O(\eps \log 1 / \eps + \eps \kappa)$, for $\kappa = \sqrt{ d \log (n / \delta)}$.
Thus, since the scores output by $\textsc{ApproxScores}$ satisfy the necessary conditions for Theorem~\ref{thm:naive-cov}, it follows that the final output satisfies $\Norm{\mu - \mu(w)}_2 = O(\gamma_1 + \eps \sqrt{\log 1 / \eps} + \sqrt{\eps (\gamma_1 + \gamma_2})$, as claimed.

We now turn to runtime.
Since each epoch runs for at most $O(\log d)$ iterations, and so we run for at most $O(\log \kappa / \eps \log d)$ iterations, the total time spent running $\textsc{ApproxScores}$ is at most $\Otilde (n d \log 1 / \delta \log 1 / \eps)$.
Thus overall the algorithm runs in time $\Otilde(n d \log 1 / \delta \log 1 / \eps)$, as claimed.
\end{proof}

\section{Outlier detection: additional experiments and fast implementation}
\label{sec:local-methods}

In this section we compare \QUE{} scoring against some additional outlier detection methods from prior literature.
We also discuss and describe experiments involving our nearly-linear time implementation of \QUE{} scoring.

\subsection{QUE scoring versus local methods}

The work \cite{campos2016evaluation} compares a number of outlier detection methods (mainly those based on $k$-NN distances) on several datasets, both low and high dimensional.
We evaluate \QUE{} scoring on the InternetAds dataset from \cite{campos2016evaluation}, with a $0.1$-fraction of outliers.
Unlike the experiments on our CIFAR-10 and text embedding data, to replicate the experimental setting of prior work as closely as possible we perform no whitening or other preprocessing.

We find that \QUE{} scoring is outperformed by LOF/$k$-NN-based methods on the InternetAds dataset.
Choosing $\alpha = 4$ for \QUE{}, we find the ROCAUC scores in the below table.

\begin{figure}[ht]
\centering
\begin{tabular}{c | c}
  method & ROCAUC \\ \hline
  naive spectral & $0.539$ \\
  \QUE{} & $0.626$ \\
  $\ell_2$ & $0.723$ \\
  isolation forest & $0.702$ \\
  local outlier factor & $0.853$ \\
  $k$-NN & $0.855$
\end{tabular}
\caption{ROCAUC scores on InternetAds dataset from \cite{campos2016evaluation}. Note that \QUE{} still improves on naive spectral methods, but because outliers are individually identifiable in the InternetAds data set, local and $\ell_2$ methods improve on any spectral method. We use $\alpha = 4$ in \QUE{}. See \cite{knn2000, lof2000, campos2016evaluation} for definitions of local outlier factor and $k$-NN methods.}
\end{figure}

To elucidate the difference between the InternetAds setting where $k$-NN methods perform well and the other experimental settings in this paper, we offer the following histograms demonstrating that the distribution of nearest-neighbor distances is markedly distinct for inliers and outliers in both data sets, but only in the InternetAds dataset do inliers have smaller $k$-NN distances.

\begin{figure}[ht]
\label{fig:high-dims}
	\centering
	\begin{subfigure}{.3\textwidth}
		\includegraphics[width=\textwidth]{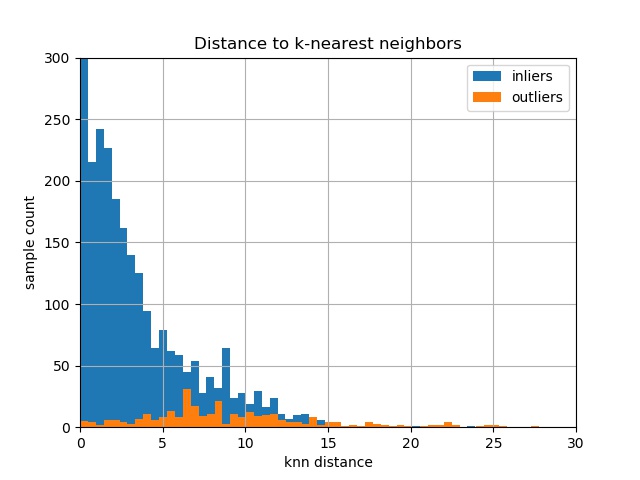}
		\caption{InternetAds, $\e = 0.1$}
	\end{subfigure}
	\begin{subfigure}{.3\textwidth}
		\includegraphics[width=\textwidth]{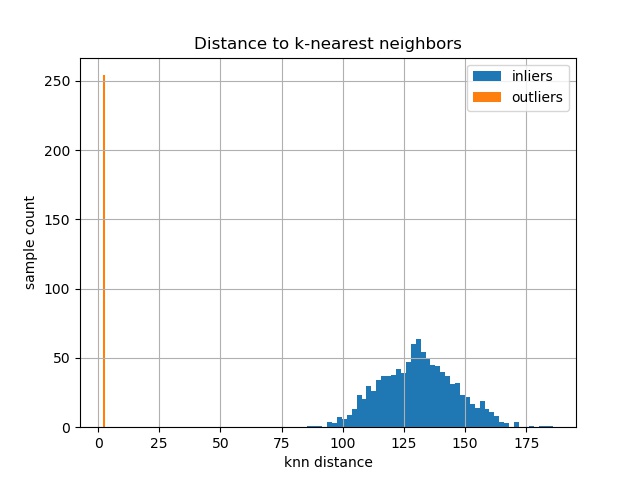}
		\caption{synthetic, $\e = 0.2$, }
	\end{subfigure}
    \caption{We plot histograms of the following collection $\{ d_k(X_i,S) \, : \, X_i \in S \}$ for $S = \text{InternetAds,synthetic}$, where $d_k(X_i,S)$ is the average squared-$\ell_2$ distance of $X_i$ and its $k$ nearest neighbors, with $k=10$.
    Observe that in the InternetAds dataset essentially only outliers have $d_k(X_i,S) > 15$, while for the synthetic dataset \emph{all inliers have $d_k(X_i,S)$ much greater than that of every outlier}.}
\end{figure}

\FloatBarrier

\subsection{Scaling up: a nearly-linear time implementation of QUE scoring}
\label{sec:scaling-up}

Most of the experiments we present involving \QUE{} scores employ the following approach to compute them.
Given $X_1,\ldots,X_n \in \R^d$, explicitly form the empirical covariance $\overline{\Sigma}$ in memory.
Use SciPy's \verb+expm+ function to compute the matrix exponential $U = \exp(\alpha \overline{\Sigma})$, then compute $\tau_i = (X_i - \overline{\mu})^\top U (X_i - \overline{\mu})$.
(This in turn uses the scaling and squaring algorithm for the matrix exponential of Al-Mohy and Higham \cite{al2009new}.)

While we are already able to run experiments in $1000$ or more dimensions using this approach, it requires at least $d^2$ memory to store the covariance, and somewhat more time to form and exponentiate it.
We also implement an approximate method to compute \QUE{} scores, whose running time is $\tO(nd)$.
We demonstrate in this section that outlier detection from approximate \QUE{} scores still improves over baseline methods on several data sets.
The technique here is very similar to the one used in Section~\ref{sec:approx-scores} to approximate the scores used in the fast robust mean estimation algorithm.
At a high level, the idea is the same: approximate the exponential with a low-degree polynomial, and sketch this using Johnson-Lindenstrauss matrices.
However, we make a couple of additional optimizations here.

\paragraph{\QUE{} scores in nearly-linear time}
We use the following approximate method, inspired by our sketching approach to compute \QUE{} scores from our nearly linear time algorithm for robust mean estimation.

Given $X_1,\ldots,X_n$ and $\alpha > 0$, our goal is to compute approximations $\tilde{\tau_i}$ to the \QUE{} scores.
At a high level, our approach employs two main tricks:
\begin{compactenum}
\item Approximate the matrix exponential $\exp(M)$ by Chebyshev polynomials of degree $O(\log d)$, along with scaling and squaring when $\|M\|_2 > 1$ \cite{al2009new}.
\item For $\overline{\Sigma}$ the empirical covariance of $X_1,\ldots,X_n$, use fast versions of the Johnson-Lindenstrauss method to approximate $\iprod{X_i, \overline{\Sigma}^j X_i}$ for all $i \leq n$ and $j \leq O(\log d)$.
\end{compactenum}

Since for most applications only the \emph{order} of the \QUE{} scores matters, we will actually compute approximations to \emph{non-normalized} scores, involving the matrix $U = \exp(\alpha \overline{\Sigma} / \|\overline{\Sigma}\|_2)$, rather than normalizing by $\tr(U)$.
For notational simplicity, let us assume $\overline{\mu} = 0$.

We first rewrite $\tau_i \propto \|\exp(\alpha \overline{\Sigma}/2\|\overline{\Sigma}\|_2) X_i\|_2^2$ where $\propto$ hides the normalization $\tr \exp(\alpha \overline{\Sigma})$.
If we approximate the matrix exponential by a degree $O(\log d)$ Chebyshev approximation $P$, the goal is now to approximate $\|P(\alpha \overline{\Sigma}/2\|\overline{\Sigma}\|_2) X_i\|^2$ for all $i$.
Suppose $S$ is an $O(\log(d + n)) \times d$ sketching matrix.
Then it will suffice to compute $M = SP(\alpha \overline{\Sigma}/2\|\overline{\Sigma}\|_2)$, since then in $\tO(nd)$ time we can compute all the matrix-vector products $MX_i$ and their norms, and
$\|S P(\alpha \overline{\Sigma}/2\|\overline{\Sigma}\|_2) X_i \|_2^2 \approx \|P(\alpha \overline{\Sigma}/2\|\overline{\Sigma}\|_2) X_i\|_2^2$.

Note that using the expansion into powers $\overline{\Sigma}^j$, right or left matrix-vector multiplication by $P(\alpha \overline{\Sigma}/2\|\overline{\Sigma}\|_2)$ can be accomplished in $O(nd \log d)$ time.
As in the fast JL transform \cite{ailon2009fast}, we take $S = S' \cdot D \cdot H$ where $S'$ is a sparse random matrix, $D$ is a diagonal matrix with random $\pm 1$ entries, and $H$ is a Hadamard matrix.
Fast Fourier transform methods may be used to compute matrix-vector multiplications $SX$ in $d (\log d)^{O(1)}$ time \cite{andoni2015practical}, leading to nearly-linear running time of this approach in theory.
In practice, we use standard matrix multiplication; this still allows for experiments in thousands of dimensions. 

Experiments show that the approximate \QUE{} scores are consistent with the exact \QUE{} scores, in that higher $k$ leads to larger maximum improvement over naive spectral scores; furthermore approximate \QUE{} consistently interpolates between $\ell_2$ scores (when $\alpha=0$) and naive spectral scores (when $\alpha=\infty$).

\begin{figure}[ht]
\label{fig:high-dims}
	\centering
	\begin{subfigure}{.3\textwidth}
		\includegraphics[width=\textwidth]{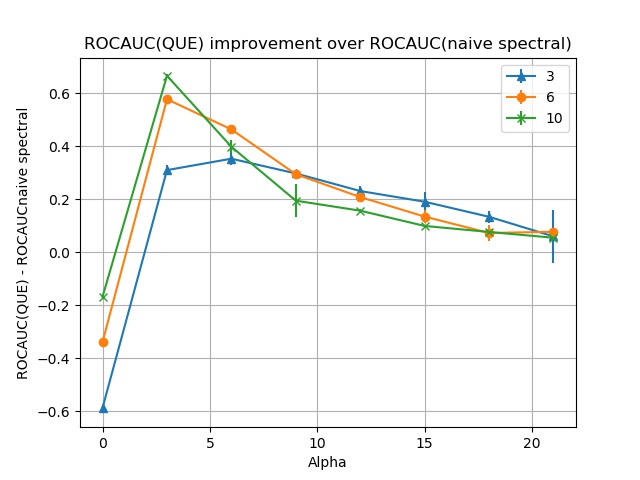}
		\caption{synthetic $128$-dimensions}
	\end{subfigure}
	\begin{subfigure}{.3\textwidth}
		\includegraphics[width=\textwidth]{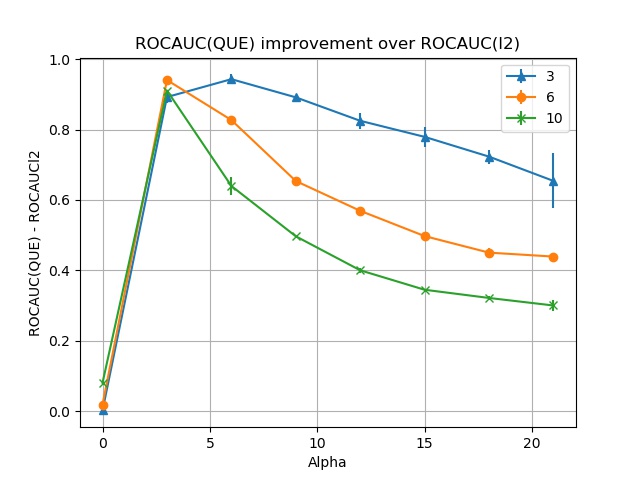}
		\caption{synthetic $128$-dimensions}
	\end{subfigure}
    \caption{We plot the improvement of ROCAUC scores of our approximate \QUE{} scoring implementation, over baseline methods $\ell_2$ scoring as well as naive spectral scoring, on $\mathbf{128}$-\textbf{dimensional} synthetic data, for 3, 6, and 10 directions of corruption. We use the degree-$5$ Chebyshev approximation to the matrix exponential function, along with scaling and squaring for approximating outside the interval $[-1,1]$.}
\end{figure}

\begin{figure}[ht]
\label{fig:high-dims}
	\centering
	\begin{subfigure}{.3\textwidth}
		\includegraphics[width=\textwidth]{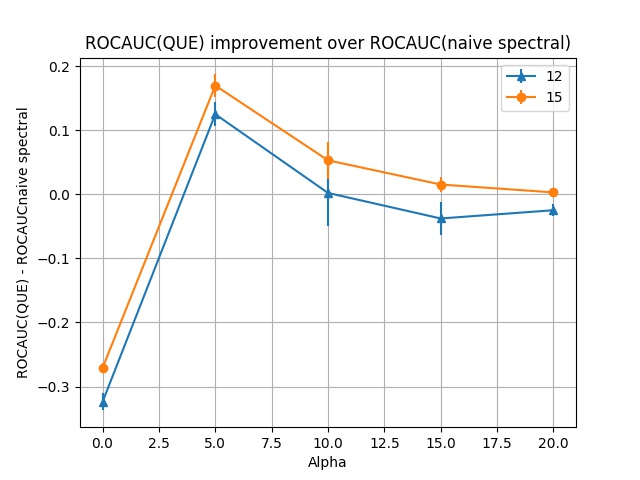}
		\caption{synthetic $8192$-dimensions}
	\end{subfigure}
	\begin{subfigure}{.3\textwidth}
		\includegraphics[width=\textwidth]{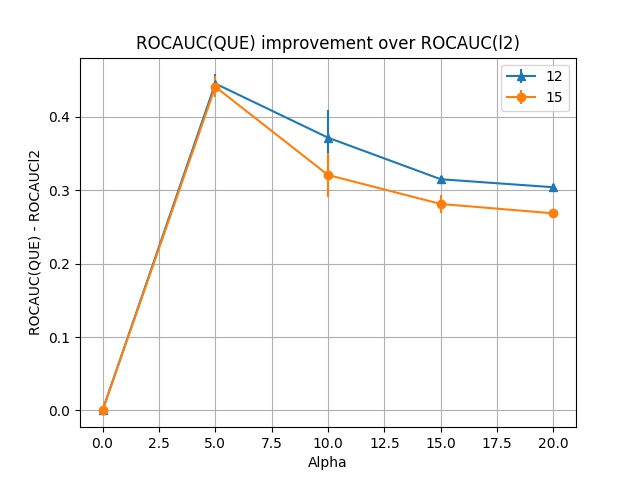}
		\caption{synthetic $8192$-dimensions}
	\end{subfigure}
    \caption{Similar to above, these show the improvement of ROCAUC scores of our approximate \QUE{} scoring implementation, over baseline methods $\ell_2$ scoring and naive spectral scoring, on $\mathbf{8192}$-\textbf{dimensional} synthetic data, for 12 and 15 directions of corruption. These approximate \QUE{} scores run in nearly-linear time and improve over the performance of naive spectral and $\ell_2$-based scoring.}
\end{figure}

\FloatBarrier

\subsection{Approximate whitening}
\label{sec:appx-whiten}
Using approximate \QUE{} scores reduces the running time of our outlier detection algorithm from quadratic to nearly-linear \text{if whitened data is already available or there is no desire to preprocess/whiten the data.}
However, as we discussed in Section~\ref{sec:experiments}, \QUE{} scoring works best with whitened data.
If given a corrupted dataset $X = X_1,\ldots,X_n$ and a clean dataset $Y = Y_1,\ldots,Y_m$ which is distributed similarly to the inliers of the dataset $X$, we would like to compute whitened data $X_i' = (\E(Y_i - \mu(Y))(Y_i - \mu(Y))^\top)^{-1} \cdot X_i$.
Unfortunately, even forming the matrix $(\E(Y_i - \mu(Y))(Y_i - \mu(Y))^\top)^{-1}$ requires quadratic time (and computing the matrix inverse is slower still).

We investigate an approximate whitening procedure which avoids computing the entire matrix $(\E(Y_i - \mu(Y))(Y_i - \mu(Y))^\top)^{-1}$.
Instead, we compute the top $k$ eigenvectors and eigenvalues $\lambda_i, v_i$ of $(\E(Y_i - \mu(Y))(Y_i - \mu(Y))^\top)$ and approximate the inverse as $\sum_{i \leq k} (1/\lambda_i) v_i v_i^\top + \Pi_\perp$, where $\Pi_\perp$ is the projector to the orthogonal complement of $span \{v_1,\ldots,v_k\}$.
We use $k=0.3d$ and demonstrate that even in conjunction with our approximate \QUE{} scoring algorithm we still obtain a nonnegligible improvement over baseline methods.

\begin{figure}[ht]
\label{fig:cifar_fw}
	\centering
	\begin{subfigure}{.5\textwidth}
		\includegraphics[width=\textwidth]{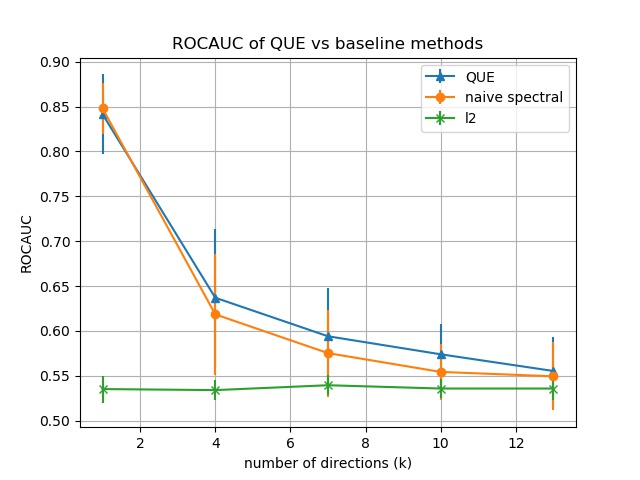}
		\caption{Scoring on CIFAR-10 data with approximate whitening}
	\end{subfigure}
    \caption{Scores under approximate whitening with $30\%$ of eigenvalues and eigenvectors, as well as using a fresh set of randomly sampled CIFAR-10 images, for computing the whitening transformation $W$.}
\end{figure}

\bibliographystyle{unsrt}
\bibliography{refs}

\appendix


\section{Deferred details from Section~\ref{sec:prelims}}
\label{sec:app-prelims}

\subsection{Proof of Lemma~\ref{lem:naive-prune}}
The algorithm is straightforward: choose a random point in $S$, and check if strictly more than $n/2$ points lie within a ball of radius $2r$ around this point.
If so, include all points with distance at most $4r$ from this point.
If not, repeat, and run for $O(\log 1 / \delta)$ iterations.
We now prove correctness.
\begin{proof}[Proof of Lemma~\ref{lem:naive-prune}]
By the triangle inequality, if we ever randomly select a point from $S'$, then we terminate, and in this case it is easy to see that the output satisfies the desired property.
Thus, it is easy to see that the probability we have not terminated after $t$ iterations is at most $2^{-t}$.
Suppose we have terminated.
Then in that iteration, we selected a point $X \in S$ that has distance at most $2r$ to more than $n/2$ other points in $S$.
This implies that it has distance at most $2r$ to some point in $S'$.
By triangle inequality, this implies that all points in $S'$ are at distance at most $4r$ from $X$, and so the output in this iteration must satisfy the claims of the Lemma.
\end{proof}

\noindent
We note that if one wishes to obtain a deterministic linear-time algorithm for this problem, it is also possible to do so, albeit using radius $O(r \sqrt{d})$.
The algorithm is again simple: simply take the coordinate-wise median of all the data points, and take all points with distance at most $O(r \sqrt{d})$ from this point.
It is not hard to see that the coordinate-wise median can differ in each coordinate from the points in $S'$ by at most $r$, and so its distance to each point in $S'$ can be at most $r \sqrt{d}$.
While this is worse by a polynomial factor than the guarantee obtained above, since in the end our overall guarantees depend only logarithmically on $r$, this does not change our runtime guarantees by more than a logarithmic factor.


\subsection{Omitted details from Section~\ref{sec:1dfilter}}
\label{app:1dfilter}
The algorithm~\textsc{1DFilter} is quite simple.
For $i = 1, \ldots, m$, and for any positive integer $t$, define
\begin{equation}
\label{eq:1D-update}
w^{\Paren{t}}_i = \Paren{1 - \frac{\tau_i}{\tmax}}^t w_i \; \mbox{, and } \; F_t = \sum_{i = 1}^n w^{\Paren{t}}_i \tau_i \; .
\end{equation}
Observe that the $F_t$ form a monotone decreasing sequence.
The algorithm will simply find the smallest $t \in \{1, \ldots, \frac{\tmax}{e b \sigma}\}$ so that $F_t \leq b \sigma$ via binary search, and outputs $w^{\Paren{t}}$.
The formal pseudocode for the algorithm is given in Algorithm~\ref{alg:1d-filter}.

\begin{algorithm}[ht]
\caption{Improved univariate score downweighting}
  \label{alg:1d-filter}
  \begin{algorithmic}[1]
  \STATE \textbf{Input:} nonnegative scores $\tau_1, \ldots, \tau_m$, weights $w_1, \ldots, w_m$, parameters $b, \eta$
  \STATE Let $\tmax = \max_{i \in [m]} \tau_i$.
  \STATE Let $\sigma = \sum_{i = 1}^m w_i \tau_i$.
  \STATE By binary search, find the smallest $t \in \{1, \ldots, \frac{\tmax}{e b \sigma}\}$ satisfying $F_t \leq b \sigma$, where $F_t$ is defined in~\eqref{eq:1D-update}
  \RETURN The weights $w_1^{\Paren{t}}, \ldots, w_m^{\Paren{t}}$, where $w_i^{\Paren{t}}$ is defined as in~\eqref{eq:1D-update}.
  \end{algorithmic}
\end{algorithm}
\noindent
We now prove that this algorithm satisfies Theorem~\ref{thm:1D}.
We say any set of weights $w'$ satisfying $\sum_{i \in S_g} w_i - w_i' \leq \sum_{i \in S_b} w_i - w_i'$ is \emph{admissible}.
We first show that the sequence of weights we produce is always admissible, under some mild conditions:
\begin{lemma}
\label{lem:1D-inv1}
Let $t$ be an integer so that $w^{\Paren{t}}$ is admissible, and $F_t > 2 \eta \sigma$.
Then $w^{\Paren{t + 1}}$ is admissible.
\end{lemma}
\begin{proof}[Proof of Lemma~\ref{lem:1D-inv1}]
Because $w^{\Paren{t}} \leq w$, we have that $\sum_{i \in \Sgood} w^{\Paren{t}}_i \tau_i \leq \eta \sigma$.
As a result, if $F_t \geq 2 \eta \sigma$, we must have $\sum_{i \in \Sbad} w^{\Paren{t}}_i \tau_i > \sigma / 2$.
Therefore, we have the following two inequalities:
\begin{align*}
\sum_{i \in \Sgood} w^{\Paren{t}}_i - w^{\Paren{t + 1}}_i &= \frac{1}{\tmax} \sum_{i \in \Sgood} w^{\Paren{t}} \tau_i \leq \frac{\sigma}{2 \tmax} \\
\sum_{i \in \Sbad} w^{\Paren{t}}_i - w^{\Paren{t + 1}}_i &= \frac{1}{\tmax} \sum_{i \in \Sbad} w^{\Paren{t}} \tau_i > \frac{\sigma}{2 \tmax} \; .
\end{align*}
Consequently, we remove more mass from the weights in $\Sbad$ than from $\Sgood$ in going from $w^{\Paren{t}}$ to $w^{\Paren{t + 1}}$.
Since by assumption $w^{\Paren{t}}$ is admissible, this immediately implies that $w^{\Paren{t + 1}}$ is admissible as well.
\end{proof}

\noindent We are now ready to prove Theorem~\ref{thm:1D}.
\begin{proof}[Proof of Theorem~\ref{thm:1D}]
By induction, Lemma~\ref{lem:1D-inv1} guarantees that the output weights remain admissible, and the termination condition of the algorithm guarantees that the output satisfies~\eqref{eq:1D-final}.
It suffices to bound the runtime of the algorithm.

First, observe that there must exist a valid $T$ in the range we are searching.
We first observe that
\begin{align*}
w_i^{\Paren{t}} \tau_i &= w_i \Paren{1 - \frac{\tau_i}{\tmax}}^t \tau_i \\
&\leq w_i \exp \Paren{- \frac{t \cdot \tau_i}{\tmax}} \tau_i \; .
\end{align*}
For any constant $A > 0$, the maximizer of the function $g(x) = x \exp (- Ax)$ in the range $x \in [0, \infty)$ is achieved by $x = \frac{1}{A}$, so $g(x) \leq \frac{1}{e A}$ for all $x \in [0, \infty)$.
Setting $A = t / \tmax$, and letting $t = \frac{\tmax}{e b \sigma}$, we conclude that
\[
\sum_{i = 1}^m w_i^{\Paren{t}} \tau_i &\leq \sum_{i = 1}^m w_i \cdot \frac{\tmax}{e t} \leq b \sigma \; .
\]
Thus, there exists some $t$ within our specified range which satisfies the conclusion.
Finally, to bound the runtime, observe that every iteration runs in $O(n)$ time,\footnote{This is true in the real RAM model; in practice we can run any iteration in $O(m \log (\tmax / \sigma) )$ time by exponentiation via repeated doubling to compute all the $w^{\Paren{t}}_i$, so we pay at most an additional log factor} and we can run for at most $O(\log \tmax / (b \sigma))$ iterations, which completes the proof.
\end{proof}

\subsection{The randomized hard filter}
\label{app:random-filter}
In this section we show that a randomized outlier removal method, rather than soft downweighting, can achieve the more or less the same guarantees as~\textsc{1DFilter}.
Formally, we show:
\begin{theorem}
\label{thm:1D-random}
Let $\eta \in (0, 1/2)$, let $b \geq 2 \eta$, and let $s, \delta > 0$.
Let $m$ satisfy
\[
m = \widetilde{\Omega} \Paren{ \frac{s \log^2 (1 / \delta) \log \frac{\tmax}{b \sigma}}{\eps} } \; .
\]
Let $\tau_1, \ldots, \tau_m$ be non-negative scalars, and let $\tmax = \max_{i \in [m]} \tau_i$.
Suppose there exist two disjoint sets $S_g, S_b$ so that $S_g \cup S_b = [m]$, and moreover,
\[
\sum_{i \in \Sgood} \tau_i \leq \eta \sigma\; \mbox{, where } \; \sigma = \sum_{i = 1}^n \tau_i \; .
\]
Then $\textsc{RandomFilter} (w, \tau)$ runs in time $O\Paren{\Paren{1 + \log \frac{\tmax}{b \sigma}} m}$ and outputs $S' \subseteq S$ so that with probability $1 - \delta$, we have:
\begin{itemize}
\item not too many more points from $S_g$ are removed than from $S_b$ i.e. 
\[
|(S \setminus S') \cap S_g| \leq |(S \setminus S') \cap S_b| + \frac{\eps m}{s} \; ~\mbox{, and}
\]
\item the sum of the $\tau$ has decreased, i.e. $S'$ satisfies
\begin{equation}
\label{eq:1D-final-random}
\sum_{i \in S'} \tau_i \leq b \sigma  \; .
\end{equation}
\end{itemize}
\end{theorem}
\noindent
The algorithm itself is very easy to describe.
First, let $T = [m]$.
Then, while $\sum_{i \in T} \tau_i > b \sigma$, throw away each point from $T$ with probability $\tau_i / \tmax(T)$, where $\tmax(T) = \max_{i \in T} \tau_i$, and let $T$ be the set of remaining points.
At termination, we simply output the set $S' = T$.
The formal pseudocode of this algorithm is given in Algorithm~\ref{alg:1d-filter-random}.

\begin{algorithm}[ht]
\caption{Improved randomized univariate filtering}
  \label{alg:1d-filter-random}
  \begin{algorithmic}[1]
  \STATE \textbf{Input:} nonnegative scores $\tau_1, \ldots, \tau_m$, parameters $b, \eta$
  \STATE Let $T = [m]$
  \STATE Let $\sigma = \sum_{i = 1}^n \tau_i$.
  \WHILE{$\sum_{i \in T} \tau_i \geq b \sigma$}
  	\STATE Let $\tmax = \max_{i \in T} \tau_i$
  	\STATE Remove each point $i \in T$ from $T$ with probability $\tau_i / \tmax$
  \ENDWHILE
  \RETURN The set $T$
  \end{algorithmic}
\end{algorithm}

As a brief aside, we note that this differs slightly from the algorithm presented in~\cite{diakonikolas2017being}, as there the algorithm randomly selects a threshold, and throws away all points above this threshold.
However, the key property which the previous algorithm used of this random threshold was that for all $i \in T$, we had that $\Pr [\mbox{$i$ is thrown out}] = \tau_i / \tmax(T)$.
Therefore a very similar analysis can be adapted for either case.
However, the algorithm in~\cite{diakonikolas2017being} only succeeds with constant probability, and a martingale-style argument (as in~\cite{diakonikolas2017being}) is needed to ensure that it works.

The remainder of this section is dedicated to the proof of Theorem~\ref{thm:1D-random}.
Our first lemma is similar to Lemma~\ref{lem:1D-inv1}.
\begin{lemma}
\label{lem:random-inv}
Suppose that $\sum_{i \in T} \tau_i > b \sigma$.
Then, if we let $T'$ be the random set obtained by throwing away each point from $T$ with probability $\tau_i / \tmax(T)$, then:
\begin{equation}
\label{eq:random-filter-exp}
\E \Brac{|(T \setminus T') \cap S_g|} < \E \Brac{|(T \setminus T') \cap S_b|} \; ,
\end{equation}
and moreover, for any $s > 0$, we have
\begin{equation}
\label{eq:random-filter-whp}
\Pr \Brac{|(T \setminus T') \cap S_g| > |(T \setminus T') \cap S_b| ~ \mbox{and} ~ |T \setminus T'| > \frac{\eps m}{s}} \leq \exp \Paren{- \Omega \Paren{\frac{\eps m}{s} \cdot \min \Paren{(b - 2 \eta)^2, 1} }} \; .
\end{equation}
\end{lemma}
\begin{proof}
For $i \in T$, let $Y_i$ be the random variable which is $1$ if we throw out $i$ in $T'$ and $0$ otherwise.
Then
\begin{align*}
\E \Brac{|(T - T') \cap S_g|} &= \E \Brac{\sum_{i \in S_g \cap T} Y_i} = \sum_{i \in \Sgood \cap T} \frac{\tau_i}{\tmax (T)} \leq \sum_{i \in \Sgood} \frac{\tau_i}{\tmax (T)} \leq \frac{1}{\tmax(T)} \eta \sigma \; ,\\
\E \Brac{|(T - T') \cap S_b|} &= \E \Brac{\sum_{i \in S_b \cap T} Y_i} = \sum_{i \in \Sbad \cap T} \frac{\tau_i}{\tmax (T)} \geq \frac{1}{\tmax(T)} (\beta - \eta) \sigma \; ,
\end{align*}
which proves~\eqref{eq:random-filter-exp}, since $\beta > 2 \eta$.

To prove~\eqref{eq:random-filter-whp}, we break into two cases depending on $\sigma / \tmax$.
Suppose that $\sigma / \tmax \geq \eps m / s$.
Then by Bernstein's inequality, we have
\begin{align*}
\Pr \Brac{|(T \setminus T') \cap S_g| > |(T \setminus T') \cap S_b| ~ \mbox{and} ~ |T \setminus T'| > \frac{2 \eps m}{s}} &\leq \Pr \Brac{|(T \setminus T') \cap S_g| > |(T \setminus T') \cap S_b|} \\
&= \Pr \Brac{\sum_{i \in \Sgood \cap T} Y_i - \sum_{i \in \Sbad \cap T} Y_i > 0} \\
&\leq \exp \Paren{- \Omega \Paren{\frac{\sigma}{\tmax} \cdot \min \Paren{(b - 2 \eta)^2, b - 2 \eta} }} \\
&\leq \exp \Paren{- \Omega \Paren{\frac{\eps m}{s} \cdot \min \Paren{(b - 2 \eta)^2, b - 2 \eta} }}
\end{align*}
Otherwise, we observe that 
\begin{align*}
\Pr \Brac{|(T \setminus T') \cap S_g| > |(T \setminus T') \cap S_b| ~ \mbox{and} ~ |T \setminus T'| > \frac{2 \eps m}{s}} &\leq \Pr \Brac{|T \setminus T'| > \frac{\eps m}{s}} \\
&= \Pr \Brac{\sum_{i \in T} Y_i > \frac{2 \eps m}{s}} \\
&\leq \exp \Paren{-\Omega \Paren{\frac{\eps m}{s}}} \; .
\end{align*}
Combining these two cases, and simplifying yields the desired claim.
\end{proof}
We now show that with high probability, we do not need to repeat this procedure too many times before the sum decreases by a constant factor.
\begin{lemma}
\label{lem:random-runtime}
Let $\delta > 0$, and let $t = \widetilde{\Omega} \Paren{\log \Paren{\frac{\tmax m}{b \sigma}} \log (1 / \delta) }$.
Then, the probability that Algorithm~\ref{alg:1d-filter-random} runs for more than $t$ iterations is at most $\delta$.
\end{lemma}
\begin{proof}
Let $J = \log \frac{\tmax m}{b \sigma}$.
For $j = 1, \ldots, J$, let 
\[
A_j = \left\{i \in T: \tau_i \in \left( \tmax \cdot 2^{-j}, \tmax \cdot 2^{-j - 1} \right] \right\} \; .
\]
For all $j = 1, \ldots, J$, we claim that conditional on the event that the algorithm has not terminated yet, and all points from $A_{j'}$ have been removed, for $j' < j$, then after $t'$ iterations, all points from $A_j$ have been removed with probability at least $1 - m 2^{-t'}$.
Indeed, for all $i \in A_j$, in every iteration, if it has not been already removed, then it is removed with probability at least $1/2$.
Thus after $t'$ iterations, the probability that any point from $A_j$ remains is at most $n 2^{-t'}$.
Therefore, by a union bound, after $J t'$ iterations, conditioned on the event that the algorithm hasn't terminated yet, the probability that any point from $A_j$ for any $j$ is at most $J m 2^{-t'}$.
However, if all points from $A_j$ are removed, for all $j$, then if $T'$ is the remaining set, we have $\sum_{i \in T'} \tau_i \leq b \sigma$, so if all such points are removed, then the algorithm must either terminate or have already terminated.
This proves the claim by setting $t' = \log (J m / \delta)$.
\end{proof}
\noindent
Theorem~\ref{thm:1D-random} follows from Lemma~\ref{lem:random-inv} and Lemma~\ref{lem:random-runtime}, by appropriately adjusting parameters.

\paragraph{The full algorithm using the randomized filter}
It is straightforward for the full matrix multiplicative weights algorithm to use the randomized filter rather than the downweighting-based method.
Given an $\eps$-corrupted dataset of size $n$ initally, we do the following.
At every instance where we pass to the filter, simply run the randomized filter instead of the downweighting-based method, and output the set of weights which is $1/n$ for every point that remains after running the randomized filter, and $0$ otherwise.

The guarantee of the randomized filter is slightly weaker than the guarantee of the downweighting-based method, so we cannot use black-box use the analysis presented beforehand to also analyze the algorithm instantiated with the weights given by the randomized filter.
This is because our guarantee allows for slightly more good points than bad points to be removed per run of the algorithm.
However, since the matrix multiplicative weights routine runs for at most polylogarthmically many iterations, by setting $s = \poly \log (n d)$, we can guarantee that at the end of all of the runs, we have removed at most $2 \eps n$ data points from $\Sgood$.
It is straightforward to verify that (up to a factor of 2), the same analysis for matrix multiplicative works with this slightly weaker guarantee, for $\eps$ sufficiently small.
For conciseness, we omit the proof.


\section{Omitted proofs from Section~\ref{sec:ideal-bounded-cov}}
\label{app:ideal-bounded-cov}

\subsection{A reduction to the constant $\eps$ regime for bounded second moments}
\label{app:reduction}
\emph{The contents of Section~\ref{app:reduction} were added after we were made aware of \cite{lecue2019robust} and hence do not represent independent contributions of the present paper.}
In this section, we demonstrate the following reduction, which is also implicit in~\cite{lecue2019robust}:
\begin{lemma}
\label{lem:reduction}
Let $n \in \N$ and $\eps \in \R$ be so that $\eps > 1/n$.
Let $D$ be a distribution over $\R^d$ with mean $\mu$ and covariance $\Sigma \preceq \Id$.
Let $S$ be an $\eps$-corrupted set of samples from $D$ of size $n$.
Then there is an efficient algorithm which, given $S$ and $\eps$, produces an $1/10$-corrupted set of samples of size $\eps n / 10$ from a distribution $D'$ with mean $\mu$, and covariance $\Sigma' \preceq 10 \eps \cdot \Id$.
\end{lemma} 
\noindent
Before we prove this lemma, we explain the usefulness of this reduction.
As explained previously, recent work of~\cite{cheng2019high} yields an algorithm for robust mean estimation under the same conditions as Theorem~\ref{thm:main-bounded-cov} that yields the same error guarantees as us\footnote{They state the results only in the regime where the number of samples $n$ is at least $\Omega (d \log d / \eps)$, however we believe that implicit in their work is an error tradeoff in terms of $n$ that is essentially the same as we achieve in Theorem~\ref{thm:main-bounded-cov}.}, but in time $\tilde{O} (n d) / \eps^6$.
However, by combining this algorithm with this reduction, it is easy to see that we are able to reduce to the regime where $\eps$ is constant, and hence the runtime of the algorithm matches that of Theorem~\ref{thm:main-bounded-cov}, up to polylogarithmic factors.

Despite this, we believe that our algorithm is still of independent interest, for a number of reasons.
First, our algorithm is much simpler than combining this reduction with the algorithm presented in~\cite{cheng2019high}.
We believe this is of independent mathematical interest.
Additionally, it is this simplicity which allows us to design a practical outlier detection method, as presented in Section~\ref{sec:experiments}.
Second, this reduction appears to preserve statistical accuracy only in the regime where aim for error bounds as in Theorem~\ref{thm:main-bounded-cov}.
In particular, it cannot achieve error below $\Omega(\sqrt{\eps})$.
For instance, if it is combined with the result in~\cite{cheng2019high} for isotropic sub-Gaussian distributions that can achieve error $O(\eps \sqrt{\log 1 / \eps})$ in time $\Otilde (nd) / \eps^6$, then the overall algorithm will still achieve error $\Omega (\sqrt{\eps})$, which is statistically suboptimal.
In contrast, by slightly modifying our algorithm as in Theorem~\ref{thm:main-gaussian}, we are able to achieve runtime $\Otilde (nd)$ while achieving error $O(\eps \sqrt{\log 1 / \eps})$, for all $\eps > 0$ sufficiently small.

\begin{proof}[Proof of Lemma~\ref{lem:reduction}]
The reduction is straightforward: obliviously group the samples in $S$ into $10 \eps n$ buckets, each of size $\frac{1}{10 \eps}$, and produce the set $S'$ which simply takes each bucket, and takes the average the data points in that bucket.
First, assume there is no corruption.
Then, each bucket contains $1/ (10 \eps)$ i.i.d. samples from $D$, and so their average is distributed as $D'$, where the mean of $D'$ is still $\mu$, and their variance is $10 \eps \Sigma \preceq 10 \eps \Id$.
Since an adversary can only corrupt $\eps n$ of these buckets, and there are a total of $10 \eps n$ buckets, the desired conclusion follows immediately.
\end{proof}

\subsection{Proof of Lemma~\ref{lem:second-moment-conc}}
Before we prove this lemma, we require the following matrix Chernoff bound:
\begin{fact}[Theorem 5.1.1 in~\cite{tropp2015introduction}]
\label{fact:tropp}
Let $M_1, \ldots, M_n \in \R^{d \times d}$ be a sequence of independent, random, PSD matrices.
Assume that $\norm{M_i}_2 \leq L$ for all $i = 1, \ldots, n$, and suppose $\Norm{\E \Brac{\sum_{i = 1}^n M_i}}_2 \leq n$.
Then, there is some universal constant $c \leq 2 \log 2 - 1$ so that for all $t \geq 2$, we have
\[
\Pr \Brac{\Norm{\sum_{i = 1}^n M_i}_2 \geq t n} \leq d \exp \Paren{- c t n / L} \; .
\]
\end{fact}
\begin{proof}
Observe that if $X \sim D$, then $\E \Brac{\Norm{X - \mu}_2^2} = \tr (\Sigma) \leq d$. 
Let $c > 0$ be a constant to specify later.
By Markov's inequality, we have
\begin{equation}
\label{eq:chebyshev}
\Pr \Brac{\Norm{X - \mu}_2 \geq \sqrt{\frac{d}{c \eps}}} \leq c \eps \; .
\end{equation}
Let $E$ be the event $E = \left\{ X: \Norm{X - \mu}_2 < \sqrt{\frac{d}{c \eps}} \right\}$, and let $S = \left\{X_i : X_i \in E \right\}$.
We claim this set satisfies the properties claimed.

We first demonstrate that with probability $1 - \exp (-\eps n)$, we have $|S| \geq (1 - \eps) n$.
Letting $Z_i = \ind \{X_i \in E^c\}$, we have we have $|S| = n - \sum_{i = 1}^n Z_i$, where $\E \Brac{Z_i} \leq c \eps$, and hence, by Chernoff bounds,
\begin{align*}
\Pr \Brac{|S| < (1 - \eps) n} &= \Pr \Brac{\sum Z_i > \eps n} \\
&\leq \Paren{\frac{e^{1/c - 1}}{(1/c)^{1/c}}}^{c \eps n} \\
&= \exp \Paren{\eps n - c \eps n - \eps n \log 1/c} \; .
\end{align*}
In particular, we let $c = e^{-2}$, by simplifying we obtain that $\Pr \Brac{|S| < (1 - \eps) n} < \exp (-\eps n)$.
Let $E_1$ be the event that $|S| \geq (1 - \eps) n$.

We now demonstrate that with probability $1 - \delta$, our set $S$ is $(\gamma_1, \gamma_2)$-good with respect to $D$, where $\gamma_1, \gamma_2$ are given as in~\eqref{eq:conc-second-moments}.
We first prove concentration of the empirical mean.
For $i = 1, \ldots, n$, let $Y_i = (X_i - \mu) \cdot \ind \{X_i \in E\}$.
Since the $X_i$ are independent, so too are the $Y_i$.
Moreover, since multiplying by an indicator variable can only decrease variance, we have that $\Cov \Brac{Y_i} \preceq I$.
Letting $\mu' = \E [Y_i]$, we thus have
\[
\E \Brac{ \Norm{\frac{1}{n} \sum_{i = 1}^n Y_i - \mu'}_2^2} \leq \frac{\tr(\Sigma)}{n} \leq \frac{d}{n} \; .
\]
Thus, if we let $E_2$ be the event
\begin{align*}
E_1 = \left\{ \Norm{\frac{1}{n} \sum_{i = 1}^n Y_i - \mu'}_2 \leq \sqrt{\frac{2 d}{n \delta}} \right\} \; ,
\end{align*}
by Markov's inequality, we have $\Pr [E_1] \geq 1 - \delta / 2$.
We additionally have that for any unit vector $v$,
\begin{align*}
\Abs{\iprod{v, \mu'}} &= \Abs{\E \Brac{\iprod{v, X_i - \mu} \cdot \ind \{X_i \in E\}}} \\
&= \Abs{\E \Brac{\iprod{v, X_i - \mu} \cdot \ind \{X_i \in E^c\}}} \\
&\leq \E \Brac{\iprod{v, X_i - \mu}^2}^{1/2} \Pr_{X \sim D} [X \in E^c]^{1/2} \\
&= \Paren{ v^T \Sigma v}^{1/2} \cdot \Pr_{X \sim D} [X \in E^c]^{1/2} \leq \sqrt{c \eps} \; ,
\end{align*}
where the third line follows from Cauchy-Schwartz, the last line follows from~\eqref{eq:chebyshev}, and since $\Sigma \preceq \Id$.
Taking a supremum over all unit vectors $v$ yields that $\Norm{\mu'}_2 \leq \sqrt{c \eps}$.
Therefore, conditioned on both $E_1$ and $E_2$, we have
\begin{align*}
\Norm{\frac{1}{|S|} \sum_{i \in S} X_i - \mu}_2 &= \frac{n}{|S|} \Norm{\frac{1}{n} \sum_{i = 1}^n Y_i}_2 \\
&\leq \frac{n}{|S|} \Paren{\Norm{\frac{1}{n} \sum_{i = 1}^n Y_i - \mu'}_2 + \Norm{\mu'}_2} \\
&\leq \frac{n}{|S|} \cdot \Paren{ \sqrt{\frac{2d}{n \delta}} + \sqrt{c \eps} } \\
&\leq \frac{1}{1 - \eps}\cdot \Paren{ \sqrt{\frac{2d}{n \delta}} + \sqrt{c \eps} } = \gamma_1 \; . \numberthis \label{eq:cond-mean}
\end{align*}

We now turn our attention to the claimed bound on the covariance.
Let $Y_i$ be as above.
Then, by assumption we have $\Norm{Y_i Y_i^\top}_2 = \Norm{Y_i}_2^2 \leq \frac{d}{c \eps}$, and moreover we have
\[
\E \Brac{\sum_{i = 1}^n Y_i Y_i^\top} \preceq n \Cov \Brac{Y_i} \leq n \cdot \Id \; .
\]
Hence, by Fact~\ref{fact:tropp}, we have that
\begin{equation}
\label{eq:matrix-conc}
\Pr \Brac{ \Norm{\frac{1}{n} \sum_{i = 1}^n Y_i Y_i^\top}_2 > \frac{d (\log d + \log 2 / \delta)}{c' \eps n}} \leq \delta / 2 \; ,
\end{equation}
for some universal constant $c' > 0$.
Let $E_3$ be the event that~\eqref{eq:matrix-conc} holds.
Then, conditioned on both $E_1$ and $E_3$ holding, we have
\begin{align*}
\Norm{\Cov (S)}_2 &= \Norm{\frac{1}{|S|} \sum_{i = 1}^n (Y_i - \mu(S)) (Y_i - \mu_S)^\top}_2 \\
&\stackrel{(a)}{\leq} \Norm{\frac{1}{|S|} \sum_{i = 1}^n Y_i Y_i^\top}_2 \\
&\leq \frac{1}{1 - \eps} \cdot \frac{d (\log d + \log 2 / \delta)}{c' \eps n} = \gamma_2 \; , \numberthis \label{eq:cond-cov}
\end{align*}
as claimed, where (a) follows since centering the second moment matrix can only decrease its top eigenvalue.
Thus,~\eqref{eq:cond-mean} and~\eqref{eq:cond-cov} imply that, conditioned on events $E_1, E_2, E_3$ simultaneously, the set $S$ is $(\gamma_1, \gamma_2)$-good with respect to $D$.
By a union bound, these three events happen simultaneously with probability at least $1 - \delta - \exp(-\eps n)$, as claimed.
\end{proof}


\section{Omitted proofs from Section~\ref{sec:subgaussian}}
\label{sec:app-subgaussian}

\subsection{Proof of Lemma~\ref{lem:conc-gaussian}}

\begin{proof}[Proof of Lemma~\ref{lem:conc-gaussian}]
The two concentration bounds in~\eqref{eq:conc-gaussian} are standard (see e.g.~\cite{rudelson2013hanson}).
Thus in this section, we focus on proving the concentration bounds corresponding to the bounds in~\eqref{eq:conc-gaussian-2}.
Since the two proofs are very similar, we will only prove the bound on $\beta_2$; the bound on $\beta_1$ will follow almost identically by substituting the appropriate Chernoff bound.

Let $X_1, \ldots, X_n$ be i.i.d. from an isotropic subgaussian distribution over $\R^d$ with variance proxy $1$.
Without loss of generality assume that the mean of the distribution is $0$.
It follows from the Hanson-Wright inequality and a standard net argument (see e.g. Lemma 2.1.7 in ~\cite{li2018principled}) that there exist universal constants $A, B > 0$ so that for all $t \geq 0$, we have:
\begin{equation}
\label{eq:hanson-wright}
\Pr \Brac{ \Norm{\frac{1}{n} \sum_{i = 1}^n X_i X_i^\top - \Id}_2 > t } \leq 4 \exp (A d - B n \min(t, t^2)) \; .
\end{equation}
In particular, applying this bound to any fixed $S \subset [n]$ with $|S| = 2 \eps n$ yields that 
\begin{equation}
\label{eq:hanson-wright}
\Pr \Brac{ \Norm{\frac{1}{|S|} \sum_{i \in S} X_i X_i^\top - \Id}_2 > t } \leq 4 \exp (A d - B \eps n \min(t, t^2)) \; ,
\end{equation}
for a different choice of universal constant $B$.
Hence, by a union bound over all $S$ of size $2 \eps n$, we obtain that
\begin{align*}
\Pr \Brac{ \exists S: |S| = 2 \eps n \; \mbox{and} \; \Norm{\frac{1}{|S|} \sum_{i \in S} X_i X_i^\top - \Id}_2 > t } &\leq 4 \exp \Paren{A d + \log \binom{n}{2 \eps n} - B \eps n \min(t, t^2)} \\
&\leq 4 \exp \Paren{Ad + n H(2 \eps) - B \eps n \min(t, t^2)} \; ,
\end{align*}
where $H(\alpha)$ is the binary entropy of $\alpha$.
For $\eps \leq 1/4$, there exists a universal constant $C$ so that $H(2 \eps) \leq C \eps \log 1 / \eps$.
Plugging this in, and by our definition of $\beta_2$, we obtain that
\begin{align*}
\Pr \Brac{ \exists S: |S| = 2 \eps n \; \mbox{and} \; \Norm{\frac{1}{|S|} \sum_{i \in S} X_i X_i^\top - \Id}_2 > \beta_2 } &\leq 4 \exp \Paren{Ad + \eps n (C \log 1 / \eps - B \gamma_2)} \\
&\leq 4 \exp \Paren{Ad - C' \eps n \gamma_2} \\
&\leq \delta \; ,
\end{align*}
as claimed.
To obtain the corresponding bound for $\beta_1$, simply use a Chernoff style bound (e.g. Lemma 2.1.6 in~\cite{li2018principled}) instead of~\eqref{eq:hanson-wright}.
\end{proof}

\end{document}